\documentclass[12pt]{article}
\usepackage{geometry}
\date{}
\usepackage{times}
\usepackage{float}
\usepackage{multirow}
\usepackage{balance}
\usepackage{graphicx}
\usepackage{booktabs} 
\usepackage{url}
\usepackage{subfigure}
\usepackage{algorithm}
\usepackage[noend]{algpseudocode}
\usepackage{xcolor}
\usepackage{hyperref}
\hypersetup{colorlinks,linkcolor={black},citecolor={blue},urlcolor={black}}
\usepackage{amssymb}
\usepackage{amsmath}
\usepackage{amsfonts}
\usepackage{amsthm}
\usepackage{mathptmx}
\usepackage{textcomp}
\usepackage{caption}
\usepackage{wrapfig}
\usepackage{setspace}
\usepackage{nccmath}
\usepackage[shortlabels]{enumitem}
\setenumerate[1]{itemsep=0pt,partopsep=0pt,parsep=\parskip,topsep=1pt}
\setitemize[1]{itemsep=0pt,partopsep=0pt,parsep=\parskip,topsep=1pt}
\setdescription{itemsep=0pt,partopsep=0pt,parsep=\parskip,topsep=1pt}

\newtheorem{definition}{\bf Definition}

\newtheorem{example}{\bf Example}

\newcommand{\ExpHead}[1]{\vspace*{4pt}\noindent\textbf{#1}\vspace*{2pt}}
\newcommand{\tabincell}[2]{\begin{tabular}{@{}#1@{}}#2\end{tabular}}
\newcommand\ExpCaption[1]{%
     \captionsetup{font=small}%
     \caption{#1}}

\newcommand{\change}[1]{{#1}}
\begin{document}
\title{An Experimental Analysis of Indoor Spatial Queries: Modeling, Indexing, and Processing}

\author{
\fontsize{12}{12}\selectfont Tiantian Liu$^\dag$ \hspace{0.1cm} Huan Li$^\dag$ \hspace{0.1cm} Hua Lu$^\ddag$ \hspace{0.1cm} Muhammad Aamir Cheema$^\S$ \hspace{0.1cm} Lidan Shou$^\P$ \\
\vspace{-2mm}
\fontsize{10}{10}\selectfont $^\dag$Department of Computer Science, Aalborg University, Denmark\\
\vspace{-2mm}
\fontsize{10}{10}\selectfont $^\ddag$Department of People and Technology, Roskilde University, Denmark\\
\vspace{-2mm}
\fontsize{10}{10}\selectfont $^\S$Faculty of Information Technology, Monash University, Australia\\
\vspace{-2mm}
\fontsize{10}{10}\selectfont $^\P$College of Computer Science, Zhejiang University, China\\
\vspace{-2mm}
\fontsize{8}{8}\selectfont\ttfamily\upshape \{liutt,lihuan\}@cs.aau.dk, luhua@ruc.dk, aamir.cheema@monash.edu, should@zju.edu.cn
}
\maketitle

\textbf{Abstract}---Indoor location-based services (LBS), such as POI search and routing, are often built on top of typical indoor spatial queries. To support such queries and indoor LBS, multiple techniques including model/indexes and search algorithms have been proposed. In this work, we conduct an extensive experimental study on existing proposals for indoor spatial queries. We survey five model/indexes, compare their algorithmic characteristics, and analyze their space and time complexities.
We also design an in-depth benchmark with real and synthetic datasets, evaluation tasks and performance metrics. Enabled by the benchmark, we obtain and report the performance results of all model/indexes under investigation. By analyzing the results, we summarize the pros and cons of all techniques and suggest the best choice for typical scenarios.

\section{Introduction}
\label{sec:intro}


Thanks to the recent advances in indoor localization and high penetration of smartphones, indoor location-based services (LBS) are becoming increasingly popular~\cite{Basiri2017, cheema2018indoor}. Indoor LBS applications, such as POI search~\cite{Lu2016,Li2018dense} and routing~\cite{Goetz2011,Costa2019,Feng2019}, are often built on top of typical spatial queries like range query, $k$ nearest neighbor query, shortest path query, and shortest distance query. Therefore, the efficiency of processing such typical indoor spatial queries plays a key role in the success of indoor LBS.

To facilitate spatial query processing for indoor LBS, multiple techniques have been proposed, including models and indexes for indoor spaces and query processing algorithms. All such proposals deal with indoor space entities such as rooms, doors, walls, and floors. Such indoor entities form distinct indoor topology that determines indoor distances and impacts indoor movement. As a result, the distances in indoor spatial queries must be measured appropriately, e.g., without involving straight line segments through walls. Also, indoor routing in shortest path/distance queries must consider connectivity and reachability between indoor locations.

To support distance computations with respect to indoor locations, existing models and indexes~\cite{Lu2012, Xie2013, Xie2015, Shao2017} employ different approaches to integrate the geometry and topology information of an indoor space. All these techniques can be used to process the aforementioned four types of indoor spatial queries. However, a comprehensive experimental study on all these proposals is still missing. Consequently, indoor LBS application developers inevitably encounter difficulties in choosing the appropriate technique for a given indoor space scenario.

To bridge this gap for LBS application development and disclose insights for further research on indoor data management, we conduct a comprehensive experimental study in this work. Our study focuses on five existing model/indexes that support typical indoor spatial queries on static indoor objects (e.g., POIs) or indoor shortest paths/distances.
%
We compare the five proposals theoretically and empirically. Our contributions are as follows.
\begin{itemize}[leftmargin=*]
    \item We survey the five proposals by scrutinizing their structures, algorithmic characteristics, and space and time complexities. 
    \item We design an in-depth benchmark with datasets, evaluation tasks, and performance metrics. The datasets consist of real and synthetic data characterized by distinctive indoor topology.
    \item Within the same framework, we conduct extensive experiments to evaluate the performance of the five proposals in terms of construction cost and query efficiency.
    \item By analyzing the results, we disclose the pros and cons of the proposals, analyze the impact of different conditions, and recommend the best choice for typical application scenarios.
\end{itemize}



The paper is organized as follows.
Section~\ref{sec:background} gives the preliminaries, defines indoor spatial queries, and briefly reviews related work.
Sections~\ref{sec:indexes} and~\ref{sec:queryProcessing} present the indoor space model/indexes and spatial query processing, respectively.
Section~\ref{sec:benchmark} proposes a benchmark for the experimental evaluation.
Section~\ref{sec:results} reports and analyzes the evaluation results.
Section~\ref{sec:extensibility} discusses the extensibility of all model/indexes. Section~\ref{sec:summary} concludes the paper. 

\section{Indoor Spatial Queries}
\label{sec:background}

Table~\ref{tab:notations} lists the frequently used notations.

\begin{table}[!htbp]
\centering
\caption{Notations}\label{tab:notations}
\footnotesize
\begin{tabular}{||l|l||}
\hline
\textbf{Symbol} & \textbf{Meaning}\\
\hline
$\mathbb{I}$ & An indoor space\\
$p, q \in \mathbb{I}$ & Indoor points\\
$o \in O$ & A static indoor object \\
$d \in D$ & A door \\
$v \in V$ & An indoor partition \\
$|p,q|_I$ & Indoor distance from $p$ to $q$ \\
$\phi = \langle p, d_i, \ldots, d_j, q \rangle$ & An indoor path \\
$L(\phi)$ & Length of a path $\phi$\\
\hline
\end{tabular}
\end{table}

\subsection{Indoor Space Concepts}
\label{ssec:concepts}

\begin{wrapfigure}{r}{10cm}
 \centering
 \includegraphics[width=0.5\columnwidth]{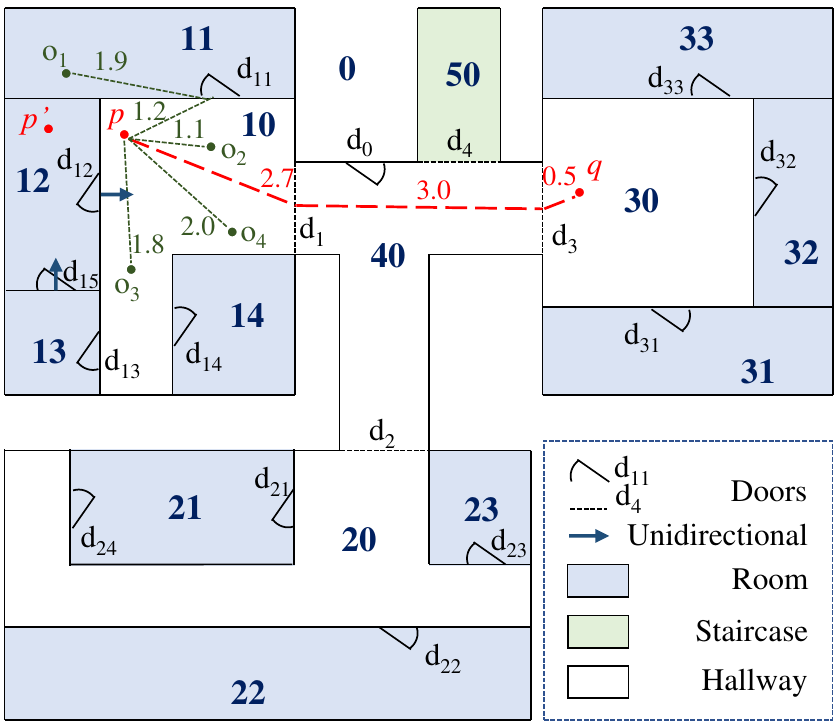}
 \caption{Example Floorplan}
 \label{fig:floorplan}
\end{wrapfigure}

Indoor space features distinct entities such as walls, doors, and rooms, which altogether form complex indoor topology that enables and constrains movements.
%
%
Naturally, an indoor space is divided by walls and doors into \textbf{indoor partitions} like rooms, hallways or staircases.
Two indoor partitions can be connected by a door or an open segment between them.
Referring to the example floorplan in Figure~\ref{fig:floorplan}, partitions 30 and 40 (denoted as $v_{30}$ and $v_{40}$, respectively) are connected by an open segment $d_{3}$,
In this paper, we refer to both doors and open segments as doors.
We do not consider the width of a door and represent a door by its center point.
In other words, each door can be generally regarded as an indoor point.
Furthermore, a door can be unidirectional such as a security checkpoint at the airport.
The door directionality makes the indoor distance between two points asymmetric.
Referring to Figure~\ref{fig:floorplan}, the shortest indoor path from $p$ to $p'$ and that from $p'$ to $p$ are different due to the unidirectionality of $d_{12}$.

Topology renders the indoor distance more complex than Euclidean distance. In Figure~\ref{fig:floorplan}, the indoor distance $|p,o_1|_I$ from $p$ to $o_1$ is not subject to the straight line segment between them; it is the total length of the polyline $p \rightarrow d_{11} \rightarrow o_1$.

Lu et al.~\cite{Lu2012} proposes mappings to capture the relationships between indoor partitions and doors.
In particular, $\mathit{D2P}_\sqsupset(d_i)$ gives the set of partitions that one can enter through door $d_i$ and $\mathit{D2P}_\sqsubset(d_j)$ gives those that one can leave through door $d_j$.
Besides, $\mathit{D2P}(d_i)$ gives a set of a partition pair $(v_j, v_k)$ such that one can go through door $d_i$ from partition $v_j$ to $v_k$.
Moreover, $\mathit{P2D}_\sqsupset(v_k)$ gives the set of enterable doors through which one can enter partition $v_k$, and $\mathit{P2D}_\sqsubset(v_k)$ gives the set of leaveable doors through which one can leave partition $v_k$.
When the door directionality is not relevant, we use $\mathit{P2D}(v_k) = \mathit{P2D}_\sqsupset(v_k) \cup \mathit{P2D}_\sqsubset(v_k)$ to denote the set of doors associated to partition $v_k$.

\begin{example}
  Referring to Figure~\ref{fig:floorplan}, given the unidirectional door $d_{12}$, we have $\mathit{D2P}_\sqsupset(d_{12}) = \{v_{10}\}$, $\mathit{D2P}_\sqsubset(d_{12}) = \{v_{12}\}$, and $\mathit{D2P}(d_{12}) = \{(v_{12}, v_{10})\}$.
  Moreover, we have $\mathit{P2D}_\sqsupset(v_{12}) = \{d_{15}\}$, $\mathit{P2D}_\sqsubset(v_{12}) = \{d_{12}\}$, and $\mathit{P2D}(v_{12}) = \{d_{15}, d_{12}\}$.
\end{example}

\subsection{Indoor Spatial Query Types}

In our study, we focus on \emph{static} indoor objects such as POIs and facilities. Our experimental study covers four fundamental indoor spatial query types as follows.

\begin{definition}[\bf Range Query (RQ)]
 Given an indoor point $p \in \mathbb{I}$, a set $O$ of indoor objects, and a distance value $r$, a \emph{range query} $\textsf{RQ}(p, r)$ returns all indoor objects from $O$ whose indoor distance from $p$ is within $r$. Formally, $\textsf{RQ}(p, r) = \{ o \mid |p, o|_{I} \leq r, o \in O \}$.
\end{definition}

\begin{definition}[\bf$k$ Nearest Neighbor Query ($k$NNQ)]
 Given an indoor point $p \in \mathbb{I}$, a set $O$ of indoor objects, and an integer value $k$, a \emph{$k$ nearest neighbor query} $k\textsf{NNQ}(p)$ returns a set $O'$ of $k$ indoor objects whose indoor distances from $p$ are the smallest, i.e., $|O'| = k$ and $\forall o_i \in O', o_j \in O \setminus O', |p, o_i|_I \leq |p, o_j|_I$.
\end{definition}

Referring to Figure~\ref{fig:floorplan} where $O= \{ o_1, \ldots, o_4 \}$, a query $\textsf{RQ}(p, 1.9\text{m})$ returns $\{o_2, o_3 \}$ since the distances from $p$ to $o_1$ and $o_4$ both exceed $1.9$m.\footnote{Meter is the distance unit in all examples in this paper.}
Further, a query $3\textsf{NNQ}(p)$ returns $\{ o_2, o_3, o_4 \}$, since $o_1$'s distance from $p$ is the longest among all.

\begin{definition}[\bf Shortest Path Query (SPQ)]
Given a source point $p \in \mathbb{I}$, a target point $q \in \mathbb{I}$, a \emph{shortest path query} $\textsf{SPQ}(p, q)$ returns the shortest path $\phi = \langle p, d_i, \ldots, d_j, q \rangle$ from $p$ to $q$ such that 1) $d_i, \ldots, d_j$ are door sequences and each two consecutive doors are associated to the same partition,
2) $p$ is in the partition having $d_i$ as a leavable door,
3) $q$ is in the partition having $d_j$ as an enterable door, and
4) $\forall \phi'$ from $p$ to $q$, $L(\phi) \leq L(\phi')$.\footnote{$L(\phi) = \Sigma_{k=0}^{k=j}|d_k, d_{k+1}|_I$ where $d_0=p$ and $d_{j+1}=q$.}
\end{definition}

\begin{definition}[\bf Shortest Distance Query (SDQ)]
 Given a source point $p \in \mathbb{I}$, a target point $q \in \mathbb{I}$, a \emph{shortest distance query} $\textsf{SDQ}(p, q)$ returns the shortest indoor distance from $p$ to $q$, i.e., the length of $\textsf{SPQ}(p, q)$.
\end{definition}

As indicated by the red dashed polyline in Figure~\ref{fig:floorplan}, a query $\textsf{SPQ}(p, q)$ returns $\phi = \langle p, d_1, d_3, q \rangle$ as the shortest path from $p$ to $q$, and the result of $\textsf{SDQ}(p, q)$ is $2.7$m + $3.0$m + $0.5$m = $6.2$m.

\subsection{Related Work}
\label{ssec:related}

Multiple studies~\cite{Lee2004, Whiting2007, Becker2009, Jensen2009, Worboys2011} have been proposed to model indoor spaces, focusing on symbolic modeling of topological relationships between 3D spatial cells or space partitions.
However, these works do not support indoor distances and thus cannot process the distance-aware queries evaluated in this study.


Other studies focus on querying indoor moving objects.
In the context of RFID indoor tracking, Yang et al. study continuous range monitoring queries~\cite{Yang2009} and probabilistic $k$ nearest neighbor queries~\cite{Yang2010}.
To improve the query result, Yu et al.~\cite{Yu2013} propose a particle filter-based method to infer the undetected locations indoor moving objects.
Assuming a probabilistic sample based location data format, Xie et al.~\cite{Xie2013, Xie2015} process $k$NN query and range query for indoor moving objects.
Considering the uncertain object movements between observed time and query time, Li et al.~\cite{Li2018dense} study the search of the current top-$k$ indoor dense regions.
These works consider indoor moving objects with uncertain positions at a particular time.
In contrast, the range and $k$NN queries evaluated in this study concern static indoor objects.

Jensen et al.~\cite{Jensen2009} study historical trajectories of RFID-tracked indoor objects.
Delafontaine et al.~\cite{Delafontaine2012} find sequential visiting patterns within historical Bluetooth tracking data.
Given a past time or a time interval, Lu et al. define spatio-temporal joins~\cite{Lu2011} to find moving object pairs in the same indoor partition, and top-$k$ queries~\cite{Lu2016} to find the most frequently visited indoor POIs.
Ahmed et al.~\cite{Ahmed2014, Ahmed2017} define threshold density query to find dense indoor semantic locations in a historical time interval.
Assuming probabilistic sample based location records, Li et al.~\cite{Li2019flow} find the top-$k$ most popular indoor semantic regions with the highest object flow values.
Unlike these works, the queries studied in this paper focus on static objects or indoor paths.

Shortest path/distance queries have been studied in indoor contexts.
Goetz and Zipf~\cite{Goetz2011} study user-adaptive length-optimal indoor routing based on a weighted routing graph.
Salgado et al.~\cite{Salgado2018} study indoor keyword-aware skyline route query, considering the number of covered keywords and route distances.
Feng et al.~\cite{Feng2019} study indoor keyword-aware routing queries to find shortest paths covering user-specified semantic keywords.
Costa et al.~\cite{Costa2019} propose the context-aware indoor-outdoor path recommendation that minimizes the outdoor exposure and path distance.
These techniques consider additional query semantics, and thus are different from the fundamental, pure shortest path/distance queries studied in this paper.

\section{Model and Indexes}
\label{sec:indexes}

The aforementioned indoor spatial queries all involve indoor distances. To facilitate such queries, indoor distances must be considered in modeling and indexing indoor space.

\subsection{Indoor Distance-Aware Model}
\label{ssec:IDModel}

Indoor distance-aware model~\cite{Lu2012} (\textsc{IDModel}) is a graph $G_\text{dist}$ $(V, E_a, L, f_\text{dv}, f_\text{d2d})$. The first three elements capture indoor topology in an \emph{accessibility base graph} $G_\text{accs}(V, E_a, L)$, where $V$ is the set of vertexes each referring to an indoor partition, $E_a = \{ (v_i, v_j, d_k) \mid d_k \in D, v_i \in \mathit{D2P}_\sqsupset(d_k) \wedge v_j \in \mathit{D2P}_\sqsubset(d_k) \}$ is a set of labeled, directed edges, and $L$ is the set of edge labels each corresponding to a door in $D$. The additional two are mapping functions defined as follows.
\begin{equation}\label{equation:fdv}
  f_\text{dv}(d_i, v_j) =
  \begin{cases}
    \max_{p \in v_j} ||d_i, p||_{v_j}, & \text{if~}v_j \in \mathit{D2P}_\sqsupset(d_i); \\
    \infty,                            & \text{otherwise}.
  \end{cases}\nonumber
\end{equation}
Here, $||p, q||_{v_j}$ is the indoor distance from a point $p$ to a point $q$ within the partition $v_j$.
Note that $||p, q||_{v_j}$ is not necessarily a Euclidean distance because even within the same partition there may be obstacles in the line of sight between $p$ and $q$.
Specifically, \emph{door-to-partition distance mapping}  $f_\text{dv}(d_i, v_j)$ returns the longest distance one can reach within partition $v_j$ from door $d_i$, if $v_j$ is an enterable partition of $d_i$. Otherwise, it returns $\infty$.

\begin{equation}\label{equation:fd2d}
  f_\text{d2d}(v_j, d_i, d_j) =
  \begin{cases}
    ||d_i, d_j||_{v_j}, & \text{if~}d_i \in \mathit{P2D}_\sqsupset(v_j) \\
                        & \text{and~}d_j \in \mathit{P2D}_\sqsubset(v_j); \\
    0,                  & \text{if~}d_i = d_j\\
                        & \text{and~} d_i, d_j \in \mathit{P2D}(v_j); \\
    \infty,             & \text{otherwise}.
  \end{cases} \nonumber
\end{equation}

The \emph{door-to-door distance mapping} $f_\text{d2d}(v_j, d_i, d_j)$ maps a partition $v_j$ and two doors $d_i$ and $d_j$ to a distance value.
If both doors are associated to $v_j$, it returns the distance from $d_i$ to $d_j$ within $v_j$, i.e., $||d_i, d_j||_{v_j}$.
If $d_i$ and $d_j$ are identical and associated to $v_j$, we stipulate $f_\text{d2d}(v_j, d_i, d_j) = 0$.
Otherwise, $f_\text{d2d}(v_j, d_i, d_j)$ returns $\infty$, indicating that one cannot go from $d_i$ to $d_j$ via $v_j$ only.

\begin{figure}[htbp]
 \centering
 \includegraphics[width=0.8\columnwidth]{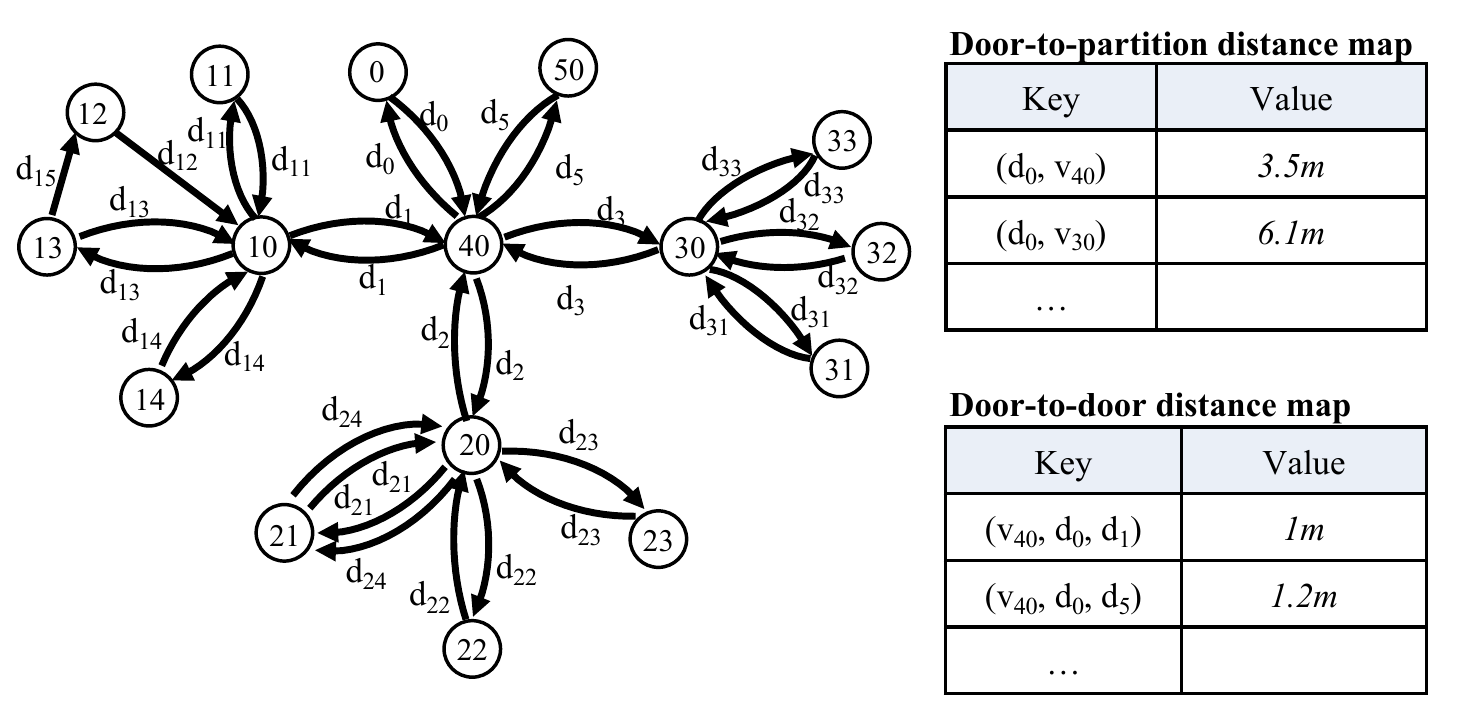}
 \caption{An Example of \textsc{IDModel}}
 \label{fig:IDModel}
\end{figure}

Figure~\ref{fig:IDModel} illustrates the \textsc{IDModel} for the example shown in Figure~\ref{fig:floorplan}.
The outdoor space is captured in a special graph vertex $v_0$. Two hashmaps implement the mappings $f_\text{dv}(d_i, v_j)$ and $f_\text{d2d}(v_j, d_i, d_j)$.
With directed edges, \textsc{IDModel} can support doors' directionality and temporal variation when needed.

%
With the two mappings $f_\text{dv}(d_i, v_j)$ and $f_\text{d2d}(v_k, d_i, d_j)$, a graph traversal algorithm~\cite{Lu2012} on \textsc{IDModel} is designed to compute the shortest door-to-door distance $\text{d2d}(d_s, d_t)$ from a source door $d_s$ to a target door $d_t$. The basic idea is to keep expanding to unvisited doors based on the current shortest path until reaching the target door.
Further, the shortest indoor distance from any point $p$ to any point $q$ can be computed by finding the minimum value of the distance summation $||p, d_p||_{v_p} + \text{d2d}(d_p, d_q) + ||d_q, q||_{v_q}$, where $v_p$ and $v_q$ are the partitions that host $p$ and $q$, respectively, $d_p \in \mathit{P2D}_\sqsubset(v_p)$, and $d_q \in \mathit{P2D}_\sqsupset(v_q)$.

However, \textsc{IDModel} does not support fast determination of the host partition of a query/source point. It boils down to sequential scanning of all partitions if no additional index, e.g., R-tree, is used for the partitions.
Also, to manage indoor static objects, \textsc{IDModel} needs additional object buckets each for a partition.

\subsection{Indoor Distance-Aware Index}
\label{ssec:IDIndex}

\textsc{IDModel} only captures the door-to-door and door-to-partition distances within a local partition, which entails extra search to compute the indoor distance for two points in different partitions.

To cut such costs, indoor distance-aware index~\cite{Lu2012} (\textsc{IDIndex}) stores extra information on top of \textsc{IDModel}, namely, precomputed global door-to-door distances and their ordering in two matrices.
The \textbf{door-to-door distance matrix} $M_\text{d2d}$ is an $\mathsf{N}$-by-$\mathsf{N}$ matrix where $\mathsf{N} = |D|$ is the total number of doors and $M_\text{d2d}[d_i, d_j]$ gives the precomputed shortest indoor distance from $d_i$ to $d_j$.
The \textbf{distance index matrix} $M_\text{idx}$ is also an $\mathsf{N}$-by-$\mathsf{N}$ matrix such that $M_\text{idx}[d_i, k]$ gives the identifier of a door whose indoor distance from $d_i$ is the $k$-th shortest among all the $N$ doors.

\begin{figure}[htbp]
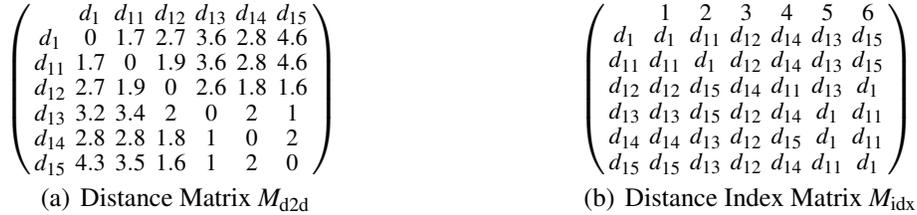

 \centering
 \subfigure[Distance Matrix $M_\text{d2d}$]{
  \begin{minipage}{0.47\columnwidth}
   \centering
   $\begin{pmatrix}
   \begin{smallmatrix}
                 &  d_{1} &  d_{11} &  d_{12} &  d_{13} & d_{14} & d_{15} \\
         d_{1}     &  0   &  1.7 &  2.7 &  3.6 & 2.8 & 4.6 \\
         d_{11}    &  1.7 &  0   &  1.9 &  3.6 & 2.8 & 4.6 \\
         d_{12}    &  2.7 &  1.9 &  0   &  2.6 & 1.8 & 1.6 \\
         d_{13}    &  3.2 &  3.4 &  2   &  0   & 2   & 1   \\
         d_{14}    &  2.8 &  2.8 &  1.8 &  1   & 0   & 2   \\
         d_{15}    &  4.3 &  3.5 &  1.6 &  1   & 2   & 0   \\
   \end{smallmatrix}
   \end{pmatrix}$
   \vspace*{4pt}
  \end{minipage}
 }
 \subfigure[Distance Index Matrix $M_\text{idx}$]{
  \begin{minipage}{0.47\columnwidth}
   \centering
   $\begin{pmatrix}
   \begin{smallmatrix}
                 &  1 &  2 &  3 &  4 & 5 & 6 \\
         d_{1}     &  d_{1}   &  d_{11} & d_{12} & d_{14} & d_{13} & d_{15}    \\
         d_{11}    &  d_{11}  &  d_{1}  &  d_{12} & d_{14} & d_{13} & d_{15}    \\
         d_{12}    &  d_{12} &  d_{15}  &  d_{14} & d_{11} & d_{13} & d_{1}    \\
         d_{13}    &  d_{13} &  d_{15}  &  d_{12}   & d_{14}   & d_{1}   &  d_{11}  \\
         d_{14}    &  d_{14} &  d_{13} &  d_{12}   & d_{15}   & d_{1}   &  d_{11}  \\
         d_{15}    &  d_{15} &  d_{13} &  d_{12} &  d_{14}   & d_{11}   &  d_{1}  \\
   \end{smallmatrix}
   \end{pmatrix}$
   \vspace*{4pt}
  \end{minipage}
 }
 \caption{An Example of \textsc{IDIndex}}\label{fig:IDIndex}
\end{figure}

The \textsc{IDIndex} matrices for the top-left part in Figure~\ref{fig:floorplan} is illustrated in Figure~\ref{fig:IDIndex}. Here, we have $M_\text{d2d}[d_1, d_{15}] = 4.6$m.
The first row of $M_\text{d2d}$ shows that $d_{15}$ has the longest indoor distance from $d_{1}$.
Accordingly, we have $M_\text{idx}[d_1, 6] = d_{15}$ in $M_\text{idx}$.\\

As the shortest indoor distances to all doors are precomputed and sorted for each door in \textsc{IDIndex}, it is faster to compute the shortest indoor distance between any two points $p$ and $q$ in the indoor space.
To support the shortest path query, in addition to the shortest distance value between any two points, \textsc{IDIndex} also keeps the first-hop door of the corresponding shortest path.
In this way, the complete shortest path between two points can be constructed by recursively concatenating the first-hop doors.


\subsection{Composite Indoor Index}
\label{ssec:CIndex}

\begin{figure}[htbp]
 \centering
 \includegraphics[width=0.65\columnwidth]{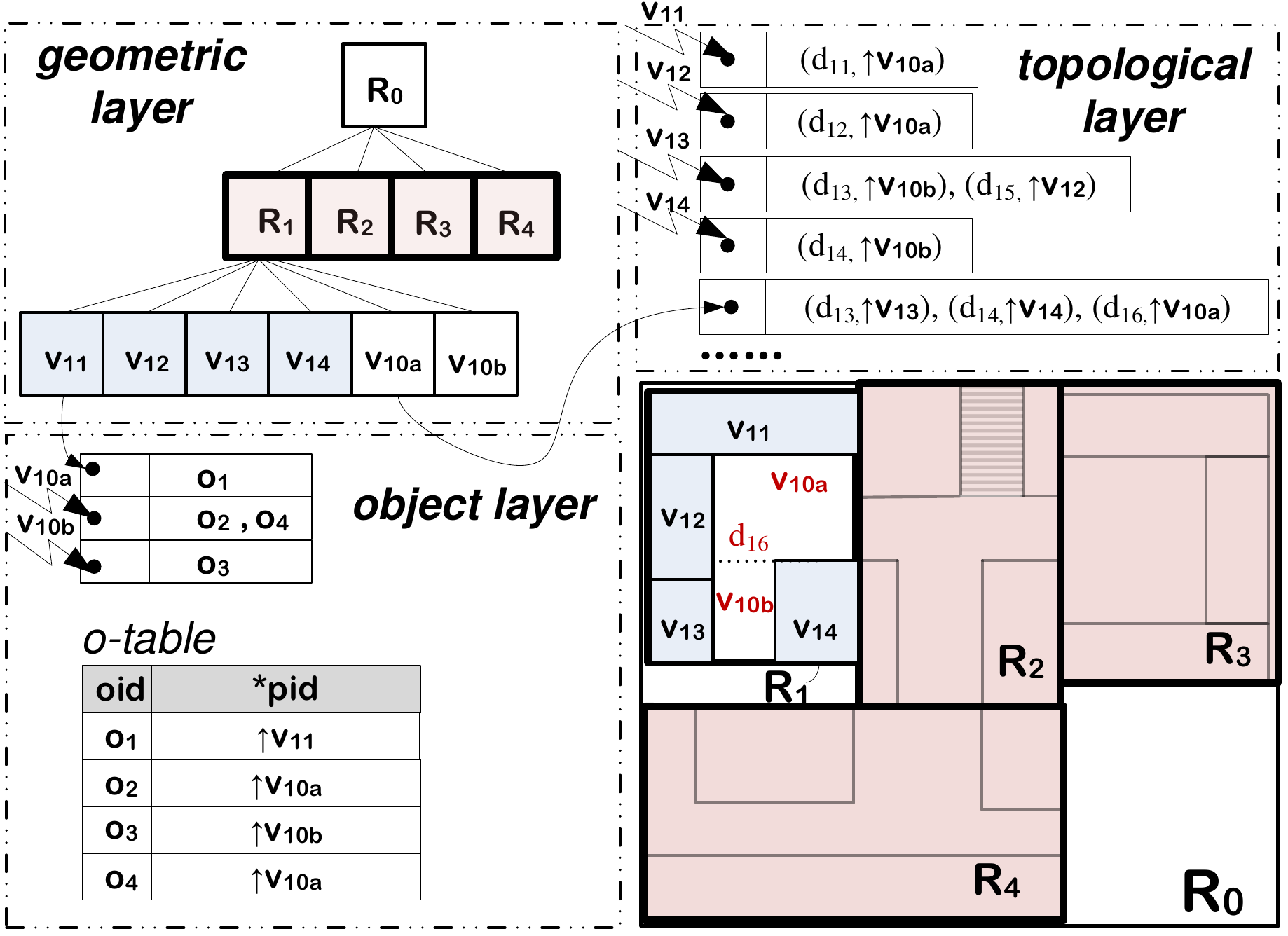}
 \caption{An Example of \textsc{CIndex} (Adapted from~\protect\cite{Xie2013})}
 \label{fig:CIndex}
\end{figure}

Composite indoor index~\cite{Xie2013} (\textsc{CIndex}) is a layered structure for indexing indoor partitions and moving objects.
It consists of three layers: geometric layer, topological layer, and object layer. In this study, we adapt the object layer to index static indoor objects.
A partial example \textsc{CIndex} for Figure~\ref{fig:floorplan} is given in Figure~\ref{fig:CIndex}.

The \textbf{geometric layer} uses an R*-tree~\cite{Beckmann1990} to index all indoor partitions, with an additional skeleton tier to maintain the distances between staircases at different floors.
To ease the geometrical computations, it decomposes each irregular partition\footnote{A partition is irregular if it is non-convex or imbalanced (long in one dimension but short in the other).} into regular ones using a decomposition algorithm~\cite{Xie2013}.
Referring to the bottom-right of Figure~\ref{fig:CIndex}, the hallway $v_{10}$ is divided into two regular indoor partitions $v_{10a}$ and $v_{10b}$ by a door $d_{16}$.
Afterwards, each regular partition is represented by a \emph{Minimum Bounding Rectangle} (MBR). The MBRs are indexed by the R*-tree.
As shown in the top-left of Figure~\ref{fig:CIndex}, a non-leaf node $R_{1}$ is composed of six partitions in the leaf level, i.e., $v_{10a}$, $v_{10b}$, and $v_{11}$-$v_{14}$.

The \textbf{topological layer} stores the connectivity information among indoor partitions, and it is integrated to the tree by inter-partition links.
In particular, a leaf node $v_i$ in the R*-tree is linked with a pointer record $(d_k, {\uparrow}v_j)$ to indicate that one can move from a partition $v_i$ to another partition $v_j$ through door $d_k$.
As shown in the top-right of Figure~\ref{fig:CIndex}, the two pointer records for $v_{13}$ mean that $v_{13}$ is adjacent to $v_{10b}$ and $v_{12}$ via $d_{13}$ and $d_{15}$, respectively.

The \textbf{object layer} maintains a number of object buckets each for an indoor partition at the leaf node level of the R*-tree. Each indoor object $o$ is kept in the bucket of the partition in which $o$ is located.
In addition, an object hashtable $\text{o-table}: O \rightarrow {}^{*}V$ maps each object to its host partition's pointer.
Unlike~\cite{Xie2013, Xie2015}, the object buckets store static objects in this study.
As shown in the bottom-left of Figure~\ref{fig:CIndex}, the leaf node $v_{10a}$ is linked to its object bucket with two static objects $o_2$ and $o_4$. Also, two corresponding records are kept in the object hashtable (o-table).

The R*-tree in \textsc{CIndex} organizes partitions hierarchically, and thus enables search space pruning for distance relevant computations. As a result, \textsc{CIndex} does not cache the precomputed door-to-door distances as \textsc{IDIndex} does. Moreover, as the topological layer maintains the links between partitions and doors, which form an implicit graph structure, \textsc{CIndex} does not need an explicit graph model to keep connectivity information.
The topological layer's dynamic link updating makes \textsc{CIndex} adaptive to possible temporal variations of doors.\\

\subsection{IP-Tree and VIP-Tree}
\label{ssec:IP-Tree}

Indoor partitioning tree~\cite{Shao2017} (\textsc{IP-Tree}) is a tree-based indoor partition index with a number of matrices each materializing the door-to-door distances within a local range.
In particular, each leaf node of \textsc{IP-Tree} covers a number of topologically adjacent indoor partitions. The adjacent leaf nodes are combined to form a non-leaf node, and adjacent non-leaf nodes are combined hierarchically until a root node is formed.
Each node $N$ has a \textbf{distance matrix} and a number of \textbf{access doors}.
An access door is a border door that connects $N$ to its external space.~$\text{AD}(N)$ denotes $N$'s access door set.
The distance matrix for a leaf node stores the shortest distance (as well as the first-hop door on the shortest path) between every door of the leaf node to every access door of the leaf node. The distance matrix for a non-leaf node only stores the shortest distances and first-hop door between each pair of access doors of its child nodes.
To compute the indoor distance from a point $p$ to a point $q$, \textsc{IP-Tree} locates the lowest common ancestor of the leaf nodes \texttt{Leaf}$(p)$ and \texttt{Leaf}$(q)$, finds the access doors constituting the shortest path in that ancestor, and connects the materialized indoor distances involving $p$, the found access doors, and $q$.

\begin{figure}[htbp]
 \centering
 \includegraphics[width=0.8\columnwidth]{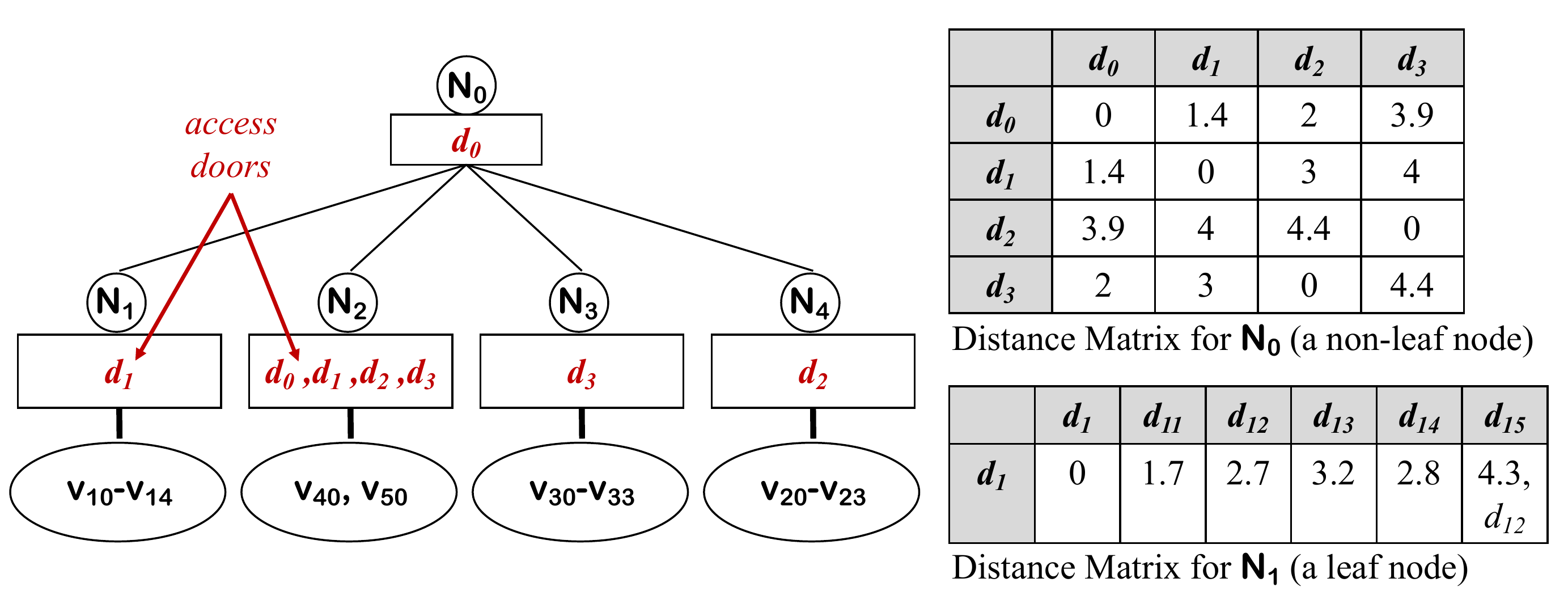}
 \caption{An Example of \textsc{IP-Tree}}
 \label{fig:IP-Tree}
\end{figure}

Figure~\ref{fig:IP-Tree} shows an example of \textsc{IP-Tree} corresponding to Figure~\ref{fig:floorplan}.

The topologically adjacent partitions $v_{10}$-$v_{14}$ form a leaf node $N_1$.
Another leaf node $N_2$ is composed of partitions $v_{40}$ and $v_{50}$.
As $N_1$ and $N_2$ are connected by a border door $d_1$, $d_1$ is put into $\text{AD}(N_1)$ and $\text{AD}(N_2)$.
For the leaf node $N_1$, the distance matrix stores the distances from each of its doors to the access door $d_1$ of $N_1$.
For instance, the distance from $N_1$'s only door $d_{15}$ to access door $d_1$ contained by $N_1$ is 4.3m.
Moreover, as the shortest path from $d_{15}$ to $d_1$ is $\langle d_{15}, d_{12}, d_1 \rangle$, the first-hop door of the path is kept as $d_{12}$ in the matrix.
Differently, for the non-leaf node $N_0$, the distance matrix only keeps the distances between each pair of access doors.
In the running example, each pair of access doors are directly connected. Therefore, no first-hop door is recorded. The storage space of each distance matrix will double when the door directionality needs to be considered.
In other words, both the distances $\text{d2d}(d_i, d_j)$ and $\text{d2d}(d_j, d_i)$ are kept in each node.

As a variant of \textsc{IP-Tree}, vivid IP-Tree (\textsc{VIP-Tree})~\cite{Shao2017} further accelerates the distance computation by materializing more precomputed information.
Specifically, each leaf node $N$ additionally maintains the shortest distance between each door contained by $N$ and each access door in $N$'s all ancestor nodes, along with the corresponding first-hop door information.

\textsc{IP-Tree} and \textsc{VIP-Tree} materialize a small number of distances only related to access doors that are critical in the overall topology of an indoor space. This design eases the on-the-fly distance related computations in spatial query processing.

\section{Query Processing}
\label{sec:queryProcessing}

All the aforementioned model/indexes can be used to process indoor spatial queries. Although query processing differs for different query types, all algorithms share a general paradigm as follows. First, an algorithm finds the initial indoor partition for a query. The initialization decides the indoor partition in which the query (or source) point $p$ is located for a given $\textsf{RQ}(p, r)$ ($k\textsf{NNQ}(p)$, $\textsf{SPQ}(p, q)$, or $\textsf{SDQ}(p, q)$).
Subsequently, an algorithm expands from the initial partition, searching adjacent partitions via doors. Finally, the expansion stops when the search range is beyond the query range $r$ for a $\textsf{RQ}(p, r)$, or $k$NNs have been found for a $k\textsf{NNQ}(p)$, or the target point $q$ is met for a $\textsf{SPQ}(p, q)$ or $\textsf{SDQ}(p, q)$.
Algorithms based on different model/indexes differ in their initializations and expansions. Below, we present a comprehensive analytical comparison of all model/indexes.

\subsection{Algorithmic Comparison}
\label{sec:feature_analysis}

Table~\ref{tab:feature_comparison} summarizes the comparison.

\begin{table}
\centering
\caption{Feature Comparison}\label{tab:feature_comparison}
\footnotesize
\resizebox{15.5cm}{1.4cm} {
\begin{tabular}{lllllllll}
\toprule
\textbf{Models} & \textbf{Precomputation} & \textbf{Structure} & \textbf{Initialization} &\textbf{Expansion} &  {\textsf{RQ}} & {$k$\textsf{NNQ}} & {\textsf{SPQ}} & {\textsf{SDQ}}\\
\midrule
\textsc{IDModel} & No & Graph+Mappings & Sequential scan & Dijkstra & $\triangle$ & $\triangle$ & \checkmark & \checkmark\\
\textsc{IDIndex} & Yes & Matrix & Sequential scan & Loop & \checkmark & \checkmark & $\triangle$ & $\triangle$ \\
\textsc{CIndex} & No & Tree+Links & R*-Tree pruning & Dijkstra  &\checkmark & \checkmark & $\triangle$ & $\triangle$\\
\textsc{IP-Tree} & Yes & Tree+Matrix & Sequential scan & LCA &  \checkmark & \checkmark & \checkmark & \checkmark\\
\textsc{VIP-Tree} & Yes & Tree+Matrix & Sequential scan & LCA &   \checkmark & \checkmark & \checkmark & \checkmark\\
\bottomrule
\end{tabular}}
\end{table}

\begin{table}
\centering
\caption{Complexity Analysis}\label{tab:complexity_analysis}
\footnotesize
\resizebox{15.5cm}{1.4cm} {
\begin{tabular}{llllll}
\toprule
 & \textbf{Space} & {\textsf{RQ}} & {$k$\textsf{NNQ}} & {\textsf{SDQ}} & {\textsf{SPQ}} \\
\midrule
\textsc{IDModel} & $\mathcal{O}(\mathtt{V} + \mathtt{D} + 2\mathtt{V}\mathtt{d} + \mathtt{V}\mathtt{d}^2)$ & $\mathcal{O}(\mathtt{o}\mathtt{V}\log_{}\mathtt{D})$ & $\mathcal{O}(\mathtt{o}\mathtt{V}\log_{}\mathtt{D})$ & $\mathcal{O}(\mathtt{V}\log_{}\mathtt{D})$ & $\mathcal{O}(\mathtt{V}\log_{}\mathtt{D} + \mathtt{w})$  \\
\textsc{IDIndex} & $\mathcal{O}(2\mathtt{D}^2)$ & $\mathcal{O}(\mathtt{o}\mathtt{d}\log_{}\mathtt{D})$ & $\mathcal{O}(\mathtt{o}\mathtt{d}\log_{}\mathtt{D})$ & $\mathcal{O}(\mathtt{d}^2)$ & $\mathcal{O}(\mathtt{d}^2 + \mathtt{w})$  \\
\textsc{CIndex} & $\mathcal{O}(\mathtt{V} + \mathtt{V}\mathtt{d} + \mathtt{O})$ & $\mathcal{O}(\mathtt{o}\mathtt{V}\log_{}\mathtt{D})$ & $\mathcal{O}(\mathtt{o}\mathtt{V}\log_{}\mathtt{D})$ & $\mathcal{O}(\mathtt{V}\log_{}\mathtt{D})$ & $\mathcal{O}(\mathtt{V}\log_{}\mathtt{D} + \mathtt{w})$  \\
\textsc{IP-Tree} & $\mathcal{O}(\rho^2\mathtt{f}^2\mathtt{L} + \rho\mathtt{D})$ & $\mathcal{O}({(\rho\log_\mathtt{f}\mathtt{L}})^2(\mathtt{V}\mathtt{o}/\mathtt{L} + \rho))$ & $\mathcal{O}({(\rho\log_\mathtt{f}\mathtt{L}})^2(\mathtt{V}\mathtt{o}/\mathtt{L} + \rho))$ & $\mathcal{O}(\rho^2 \log_\mathtt{f}\mathtt{L})$ & $\mathcal{O}((\rho^2 + \mathtt{w}) \log_\mathtt{f}\mathtt{L})$  \\
\textsc{VIP-Tree} & $\mathcal{O}(\rho^2\mathtt{f}^2\mathtt{L} + \rho\mathtt{D}\log_\mathtt{f}\mathtt{L})$ & $\mathcal{O}(\rho^2\log_\mathtt{f}\mathtt{L}(\mathtt{V}\mathtt{o}/\mathtt{L} + \rho))$ & $\mathcal{O}(\rho^2\log_\mathtt{f}\mathtt{L}(\mathtt{V}\mathtt{o}/\mathtt{L} + \rho))$ & $\mathcal{O}(\rho^2)$ & $\mathcal{O}(\rho^2 + \mathtt{w})$  \\
\bottomrule
\end{tabular} }
\end{table}

\noindent\textbf{Distance Precomputation.}~\textsc{IDModel} and \textsc{CIndex} do not precompute any indoor distances, whereas \textsc{IDIndex} and \textsc{IP-Tree}/\textsc{VIP-Tree} maintain some door-to-door distances before query processing.
In particular, \textsc{IDIndex} precomputes the shortest indoor distances between every pair of doors, but \textsc{IP-Tree}/\textsc{VIP-Tree} only keeps a small number of distances in each tree node.

\noindent\textbf{Model/Index Structure.} \textsc{IDModel} is a labeled graph with distance mapping functions, whereas \textsc{IDIndex} materializes two matrices for global door-to-door distances.
Employing a tree-based structure, \textsc{CIndex} keeps topological information incrementally by maintaining inter-partition links, whereas \textsc{IP-Tree}/\textsc{VIP-Tree} augments each tree node with a local distance matrix.
More importantly, \textsc{CIndex} forms the non-leaf tree nodes according to the geometrical proximity of partitions, whereas \textsc{IP-Tree}/\textsc{VIP-Tree} do so based on the topological proximity of partitions.

\noindent\textbf{Query Types.}
All model/indexes can support all the four query types.
However, \textsc{IDModel}~\cite{Lu2012} does not provide \textsf{RQ} and $k$\textsf{NNQ} algorithms.
In this study, we implement two algorithms as presented in Appendix. 
Also, there are no off-the-shelf \textsf{SPQ} and \textsf{SDQ} algorithms for \textsc{IDIndex} and \textsc{CIndex}.
Nevertheless, the global door-to-door distances and the corresponding last-hop door information in \textsc{IDModel} can be used to expand path searching in \textsf{SPQ} and \textsf{SDQ} algorithms for \textsc{IDIndex}.
For \textsc{CIndex}, the inter-partition links can be used to support path expansion.

\noindent\textbf{Initialization.} To decide the initial indoor partition for a query, \textsc{IDModel} and \textsc{IDIndex} sequentially scan all partitions. Enabled by the R*-tree indexing partitions, \textsc{CIndex} can quickly find the host partition of any indoor point. In contrast, \textsc{IP-Tree} and \textsc{VIP-Tree} are based on pure topological relationships among partitions, and thus they also sequentially scan all partitions.

\noindent\textbf{Expansion.} As a graph-based model, \textsc{IDModel} expands to the next unvisited door in the spirit of Dijkstra's algorithm~\cite{Bollobas1993}.
\textsc{CIndex} does so as well since the next-hop doors are captured in the inter-partition links on the topological layer.
Instead of expanding via directly connected doors, \textsc{IP-Tree}/\textsc{VIP-Tree} finds the lowest common ancestor (LCA) node of $p$ and $q$ and locates the intermediate access doors on the shortest path straightforwardly.
It is noteworthy that \textsc{IDIndex} alone cannot support topological door expansion. Instead, \textsc{IDIndex} relies on an underlying \textsc{IDModel} to loop through relevant indoor partitions' doors.

\subsection{Complexity Analysis}
\label{sec:complexity}

Let $\mathtt{V}$, $\mathtt{D}$, $\mathtt{O}$ be the total number of indoor partitions, doors, and indoor objects, respectively.
Let $\mathtt{d}$ and $\mathtt{o}$ be the average door number and average object number per partition, respectively. Let $\mathtt{w}$ be the average number of door nodes on a shortest path.
For \textsc{IP-Tree}/\textsc{VIP-Tree}, we use $\mathtt{f}$ to denote the fan-out of the tree node, $\rho$ the average access door number per node, and $\mathtt{L}$ the total number of leaf nodes.
Table~\ref{tab:feature_comparison} summarizes the space complexity of all model/indexes and their time complexity for queries.

\noindent\textbf{Space Complexity.}
\textsc{IDModel} $(V, E_a, L,$ $f_\text{dv}, f_\text{d2d})$'s space complexity is $\mathcal{O}(\mathtt{V} + \mathtt{V}\mathtt{d} + \mathtt{D} + \mathtt{V}\mathtt{d} + \mathtt{V}\mathtt{d}^2)$ = $\mathcal{O}(\mathtt{V}\mathtt{d}^2)$.
\textsc{IDIndex}'s space complexity is $\mathcal{O}(2\mathtt{D}^2)$ = $\mathcal{O}(\mathtt{D}^2)$ as it consists of two door matrices.
\textsc{CIndex}'s space complexity is $\mathcal{O}(\mathtt{V} + \mathtt{V}\mathtt{d} + \mathtt{O})$ = $\mathcal{O}(\mathtt{V}\mathtt{d} + \mathtt{O})$ where $\mathtt{V}$, $\mathtt{V}\mathtt{d}$, and $\mathtt{O}$ correspond to partition R*-tree, inter-partition links, and object hashtable, respectively.
\textsc{IP-Tree}'s space cost mainly consists of the distance matrices for leaf nodes and those for non-leaf nodes.
The former's complexity is $\mathcal{O}(\rho\mathtt{D})$ and the latter's is $\mathcal{O}((\rho\mathtt{f})^2\mathtt{L})$ where $\rho\mathtt{f}$ corresponds to the number of access doors from a child node and $\mathtt{L}$ reflects the number of non-leaf nodes.
In contrast, \textsc{VIP-Tree}'s space cost on the distance matrices for leaf nodes is $\mathcal{O}(\rho\mathtt{D}\log_\mathtt{f}\mathtt{L})$, where $\log_\mathtt{f}\mathtt{L}$ corresponds to the ancestor number of each leaf node.

\noindent\textbf{Time Complexity for \textsf{RQ} and $k$\textsf{NNQ}.}~\textsf{RQ} and $k$\textsf{NNQ} have similar time complexity as they both prune objects based on shortest distances.
\textsc{IDModel}'s search expands via qualified doors by graph traversal in $\mathcal{O}(\mathtt{V}\log_{}\mathtt{D})$ and iterates on the objects in each visited partition in $\mathcal{O}(\mathtt{o})$.
Also based on graph traversal, the search on \textsc{CIndex} obtains a subgraph in $\mathcal{O}(\mathtt{V}\log_{}\mathtt{D})$ and visits all objects in each partition of the subgraph in $\mathcal{O}(\mathtt{o})$.
\textsc{IDIndex}'s search expands to the nearest partitions based on the sorted result in $M_\text{idx}$, and loops through each object in the expanded partition. So its time complexity is $\mathcal{O}(\mathtt{o}\mathtt{d}\log_{}\mathtt{D})$.
The searches via \textsc{IP-Tree} and \textsc{VIP-Tree} work similarly.
They prune a tree node based on its distance from the query point in $\mathcal{O}(\log_\mathtt{f}\mathtt{L} \cdot \rho \cdot c)$, where $c$ is the unit \textsf{SDQ} cost.
Then, they qualify each object in the remaining nodes in $\mathcal{O}(\log_\mathtt{f}\mathtt{L} \cdot \mathtt{V}/\mathtt{L} \cdot \mathtt{o} \cdot c)$.
Given the \textsf{SDQ} complexity $\mathcal{O}(\rho^2 \log_\mathtt{f}\mathtt{L})$ for \textsc{IP-Tree} and $\mathcal{O}(\rho^2)$ for \textsc{VIP-Tree} (to be detailed below), their \textsf{RQ} and $k$\textsf{NNQ} complexities are $\mathcal{O}({(\rho\log_\mathtt{f}\mathtt{L}})^2(\mathtt{V}\mathtt{o}/\mathtt{L} + \rho))$ and $\mathcal{O}(\rho^2\log_\mathtt{f}\mathtt{L}(\mathtt{V}\mathtt{o}/\mathtt{L} + \rho))$, respectively.

\noindent\textbf{Time Complexity for \textsf{SDQ} and \textsf{SPQ}.}
For the graph traversal algorithms of \textsc{IDModel} and \textsc{CIndex}, the \textsf{SDQ} complexity is $\mathcal{O}(\mathtt{V}\log_{}\mathtt{D})$ and \textsf{SPQ} complexity is $\mathcal{O}(\mathtt{V}\log_{}\mathtt{D} + \mathtt{w})$ with additional cost to backtrack the shortest path in $\mathtt{w}$ hops.
For \textsc{IDIndex}, the only cost of \textsf{SDQ} is to loop through two door sets corresponding to $p$ and $q$ by a complexity of $\mathcal{O}(\mathtt{d}^2)$. The extra cost of \textsf{SPQ} to concatenate shortest path is of $\mathcal{O}(\mathtt{w})$.
For \textsc{IP-Tree}, \textsf{SDQ} needs to search the lowest common ancestor and then find a pair of access doors from that ancestor node, resulting in a complexity of $\mathcal{O}(\rho^2 \log_\mathtt{f}\mathtt{L})$.
In contrast, \textsc{VIP-Tree} materializes the distances from a leaf node to each access door in the ancestors. Its \textsf{SDQ} complexity is $\mathcal{O}(\rho^2)$.
The additional cost to construct shortest path in \textsf{SPQ} is $\mathcal{O}(\mathtt{w}\log_\mathtt{f}\mathtt{L})$ for \textsc{IP-Tree} and $\mathcal{O}(\mathtt{w})$ for \textsc{VIP-Tree}.

\section{Benchmark}
\label{sec:benchmark}

In this section, we detail the benchmark for evaluating the indoor spatial query techniques (model/indexes and algorithms). All code and data can be found at \url{https://github.com/indoorLBS/ISQEA}.

\subsection{Datasets}
\label{ssec:datasets}

\begin{figure}[ht]
\centering
\includegraphics[width=\columnwidth]{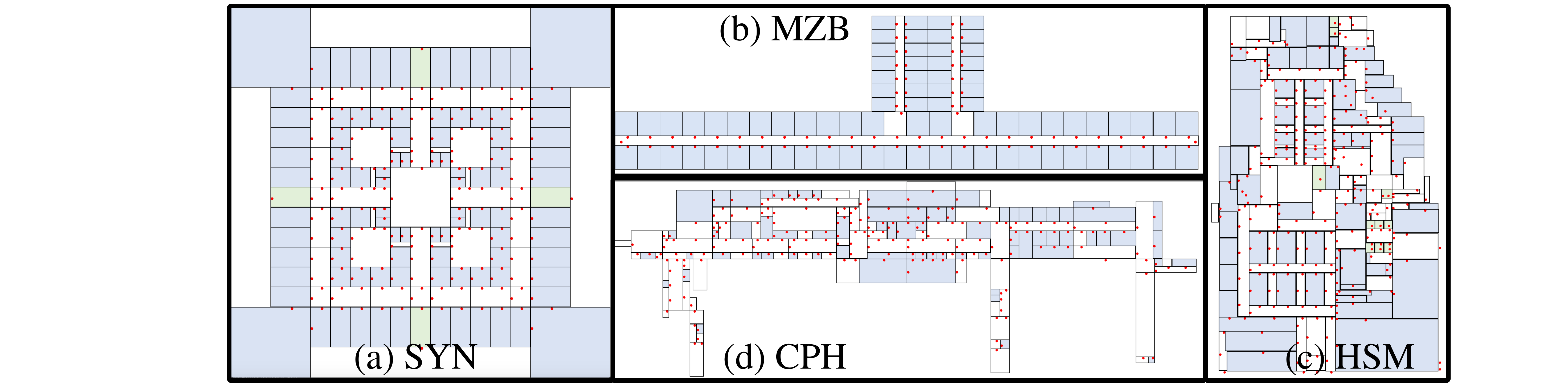}
\caption{Floorplan of Datasets.}\label{fig:dataset_floorplan}
\end{figure}

We use four very different indoor space datasets, each featuring a distinctive indoor topology. The floorplans are briefly represented and illustrated in Figure~\ref{fig:dataset_floorplan}. The data statistics are given in Table~\ref{tab:dataset}.

Synthetic Building (\textbf{SYN}) is a $n$-floor building. Its each floor is from a real-world floorplan~\cite{syndatafloorplan} of 1368m $\times$ 1368m with 141 partitions and 216 doors. Its each two adjacent floors are connected by four 20m long stairways.
By default, we set $n=5$ and get the default dataset SYN5.
To study the effect of topological changes, from SYN5 we derived SYN5$^-$ with fewer doors and SYN5$^+$ with more doors. Note that varying the door number will significantly change the connectivity and accessibility of the partitions, leading to a major topological change.
We also form SYN5$^0$ in which the hallways are not decomposed{\footnote {We precompute the door-to-door distance matrix for each hallway when it is not decomposed. The hallways are of irregular and concave shapes, and thus the door-to-door distance in a hallway can not use the Euclidean distance.}}.

Menzies Building (\textbf{MZB})~\cite{mzbfloorplan} is a landmark building at Clayton campus of Monash University.
Each floor takes approximately 125m $\times$ 35m and connects to adjacent floors by two or four stairways each being 5m long.
In total, there are 1344 partitions (including 34 staircases and 85 hallways) and 1375 doors.
By changing the hallway decomposition, we form MZB$^0$ in which the hallways are not decomposed and MZB$^\Delta$ in which the hallways are decomposed into more partitions than default.
%

Hangzhou Shopping Mall (\textbf{HSM}) is a 7-floor mall in Hangzhou, China, occupying 2700m $\times$ 2000m.
Ten stairways connect each two adjacent floors. Each floor contains 150 partitions and 299 doors on average. In total, there are 1050 partitions (including 70 staircases and 133 hallways) and 2093 doors.

Copenhagen Airport (\textbf{CPH}) refers to the ground floor of Copenhagen Airport~\cite{cphfloorplan}, taking around 2000m $\times$ 600m with 147 partitions (including 25 hallways) and 211 doors.

\begin{figure*}[ht]
\centering
\begin{minipage}[t]{0.245\textwidth}
\centering
\includegraphics[width=\textwidth]{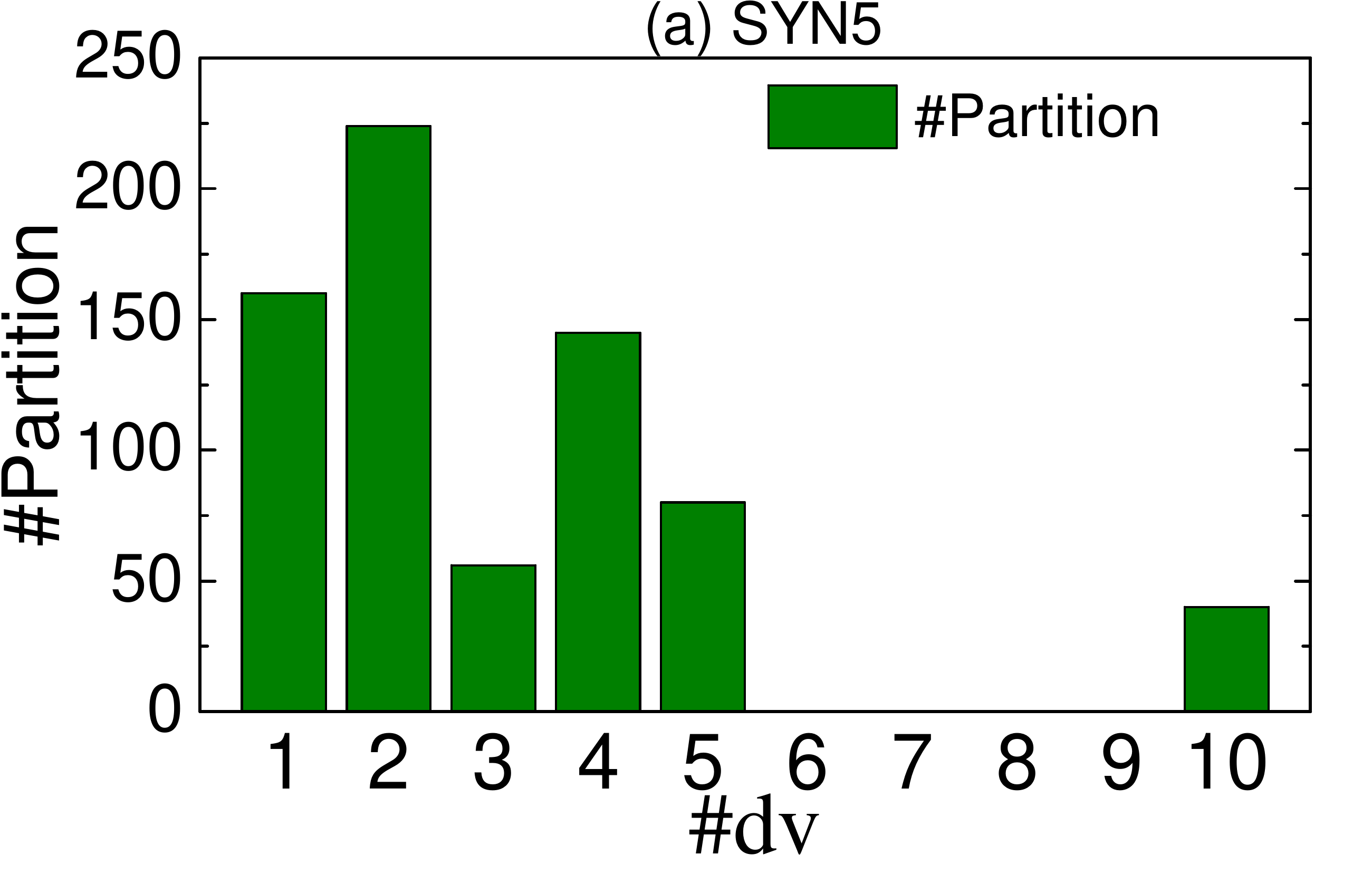}
\end{minipage}
\begin{minipage}[t]{0.245\textwidth}
\centering
\includegraphics[width=\textwidth]{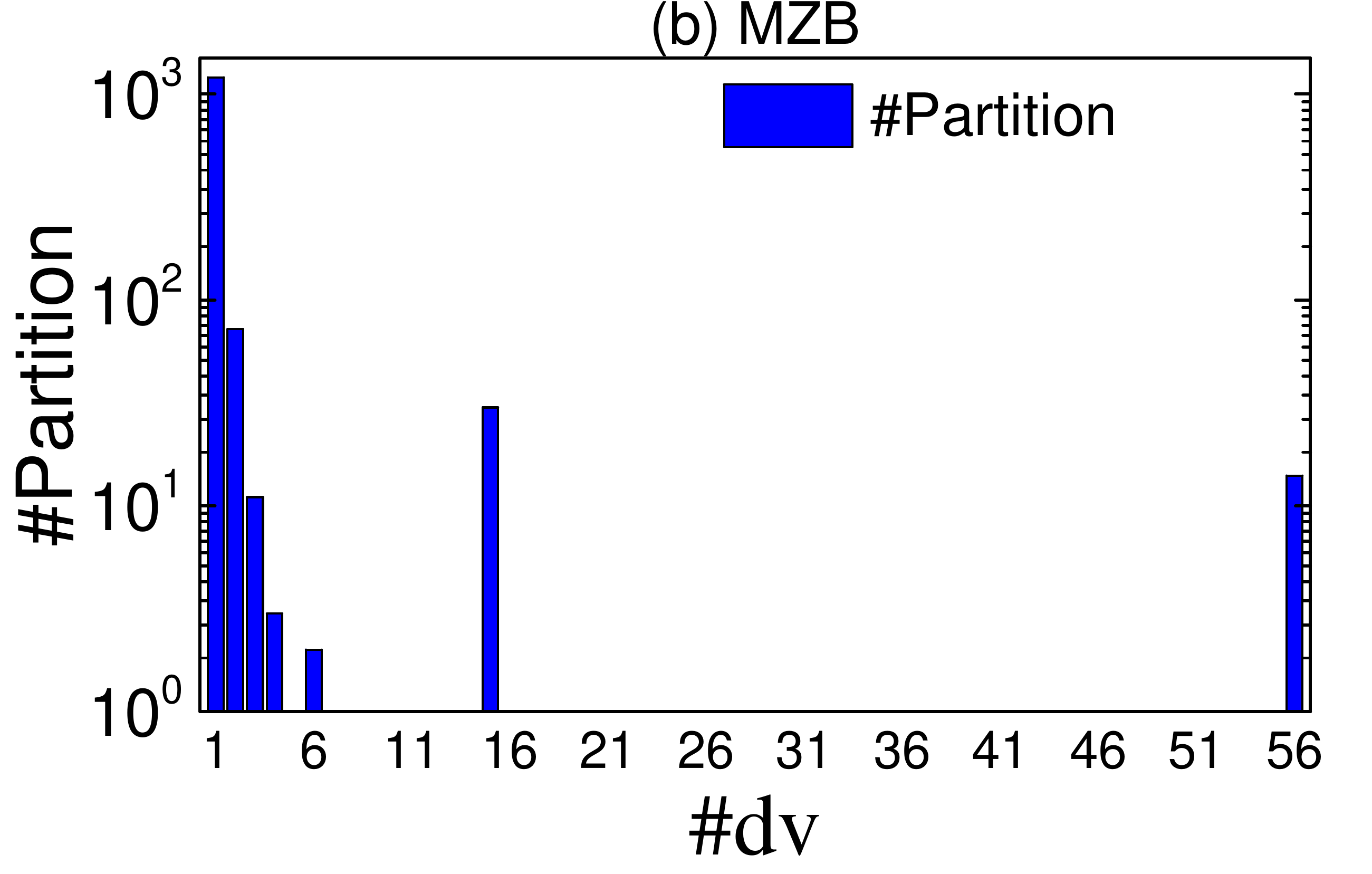}
\end{minipage}
\begin{minipage}[t]{0.245\textwidth}
\centering
\includegraphics[width=\textwidth]{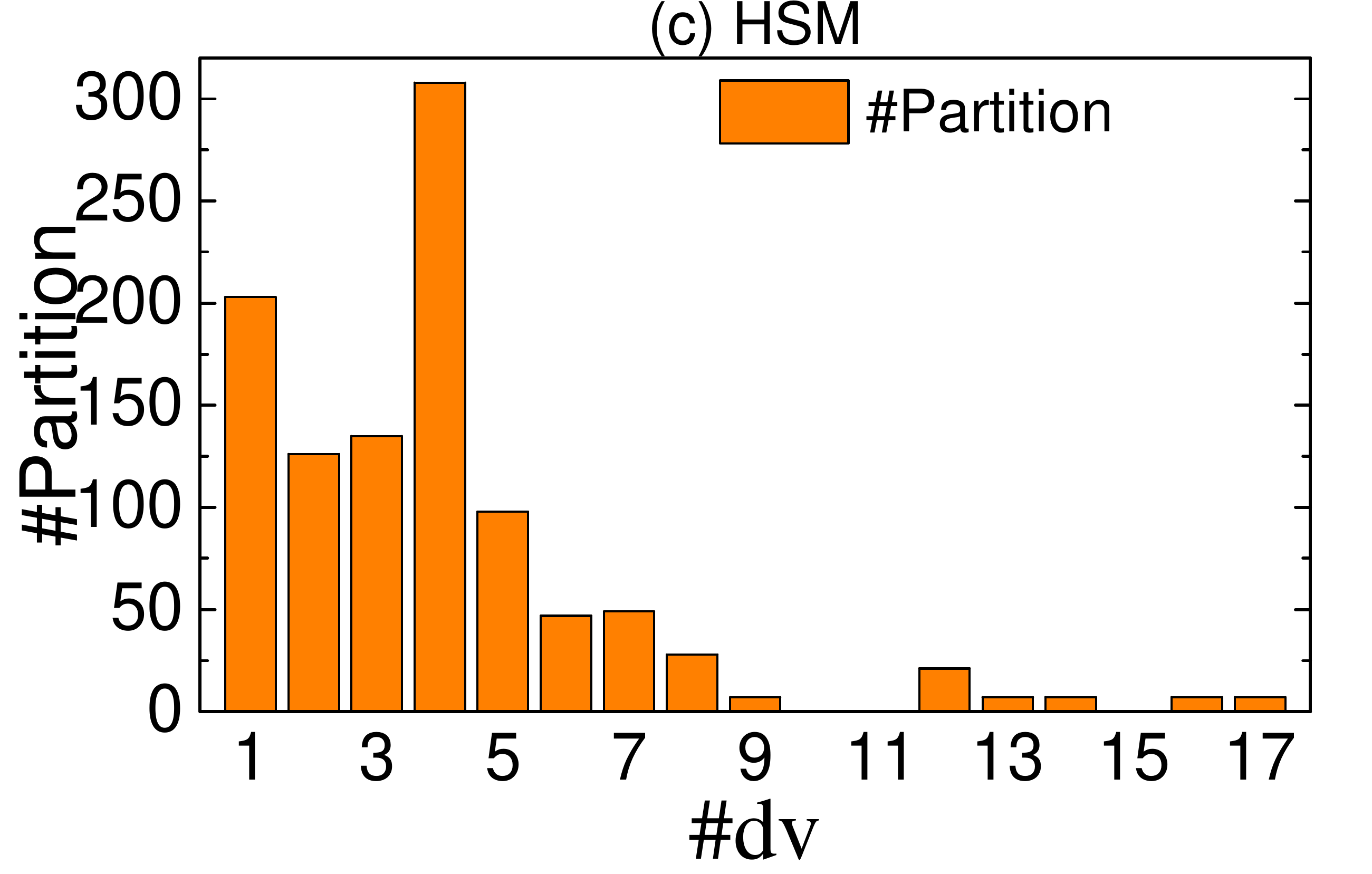}
\end{minipage}
\begin{minipage}[t]{0.245\textwidth}
\centering
\includegraphics[width=\textwidth]{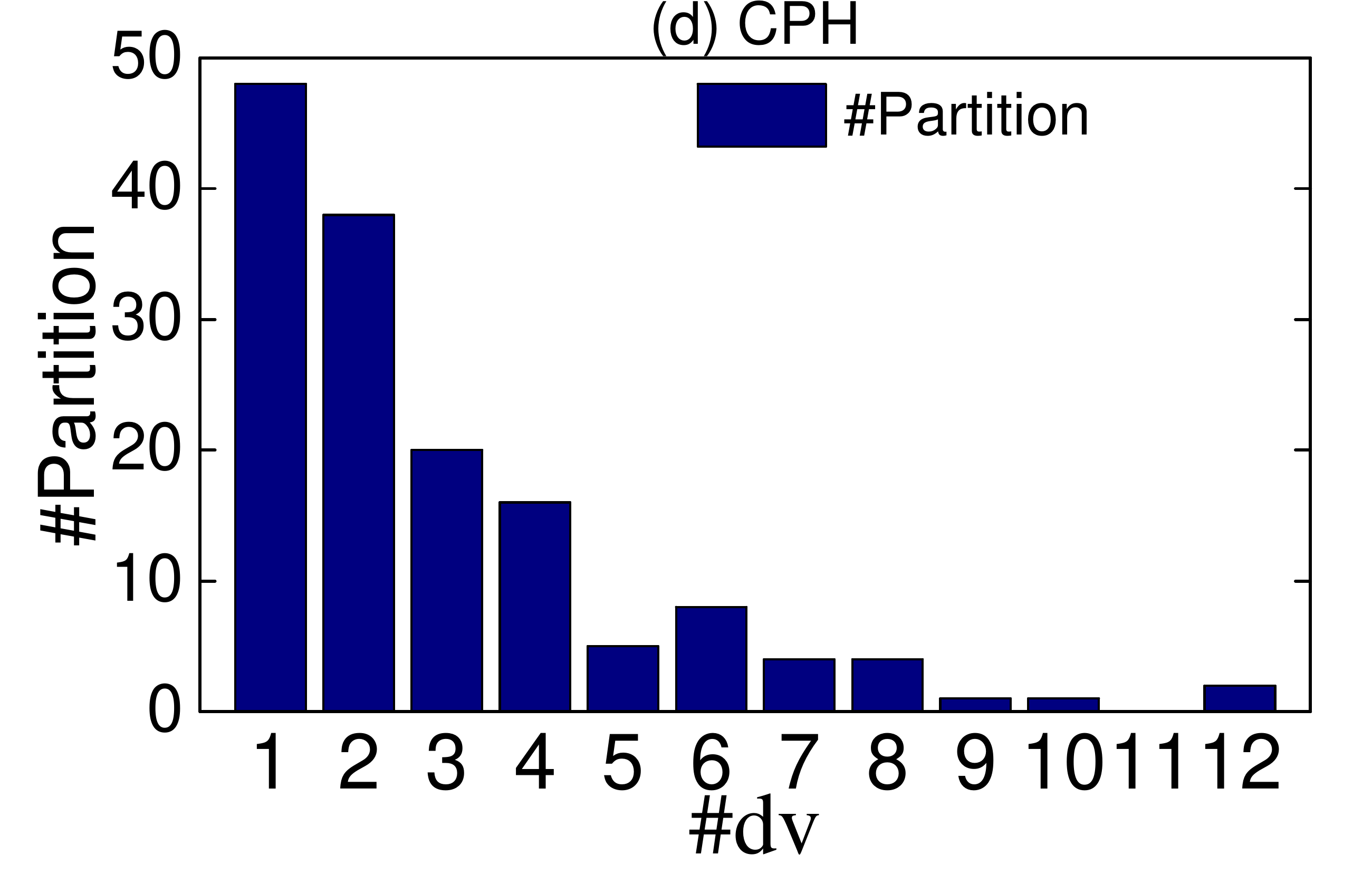}
\end{minipage}
\caption{Distribution of $\#dv$ (number of doors in a partition) on (a) SYN5, (b) MZB, (c) HSM, and (d) CPH.}\label{fig:door_distribution}
\end{figure*}

\vspace*{4pt}
\noindent\textbf{Overall Analysis of Different Datasets.}
The statistics of the datasets are given in Table~\ref{tab:dataset}.
We use $\#dv$ to denote the number of doors in a partition, and conduct quartile statistics~\cite{altman1994statistics} on $\#dv$.
In Table~\ref{tab:dataset}, $\mathsf{Q1}(\#dv)$, $\mathsf{Q2}(\#dv)$, and $\mathsf{Q3}(\#dv)$ denote the first, second, third quartiles of $\#dv$, respectively, and $\max(\#dv)$ denotes the maximum value of $\#dv$.
In addition, we also plot the distributions of $\#dv$ over all partitions in each dataset in Figure~\ref{fig:door_distribution}.

\begin{table}
\centering
\caption{Statistics of Datasets}\label{tab:dataset}
\footnotesize
\resizebox{15cm}{2.6cm}{
\begin{tabular}{c|l|llll|ll|l|ll}
\toprule
\multicolumn{2}{c|}{Datasets} & SYN & MZB & HZM & CPH & SYN5$^-$ & SYN5$^+$ & SYN5$^0$ & MZB$^0$ & MZB$^\Delta$\\
\midrule
\multirow{6}{*}{\tabincell{c}{Scale\\ of\\ Space}} & Floors & $n$ & 17 & 7 & 1 & 5 & 5 & 5 & 17 & 17 \\
& Doors  & 216$n$ & 1375 & 2093 & 211 & 840 & 1280 & 880 & 1308 & 1480\\
& Partitions & 141$n$ & 1344 & 1050 & 147 & 705 & 705 & 505 & 1276 & 1449\\
& Hallways & 41$n$ & 85 & 483 & 72 & 205 & 205 & 5 & 17 & 190\\
& Crucial Partitions & 8$n$ & 52 & 133 & 20 & 20 & 40 & 5 & 19 & 157\\
& Length(m) & 1368 & 125 & 2700 & 2000 & 1368 & 1368 & 1368 & 125 & 125 \\
& Width(m) & 1368 & 35 & 2000 & 600 & 1368 & 1368 & 1368 & 35 & 35 \\
\midrule
\multirow{4}{*}{\tabincell{c}{Door\\ per\\ Partition}} & $\mathsf{Q1}(\#dv)$ & 2 & 1 & 2 & 1 & 1 & 2 & 1 & 1 & 1 \\
& $\mathsf{Q2}(\#dv)$ & 2 & 1 & 4 & 2 & 1 & 3 & 2 & 1 & 1 \\
& $\mathsf{Q3}(\#dv)$ & 4 & 1 & 5 & 4 & 3 & 4 & 3 & 1 & 1 \\
& $\max(\#dv)$ & 10 & 56 & 17 & 12 & 10 & 10 & 132 & 82 & 47 \\
\bottomrule
\end{tabular}
}
\end{table}

\begin{table}
\centering
\caption{Evaluation Settings (Default Parameters in Bold)}\label{tab:parameters}
\footnotesize
\resizebox{15.5cm}{2.8cm} {
\begin{tabular}{l|l|ll|l|ll}
\toprule
\multicolumn{2}{c|}{\textbf{Symbol \& Meaning}} & \textbf{Task} & \textbf{Metrics}  & \textbf{Queries} & \textbf{Dataset} & \textbf{Parameter Setting} \\
\midrule
\multirow{2}*{$n$} & \multirow{2}*{floor number} & A & a1, a2 & -  & \multirow{2}*{SYN} & \multirow{2}*{3, \textbf{5}, 7, 9} \\
& & B1 & b1, b2, b3 (only for \textsf{SPDQ}) &  \textsf{RQ}, $k$\textsf{NNQ}, \textsf{SPDQ}  &  & \\
\midrule
$|O|$ & object number & B2 & b1, b2 & \textsf{RQ}, $k$\textsf{NNQ}  & all & 500, 1000, \textbf{1500}, 2000, 2500 \\
\midrule
\multirow{2}*{$r$} & \multirow{2}*{\tabincell{l}{range value}} & \multirow{2}*{B3} & \multirow{2}*{b1, b2} & \multirow{2}*{\textsf{RQ}}  & SYN5, HZM, CPH & 200, 400, \textbf{600}, 800, 1000 \\
&&&&  & MZB  & 20, 40, \textbf{60}, 80, 100 \\
\midrule
$k$ & - & B4 & b1, b2 & $k$\textsf{NNQ}  & all &1, 5, \textbf{10}, 50, 100 \\
\midrule
\multirow{2}*{s2t} & \multirow{2}*{\tabincell{l}{source-target\\ distance}} & \multirow{2}*{B5} & \multirow{2}*{b1, b2, b3} & \multirow{2}*{\textsf{SPDQ}} & SYN5, HZM, CPH & 1100, 1300, \textbf{1500}, 1700, 1900 \\
&&& & & MZB  & 30, 60, \textbf{90}, 120, 150 \\
\midrule
- & topological change & B6 & b1, b2, b3 (only for \textsf{SPDQ}) & \textsf{RQ}, $k$\textsf{NNQ}, \textsf{SPDQ}  & SYN & SYN5$^-$, SYN5, SYN5$^+$ \\
\midrule
\multirow{2}*{-}& \multirow{2}*{decomposition method} & \multirow{2}*{B7} & \multirow{2}*{b1, b2, b3 (only for \textsf{SPDQ})} & \multirow{2}*{\textsf{RQ}, $k$\textsf{NNQ}, \textsf{SPDQ}} & SYN & SYN5$^0$, SYN5 \\
&&&&  & MZB  & MZB$^0$, MZB, MZB$^\Delta$ \\
\bottomrule
\end{tabular}
}
\end{table}

Based on the space scale information and door distribution information from Table~\ref{tab:dataset} and Figure~\ref{fig:door_distribution}, we summarize the characteristics of each dataset as follows.
\begin{itemize}[leftmargin=*]
\item SYN: The overall space is {square and regular}. The number of doors and partitions in each floor is {medium} (216 doors and 141 partitions per floor). The door density within each partition is {small} (with $\mathsf{Q2}$ equals only 2).
\item MZB: The overall space is {long and narrow with large scale crucial partitions}. The number of doors and partitions {in each floor is relatively small} (80.4 doors and 76.8 partitions on average), whereas {the overall size of doors and partitions is large} due to the floor number. The planning of doors is {rather skewed} in that most partitions have only 1 or 2 doors while there are some crucial partitions that accommodate 56 doors (as shown in Figure~\ref{fig:door_distribution}(b)).
\item HSM: The overall space is {long and relatively narrow}. The number of doors and partitions {in each floor is medium} and {the overall size of doors and partitions is large}. The planning of doors is {regular} and door density in each partition is {medium} ($\mathsf{Q2}$ and $\mathsf{Q3}$ are equal to 4 and 5, respectively).
\item CPH: The space is {long, narrow} yet open, resulting in a {small} number of doors and partitions. The door distribution is {regular} and door density in each partition is {small} ($\mathsf{Q2}$ equals 2).
\end{itemize}

\subsection{Object/Query Workload Generation}
\label{ssec:query_generation}

\change{For each dataset, we randomly generated a set $O$ of valid points as static objects, each object in $O$ falling in an indoor partition.
To test the effect of different object numbers, we vary $|O|$ as 500, 1000, 1500, 2000 and 2500.}

The augment generation for each query type is detailed below.

{$\textsf{RQ}(p, r)$.}
We vary the range value $r$ according to the predefined values in Table~\ref{tab:parameters} (default values in bold).
For each $r$, we generate ten \textsf{RQ} instances with a random $p$ in the indoor space.

{$k\textsf{NNQ}(p)$.}
Similar to \textsf{RQ} generation, we generate ten random $k\textsf{NNQ}$ instances for each $k$ value given in Table~\ref{tab:parameters}.

As \textsf{SPQ} and \textsf{SDQ} can be integrated into one search procedure, we use $\textsf{SPDQ}(p, q)$ to denote the integrated query that returns the shortest path from $p$ to $q$ along with the corresponding shortest distance value.
In the following sections, we evaluate search performance of \textsf{SPDQ} only.

{$\textsf{SPDQ}(p, q)$.}
We use a parameter s2t to control the shortest distance from the source $p$ and target $q$. Its parameter values are listed in Table~\ref{tab:parameters}.
For each s2t, we generate ten different $(p, q)$ pairs to form \textsf{SPDQ} instances as follows.
First, we randomly select an indoor point $p$ and find a door $d$ whose indoor distance from $p$ approximates s2t. Next, we expand from $d$ to find a random point $q$ whose indoor distance from $p$ approximates s2t.

\subsection{Model/Index Settings}
\label{ssec:model_setting}


{\textsc{IDModel}}.
For each partition instance $v_i$, we implemented the door-to-door distance mapping $f_\text{d2d}(v_i, \cdot, \cdot)$ as a  2D array, and door-to-partition distance mapping $f_\text{dv}(\cdot, v_i)$ as an 1D array.
Besides, the partition mappings $\mathit{P2D}_\sqsupset(v_i)$ and $\mathit{P2D}_\sqsubset(v_i)$ (cf. Section~\ref{ssec:concepts}) were implemented as lists and attached to their corresponding instance $v_i$.
Further, the door mappings $\mathit{D2P}(d_i)$, $\mathit{D2P}_\sqsupset(d_i)$, and $\mathit{D2P}_\sqsubset(d_i)$ were implemented as lists associated with the door instance $d_i$.

{\textsc{IDIndex}}.
The distance matrix and distance index matrix were implemented as 2D arrays.

{\textsc{CIndex}}.
Since the partitions in the datasets rarely intersect, we used an R-tree instead of R*-tree to index partitions while preserving roughly the same spatial search performance. We set the tree fan-out to 20 as suggested in a previous work~\cite{Xie2013}. Each partition's inter-partition links were maintained in an inner list.

\textsc{IP-Tree} and \textsc{VIP-Tree}.
We set the minimum children degree to 2 when constructing non-leaf tree nodes, as suggested in~\cite{Shao2017}.
As each leaf node maintains the shortest distance for each pair of doors in it,
the computation will be complicated if a leaf node contains too many ``crucial partitions'' that each has many doors.
Following work~\cite{Shao2017}, we designate that each leaf node can only contain one crucial partition and regard a partition as crucial partition if its door number exceeds a threshold $\gamma$.
We tuned optimal $\gamma$ for different datasets, namely 6, 4, 7, and 5 for SYN, MZB, HZM, and CPH, respectively.

\subsection{Performance Evaluation Procedure}
\label{ssec:evaluation_procedure}

Concerning model construction and query processing, the following tasks are implemented to evaluate each model/index. For each task, a parameter is varied with others fixed to default. Table~\ref{tab:parameters} lists all the evaluation settings.
\begin{itemize}[leftmargin=*]
\item[A] {\bf Model Construction.} We evaluate the space and time efficiency of a model/index using two metrics: ({\bf a1}) model size and ({\bf a2}) construction time. In this task, we use synthetic datasets, varying the number of floors.
\item[B] {\bf Query Processing.} We evaluate the search efficiency of a given query type.
The metrics are ({\bf b1}) {running time}, ({\bf b2}) {memory use}, and ({\bf b3}) {number of visited doors (NVD)} for \textsf{SPDQ}.
\begin{itemize}[leftmargin=5pt]
	\item[B1] {\bf Effect of Floor Number $n$.} Using SYN with floor number $n$ varied from 3 to 9, we test the search efficiency for each indoor spatial query algorithm.
	\item[B2] {\bf Effect of Object Number $|O|$.} To test \textsf{RQ} and $k$\textsf{NNQ}, we vary  $|O|$ from 500 to 2500 in all datasets.
	\item[B3] {\bf Effect of Range Distance $r$.} We vary and test the augment $r$ of \textsf{RQ}. In particular, we vary $r$ from 200m to 1000m in SYN5, HZM and CPH, and from 20m to 100m in MZB.
	\item[B4] {\bf Effect of $k$.} We vary and test $k$\textsf{NNQ}'s augment $k$ from 1 to 100 in all datasets.
	\item[B5] {\bf Effect of Source-Target Distance s2t.} To test \textsf{SPDQ}, we vary s2t from 1100m to 1900m in SYN5, HZM, and CPH, and from 30m to 150m in MZB.
	\item[B6] {\bf Effect of Topological Change.} We vary indoor topology by changing the door number from 840 to 1280 in SYN5 and obtain SYN5$^-$ and SYN5$^+$. 
	\item[B7] {\bf Effect of Hallway's Decomposition Method.} We use SYN5 and MZB with the derived datasets, SYN5$^0$, MZB$^0$ and MZB$^\Delta$. 
\end{itemize}
\end{itemize}
\vspace*{-8pt}


\section{Results Analysis}
\label{sec:results}

This section reports and analyzes the experimental results.
All experiments are implemented in Java and run on a MAC with a 2.30GHz Intel i5 CPU and 16 GB memory.
\vspace*{-5pt}
\subsection{Model/Index Construction}

\begin{figure}
\centering
\begin{minipage}[t]{0.49\columnwidth}
\centering
\includegraphics[width=\columnwidth]{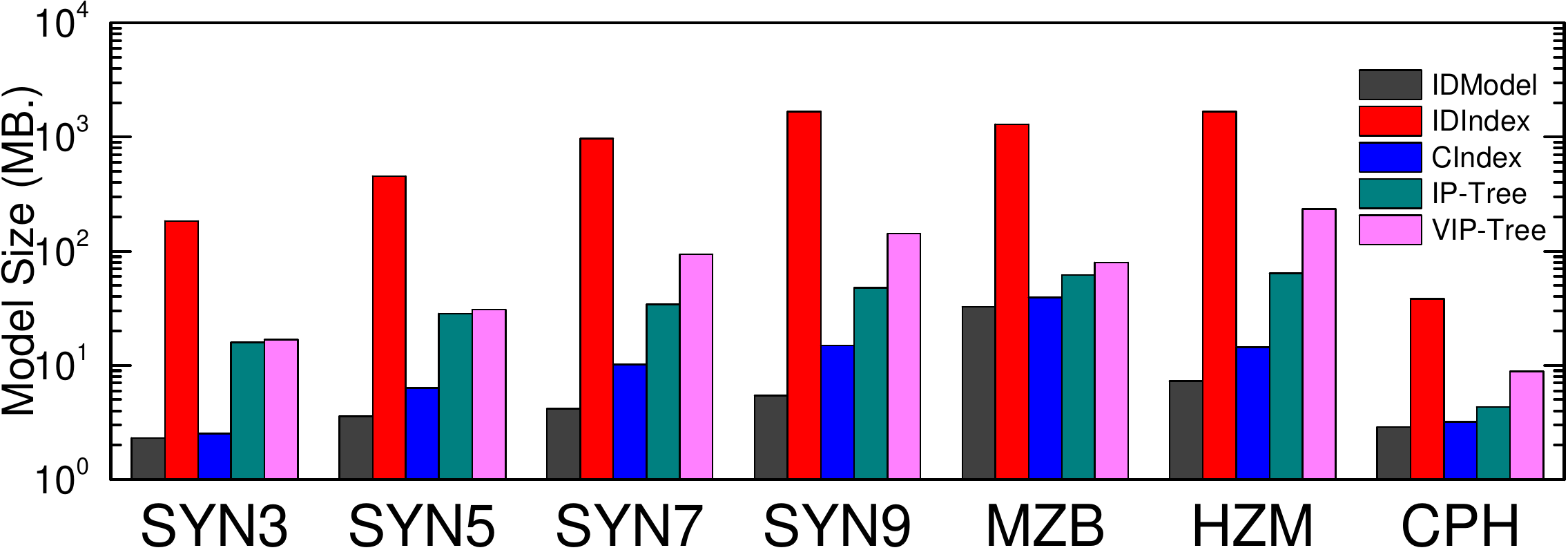}
\caption{Model Size}\label{fig:A1_size}
\end{minipage}
\begin{minipage}[t]{0.49\columnwidth}
\centering
\includegraphics[width=\columnwidth]{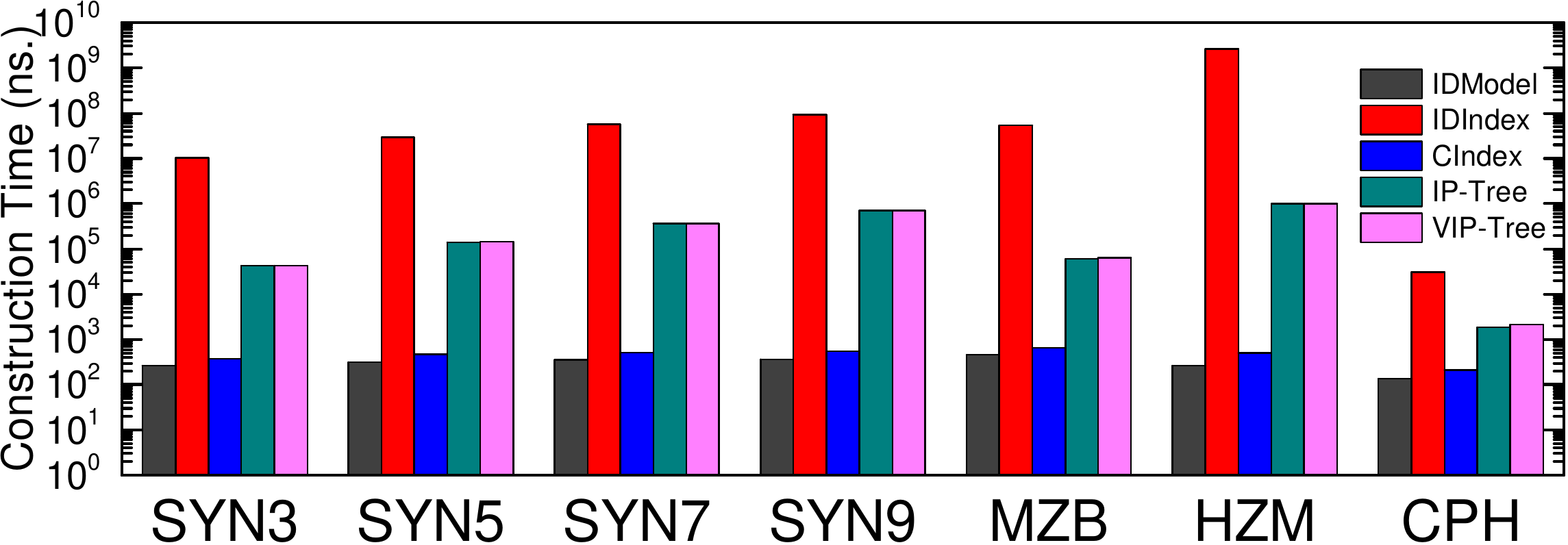}
\caption{Construction Time}\label{fig:A1_time}
\end{minipage}
\end{figure}

%
We vary the floor number $n$ on SYN and obtain four variants SYN3, SYN5, SYN7, and SYN9.
We construct the five model/indexes (cf. Section~\ref{sec:indexes}) and report their size and construction time in Figures~\ref{fig:A1_size} and~\ref{fig:A1_time}. The cost of maintaining static objects is excluded as it is the same for all model/indexes.
\begin{itemize}[leftmargin=*]
\item According to the results on SYN3 to SYN9 in Figure~\ref{fig:A1_size}, each model/index's size increases steadily with a larger floor number. When there are more doors and partitions, more storage space is needed to handle the indoor space.
\item Among all, \textsc{IDModel} construction requires the least costs on storage \change{(Figure~\ref{fig:A1_size})} and time \change{(Figure~\ref{fig:A1_time})}. This is because \textsc{IDModel} is extended based on a simple graph model and maintains only a small amount of geometric information locally.
For large-scale and complex-topology spaces (e.g., SYN9, MZB, and HZM), \textsc{IDModel} has clearer advantages over the tree-based indexes (i.e., \textsc{IP-Tree} and \textsc{VIP-Tree}).
\item As expected, \textsc{IDIndex} always takes much time and storage to construct due to its global door-to-door distance precomputation. When there are many doors, it is difficult to fit the corresponding matrices in memory.
In comparison, \textsc{IP-Tree} and \textsc{VIP-Tree} precompute less information and therefore their consumptions on time and storage are medium.
\item In addition to maintaining the topology, \textsc{CIndex} needs to construct a partition R-tree. Therefore, it incurs extra time and space overheads compared to \textsc{IDModel}.
\end{itemize}

\subsection{Query Processing}

All results are averaged over 10 queries (cf. Section~\ref{ssec:query_generation}).

\ExpHead{B1 Effect of Floor Number $n$} (using SYN)

\begin{figure}[!htbp]
\centering
\begin{minipage}[t]{0.24\columnwidth}
\centering
\includegraphics[width=\textwidth, height = 3cm]{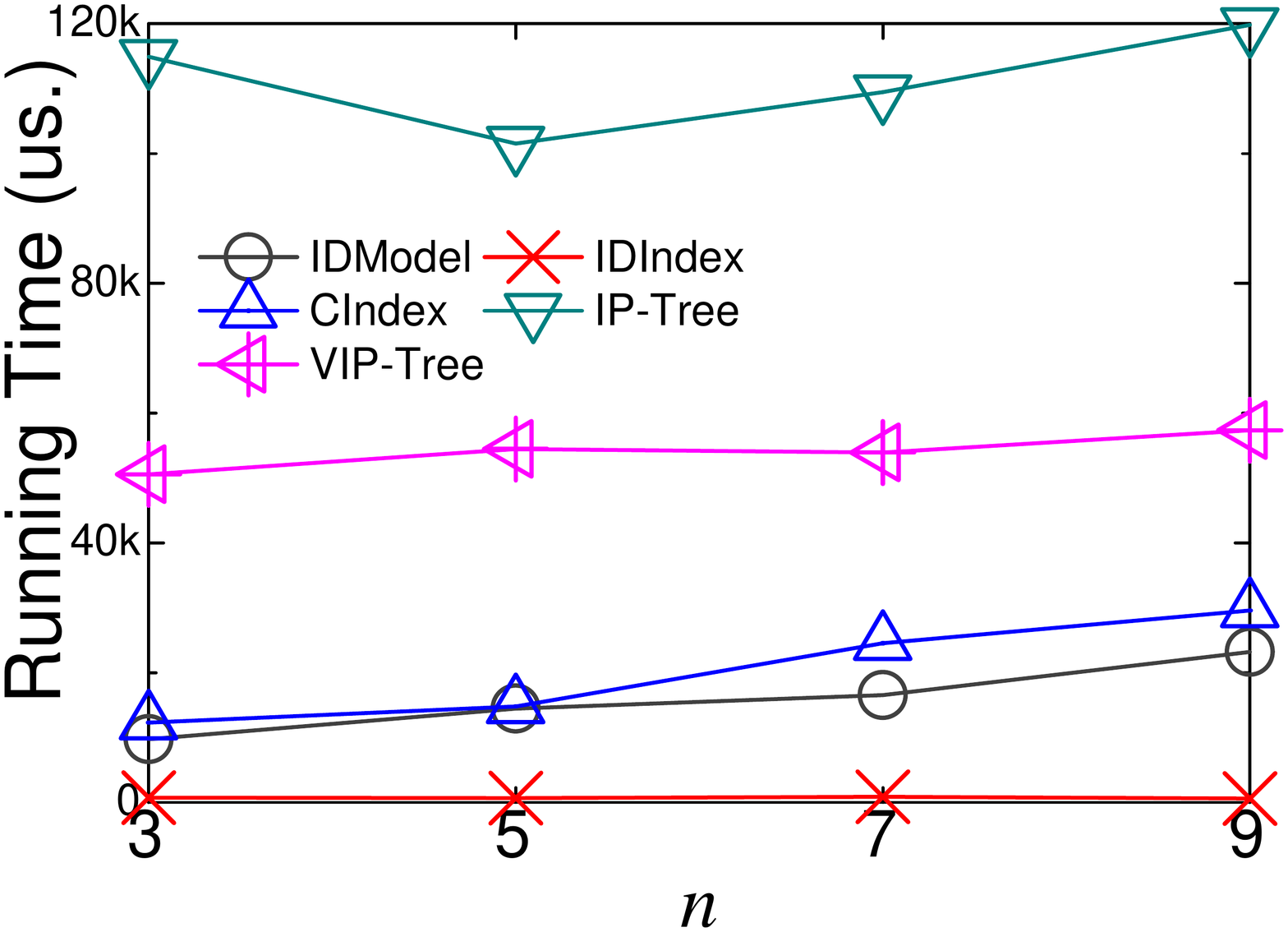}
\ExpCaption{\textsf{RQ} Time \\vs. $n$}\label{fig:B1_RQ_time}
\end{minipage}
\begin{minipage}[t]{0.24\columnwidth}
\centering
\includegraphics[width=\textwidth, height = 3cm]{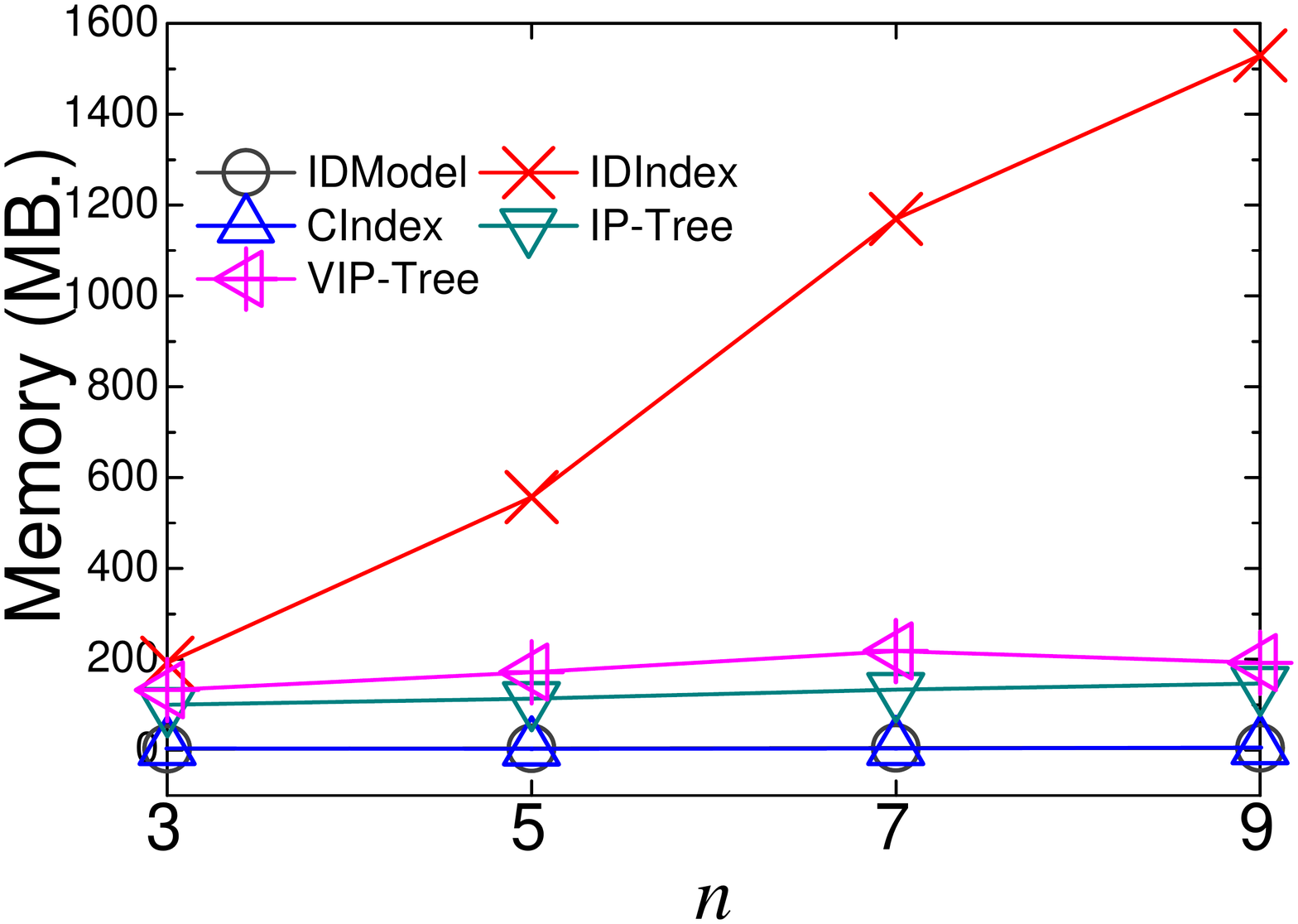}
\ExpCaption{\textsf{RQ} Memory \\vs. $n$}\label{fig:B1_RQ_mem}
\end{minipage}
\begin{minipage}[t]{0.24\columnwidth}
\centering
\includegraphics[width=\textwidth, height = 3cm]{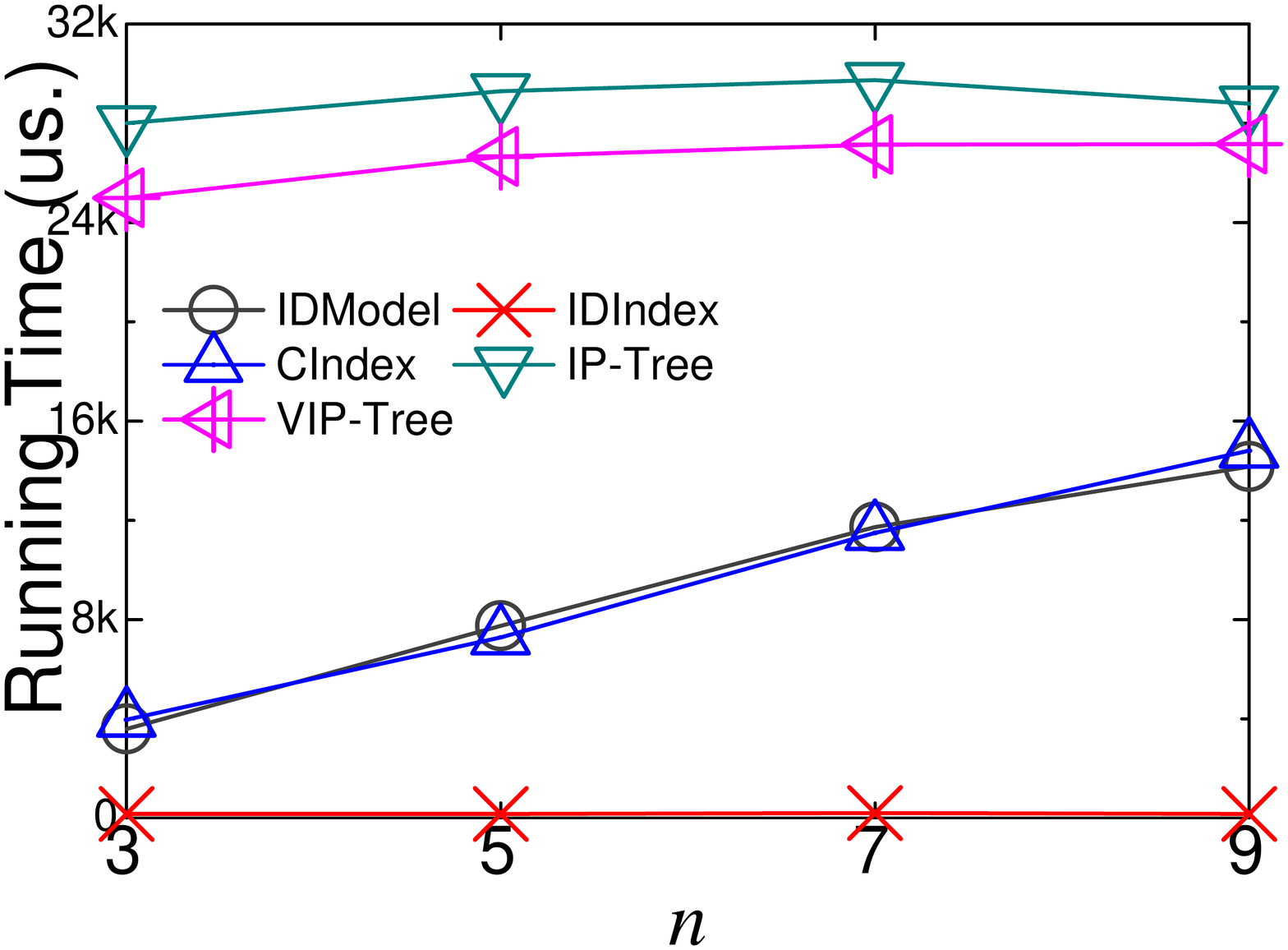}
\ExpCaption{$k$\textsf{NNQ} Time \\vs. $n$}\label{fig:B1_kNNQ_time}
\end{minipage}
\begin{minipage}[t]{0.24\columnwidth}
\centering
\includegraphics[width=\textwidth, height = 3cm]{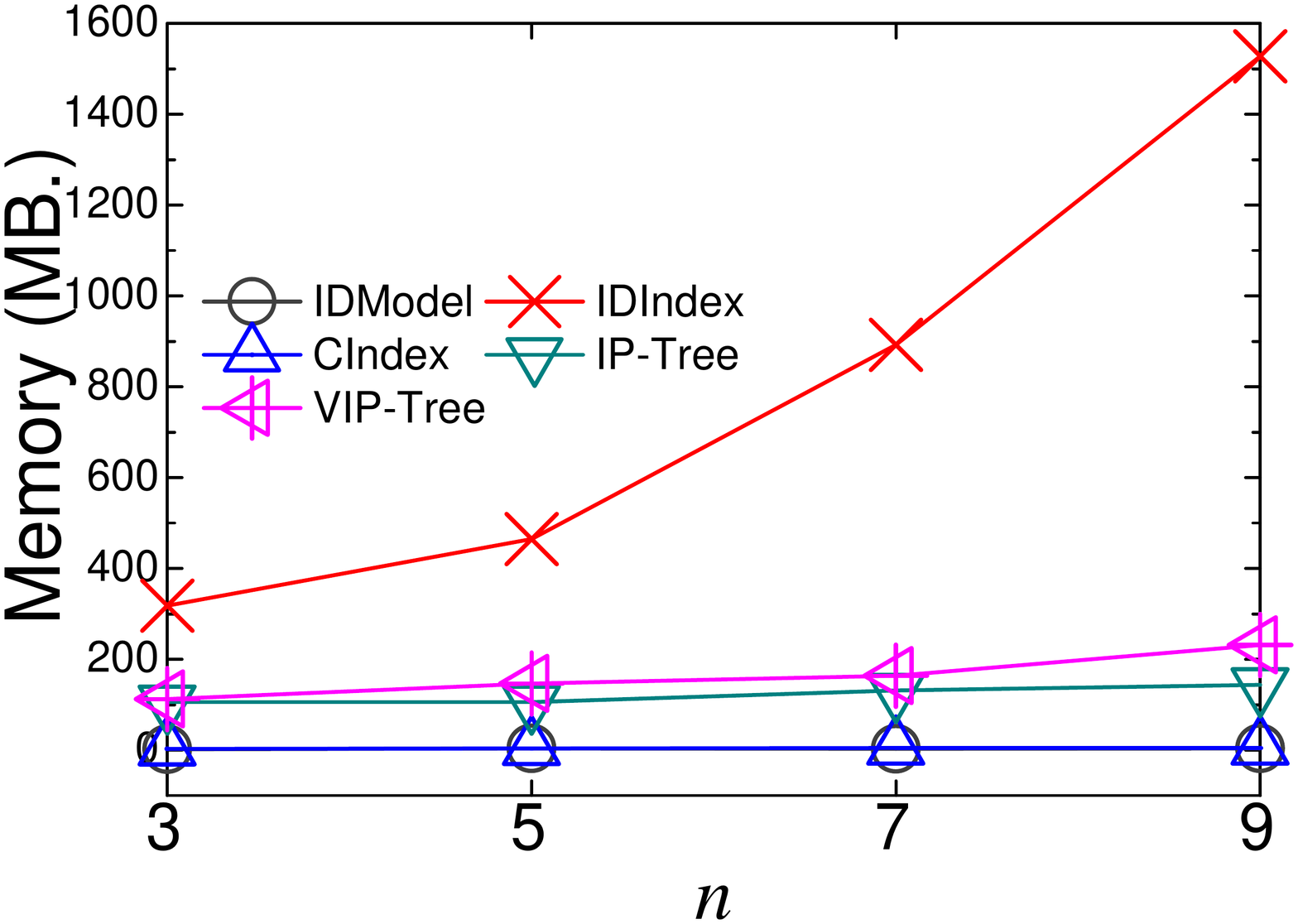}
\ExpCaption{$k$\textsf{NNQ} Memory vs. $n$}\label{fig:B1_kNNQ_mem}
\end{minipage}
\end{figure}

\noindent\textsf{RQ} and $k$\textsf{NNQ}: The query time and memory use for \textsf{RQ} are reported in Figures~\ref{fig:B1_RQ_time} and~\ref{fig:B1_RQ_mem}, respectively, and those for $k$\textsf{NNQ} are reported in Figures~\ref{fig:B1_kNNQ_time} and~\ref{fig:B1_kNNQ_mem}, respectively.
\begin{itemize}[leftmargin=*]
\item For both query types, \textsc{IDIndex} always runs fastest \change{as shown in Figures~\ref{fig:B1_RQ_time} and~\ref{fig:B1_kNNQ_time}}, unaffected by the varying floor number $n$.
The price behind this is to maintain the memory-resident distance matrices, which increases rapidly with $n$. \change{Referring to Figures~\ref{fig:B1_RQ_mem} and~\ref{fig:B1_kNNQ_mem}}, when $n$ grows to 9, \textsc{IDIndex} requires up to 1600MB of memory on both queries.
\item On each SYN dataset, \textsc{IP-Tree} and \textsc{VIP-Tree} need more time to complete the two queries. Through analysis, we found that the two indexes need to prune tree nodes when searching for qualified objects. In the absence of global door-to-door distances, they need a lot of on-the-fly calculations to get the shortest distance from a query point to a tree node.
Being consistent with the complexity analysis in Table~\ref{tab:complexity_analysis}, \textsc{VIP-Tree} outperforms \textsc{IP-Tree} for both queries.
However, due to the good scalability of the tree structure, both indexes' running time is relatively stable as shown in Figures~\ref{fig:B1_RQ_time} and~\ref{fig:B1_kNNQ_time}.
\item \textsc{IDModel} and \textsc{CIndex} perform similarly, and their execution time increases with a larger $n$ \change{(Figures~\ref{fig:B1_RQ_time} and~\ref{fig:B1_kNNQ_time})}.
When $n$ increases, \textsc{IDModel} has a slight advantage as \textsc{CIndex} costs more time in space pruning. In terms of memory overhead, the two indexes are almost the same.
\end{itemize}

\begin{figure*}[!htbp]
\begin{minipage}[t]{0.328\textwidth}
\centering
\includegraphics[width=\textwidth, height = 3cm]{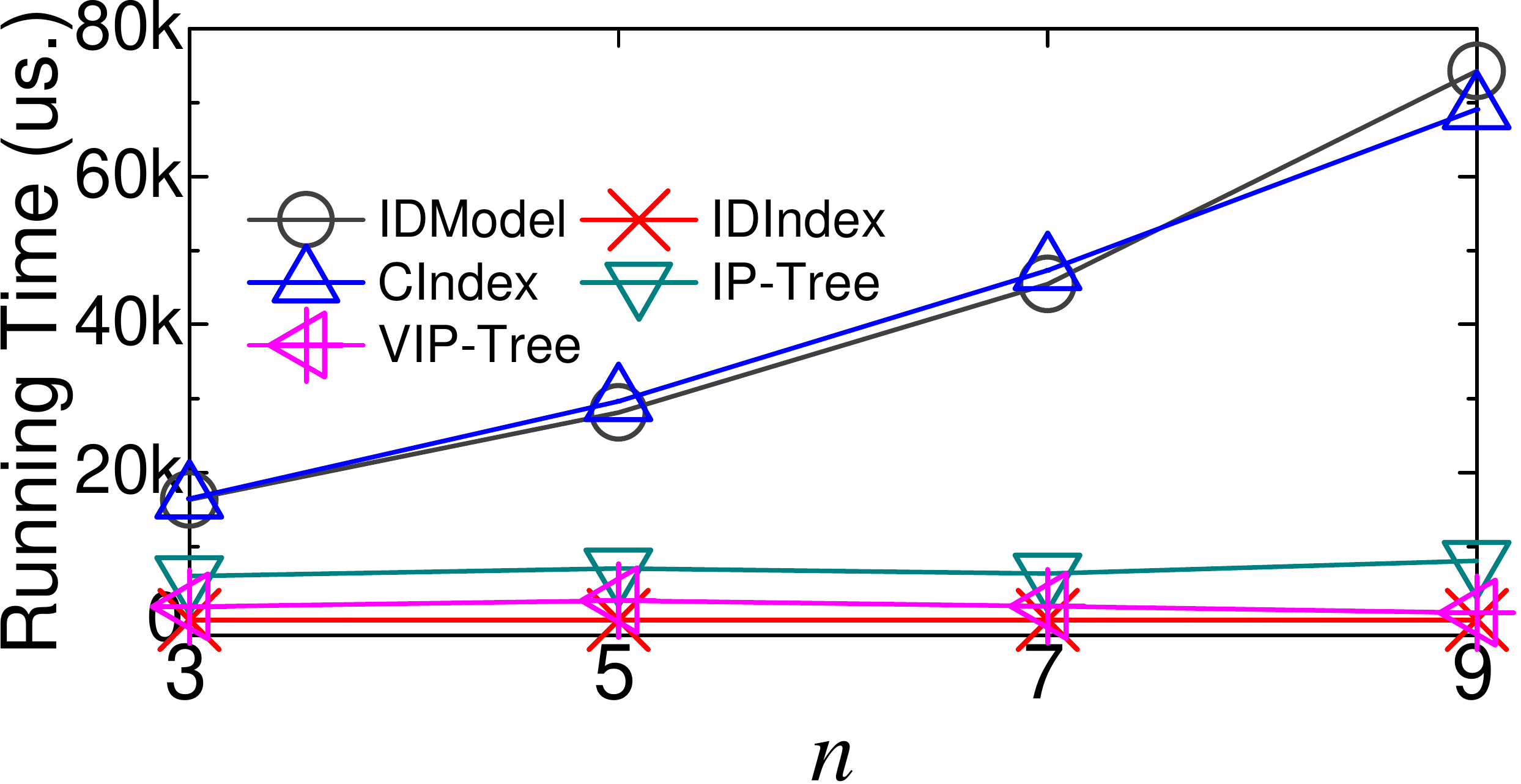}
\ExpCaption{\textsf{SPDQ} Time vs. $n$}\label{fig:B1_SPDQ_time}
\end{minipage}
\begin{minipage}[t]{0.328\textwidth}
\centering
\includegraphics[width=\textwidth, height = 3cm]{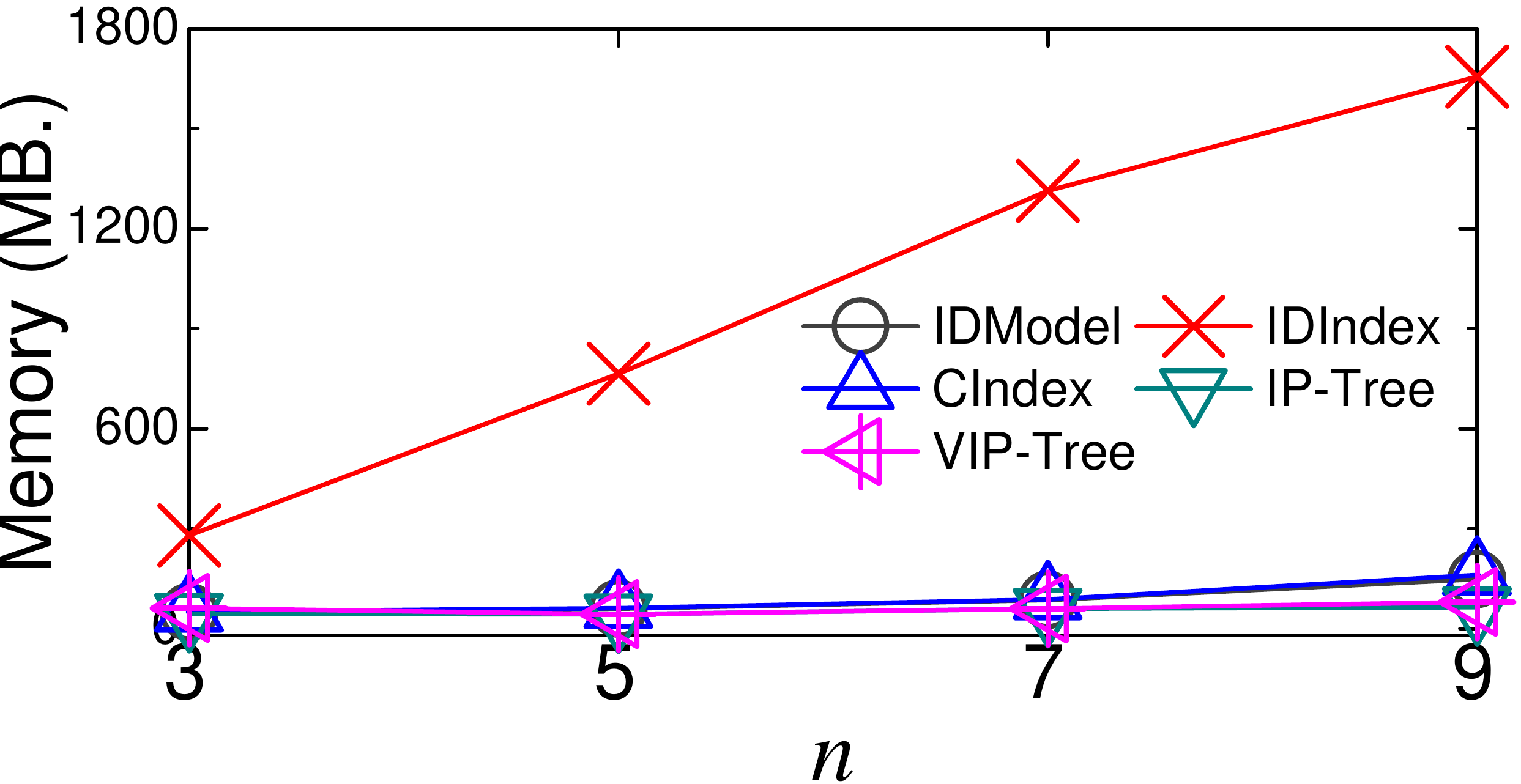}
\ExpCaption{\textsf{SPDQ} Memory vs. $n$}\label{fig:B1_SPDQ_mem}
\end{minipage}
\begin{minipage}[t]{0.328\textwidth}
\centering
\includegraphics[width=\textwidth, height = 3cm]{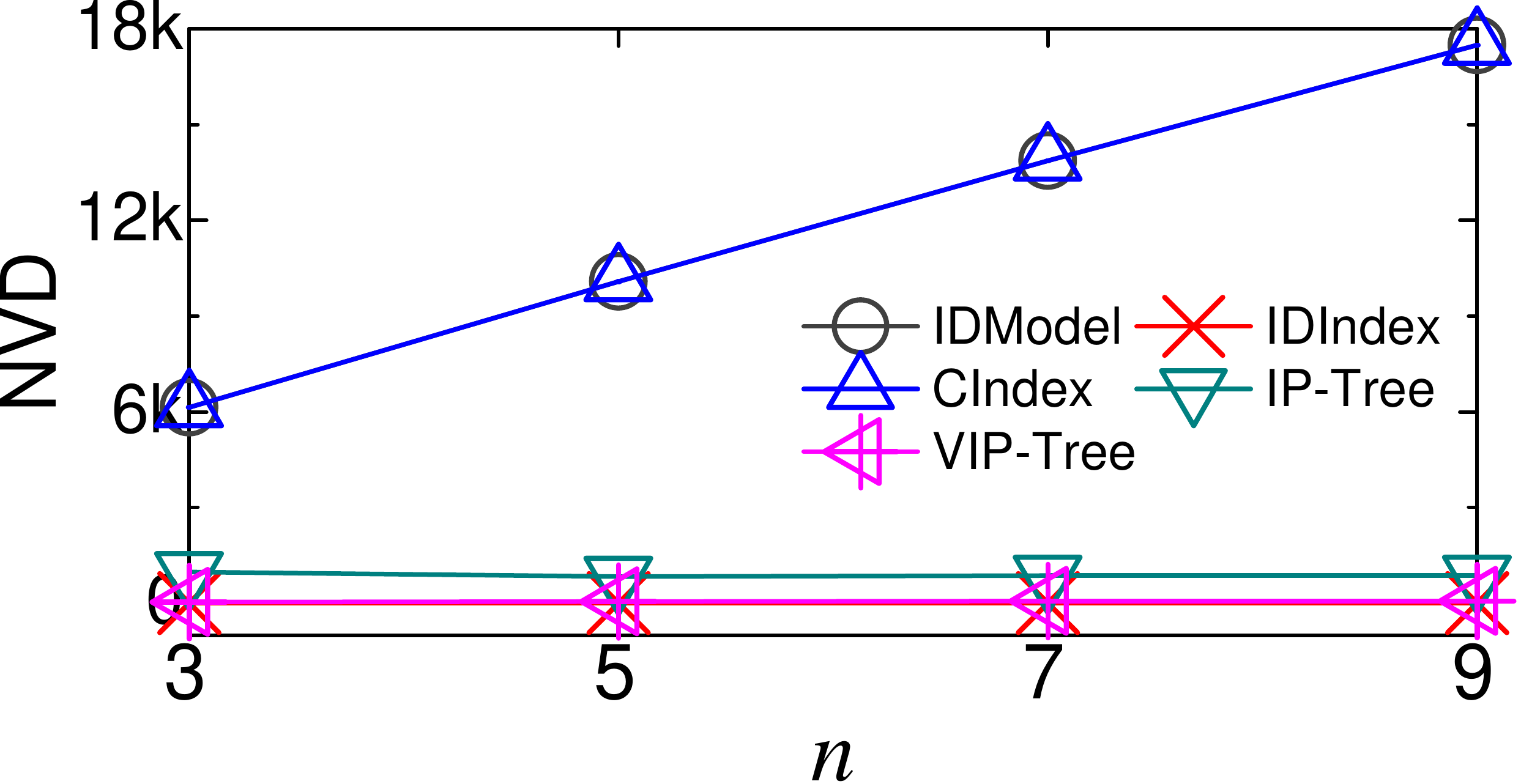}
\ExpCaption{NVD in \textsf{SPDQ} vs. $n$}\label{fig:B1_SPDQ_nvd}
\end{minipage}
\end{figure*}

\noindent\textsf{SPDQ}: The running time, memory use, and number of visited doors (NVD) are reported in Figures~\ref{fig:B1_SPDQ_time},~\ref{fig:B1_SPDQ_mem}, and~\ref{fig:B1_SPDQ_nvd}, respectively.
\begin{itemize}[leftmargin=*]
\item \textsc{IDIndex}'s running time and NVD are insensitive to the increasing floor number $n$.
However, its memory use grows moderately as $n$ increases. In the case of \textsf{SPQ} and \textsf{SDQ}, we recommend using \textsc{IDIndex} when the door size is relatively small.
\item In contrast to \textsc{IDIndex}, the memory of \textsc{IDModel} and \textsc{CIndex} is relatively stable \change{(Figure~\ref{fig:B1_SPDQ_mem})}, and their query performance deteriorates as the space scale increases \change{(Figure~\ref{fig:B1_SPDQ_time})}.
\item \textsc{IP-Tree} and \textsc{VIP-Tree} achieve clearly good performance on \textsf{SPDQ}, in both running time and memory use.
Unlike \textsc{IDIndex} that precomputes global door-to-door distances or \textsc{IDModel} and \textsc{CIndex} that compute distances on the fly, \textsc{IP-Tree} and \textsc{VIP-Tree} cache relevant distance information only for those access doors on shortest paths. Thus, without degrading query performance, they only incur slightly more memory overhead than \textsc{IDModel} and \textsc{CIndex} \change{(Figures~\ref{fig:B1_SPDQ_time} and~\ref{fig:B1_SPDQ_mem})}.
\end{itemize}

\ExpHead{B2 Effect of Object Number $|O|$}

\begin{figure*}[!htbp]
\centering
\begin{minipage}[t]{0.245\textwidth}
\centering
\includegraphics[width=\textwidth, height = 3cm]{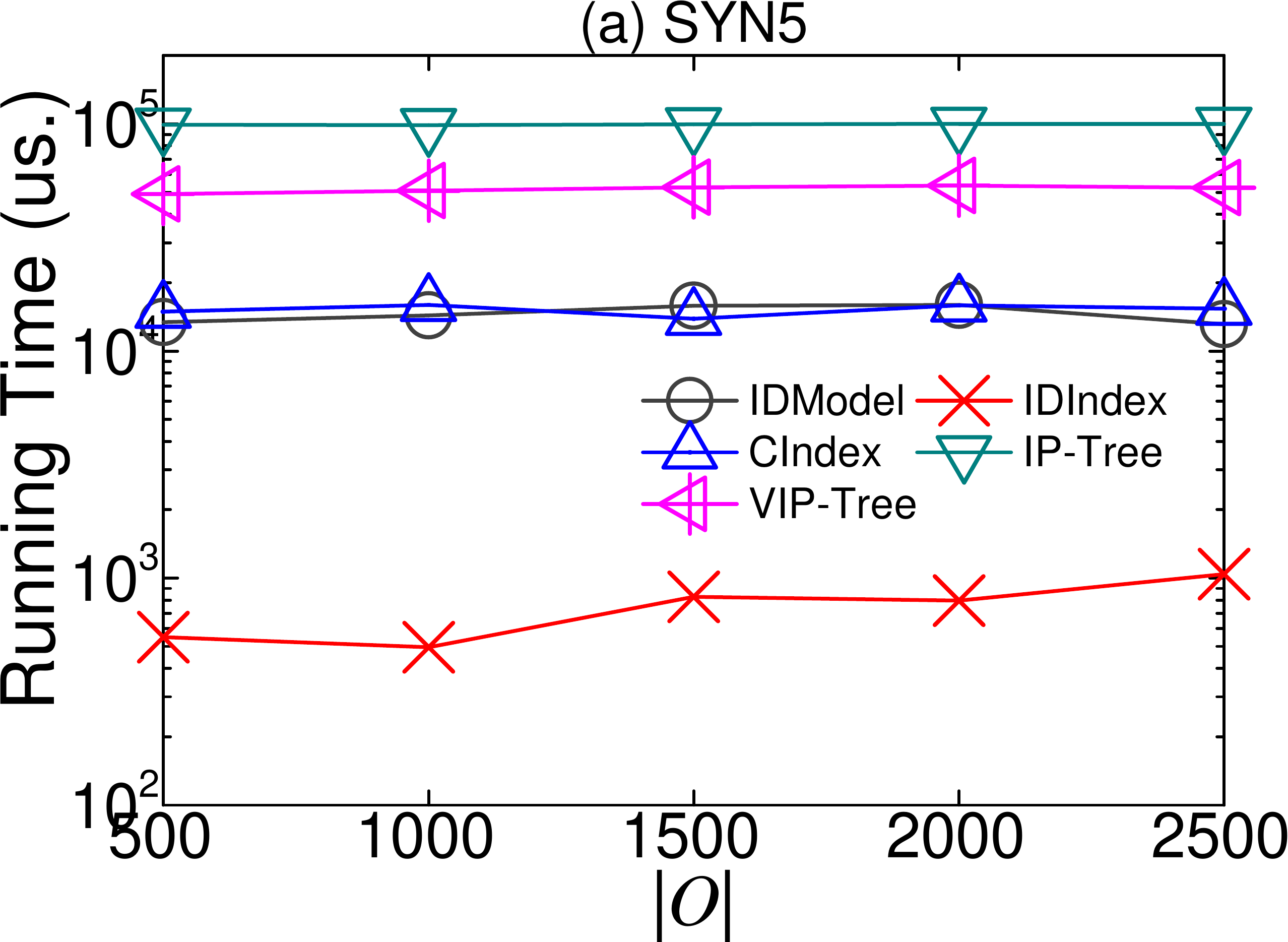}
\end{minipage}
\begin{minipage}[t]{0.245\textwidth}
\centering
\includegraphics[width=\textwidth, height = 3cm]{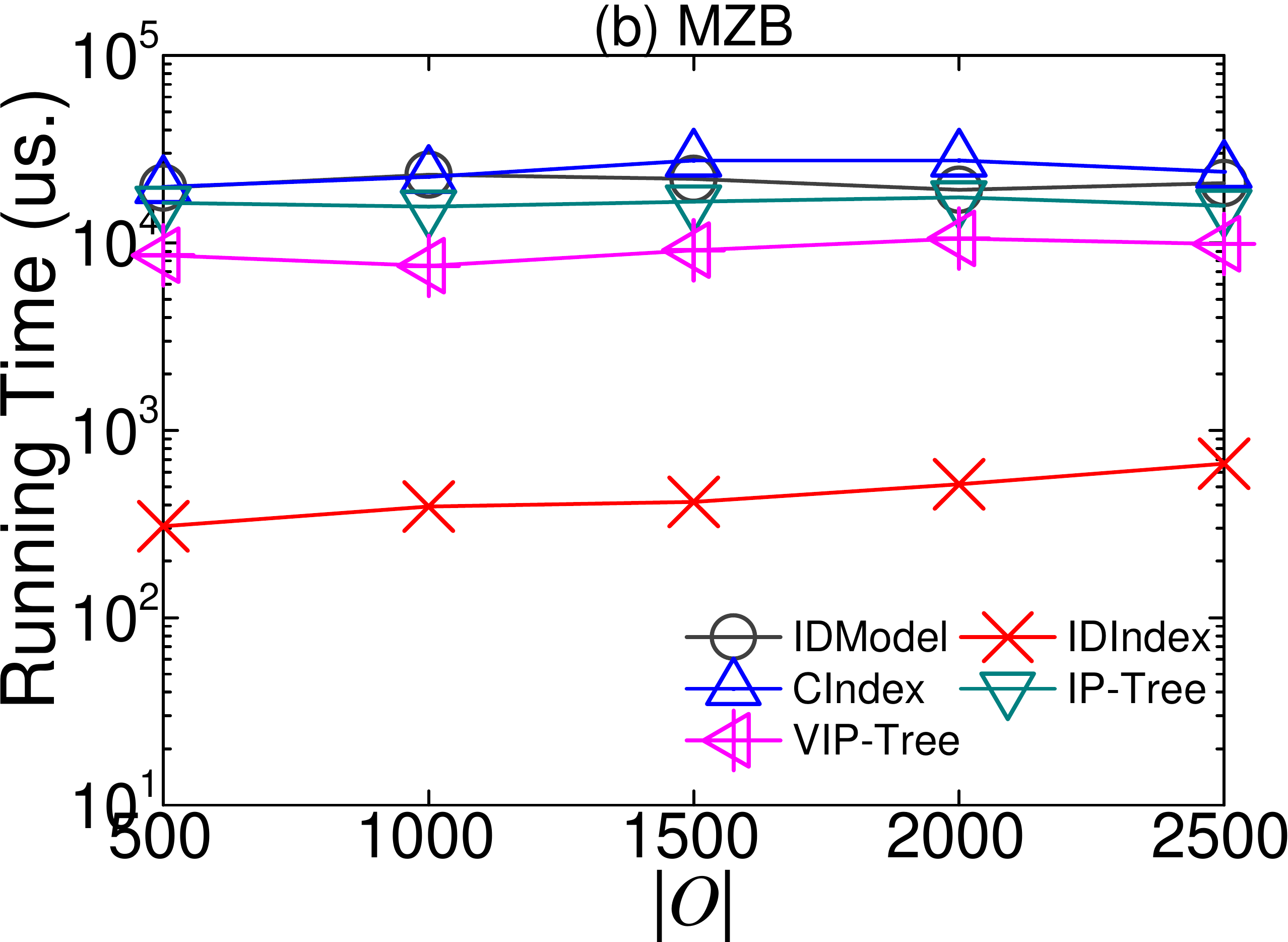}
\end{minipage}
\begin{minipage}[t]{0.245\textwidth}
\centering
\includegraphics[width=\textwidth, height = 3cm]{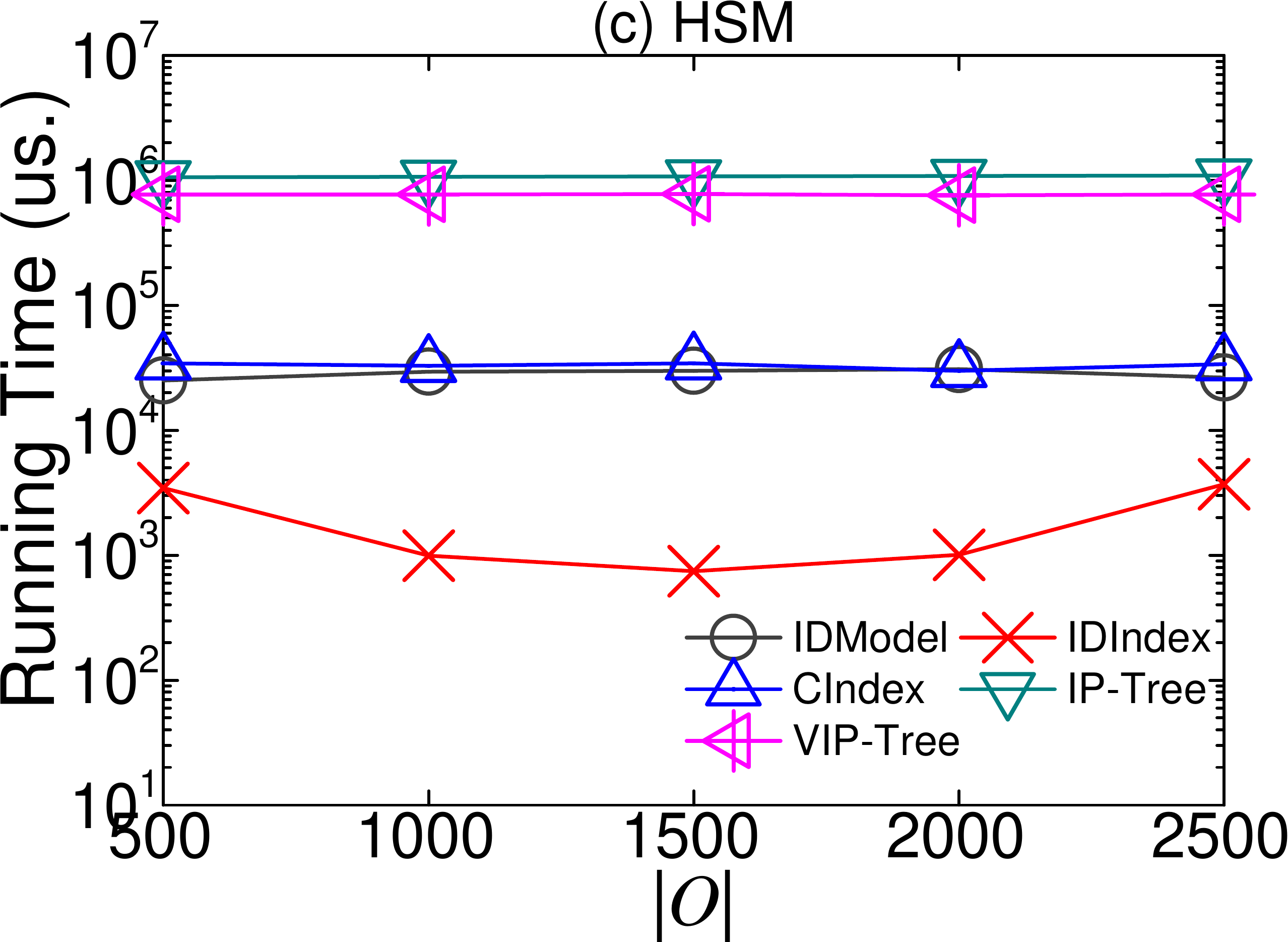}
\end{minipage}
\begin{minipage}[t]{0.245\textwidth}
\centering
\includegraphics[width=\textwidth, height = 3cm]{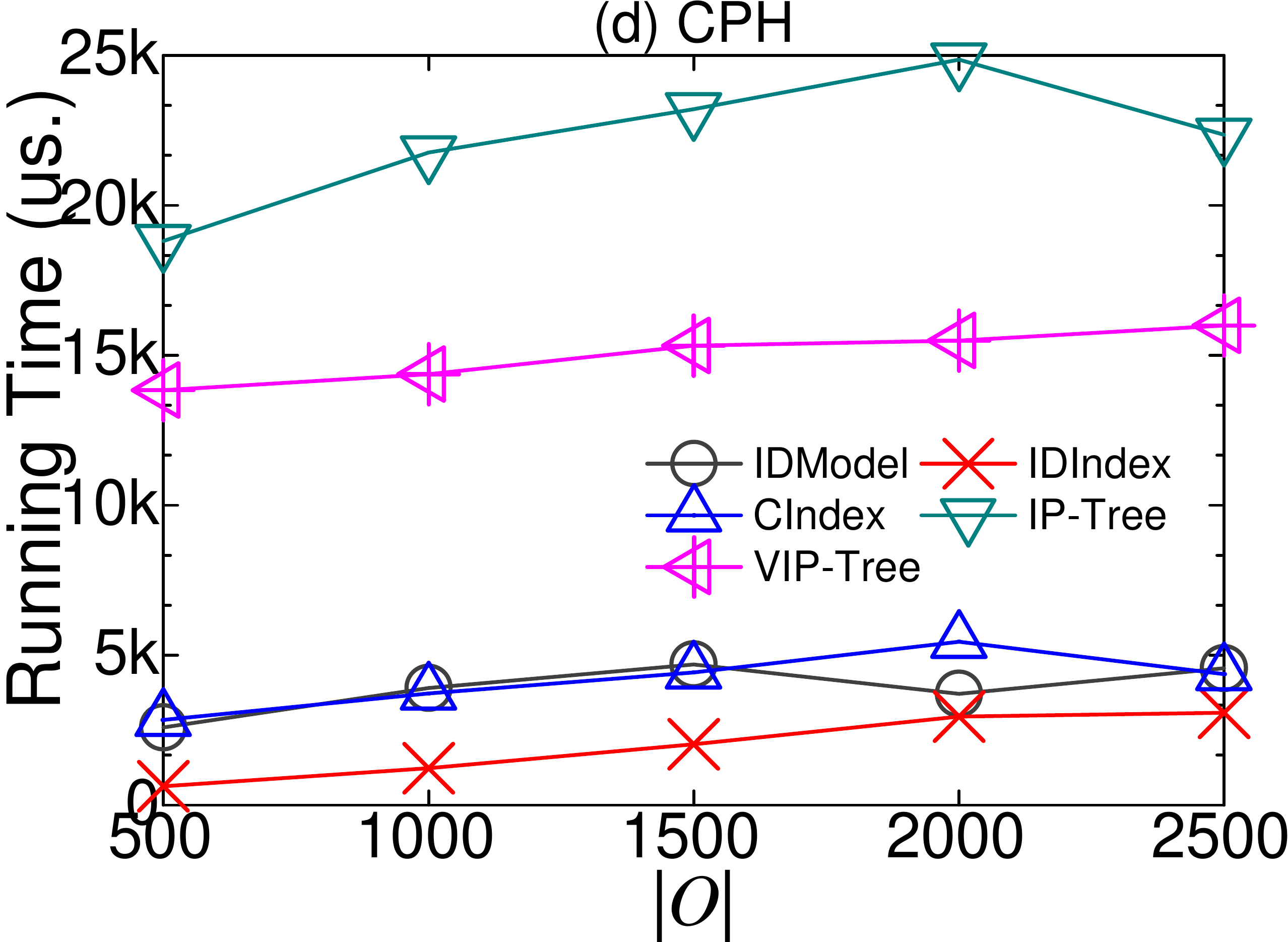}
\end{minipage}
\ExpCaption{\textsf{RQ} Time vs. $|O|$}\label{fig:B2_O_RQ_time}
\end{figure*}

\begin{figure*}[!htbp]
\centering
\begin{minipage}[t]{0.245\textwidth}
\centering
\includegraphics[width=\textwidth, height = 3cm]{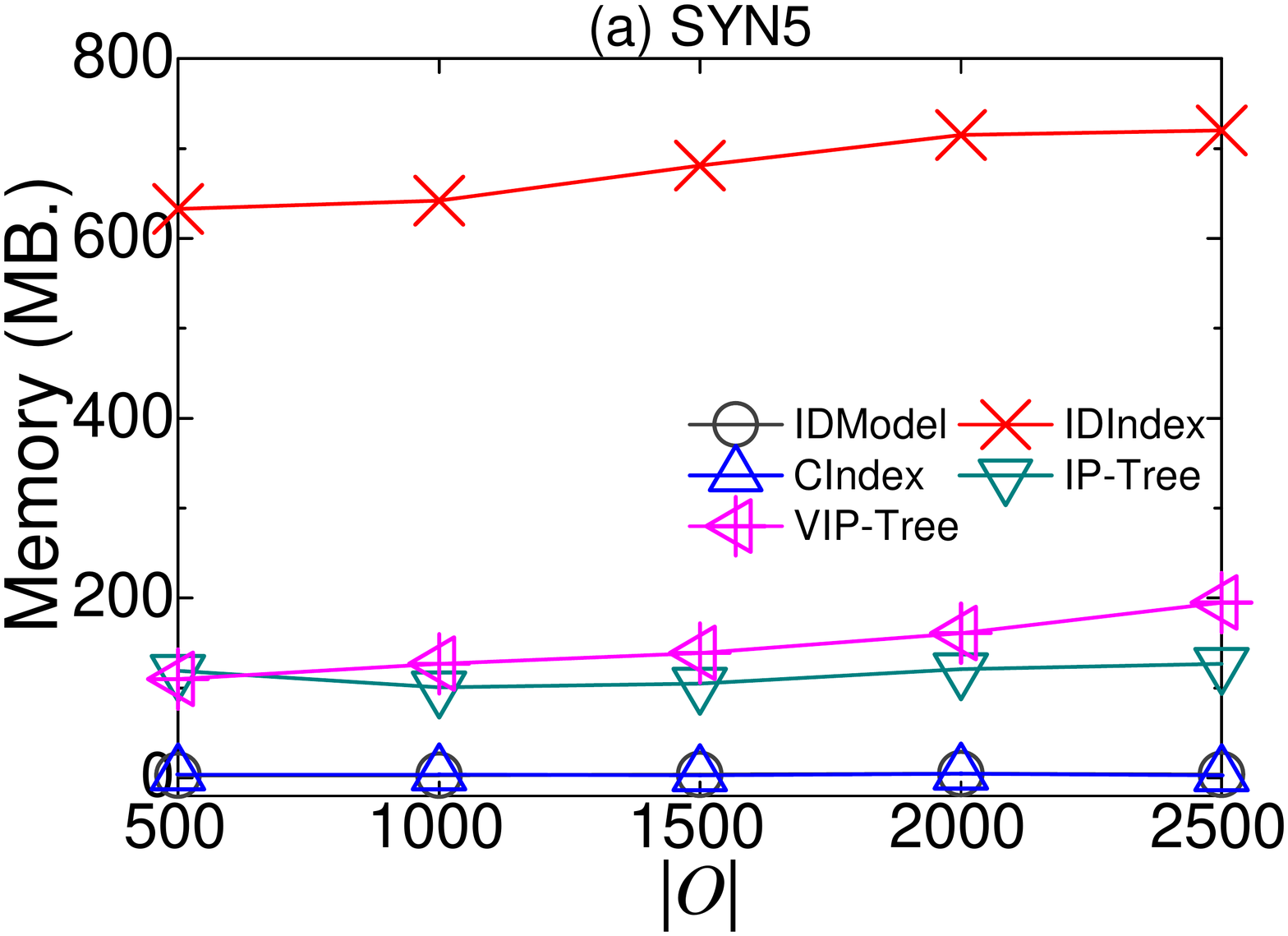}
\end{minipage}
\begin{minipage}[t]{0.245\textwidth}
\centering
\includegraphics[width=\textwidth, height = 3cm]{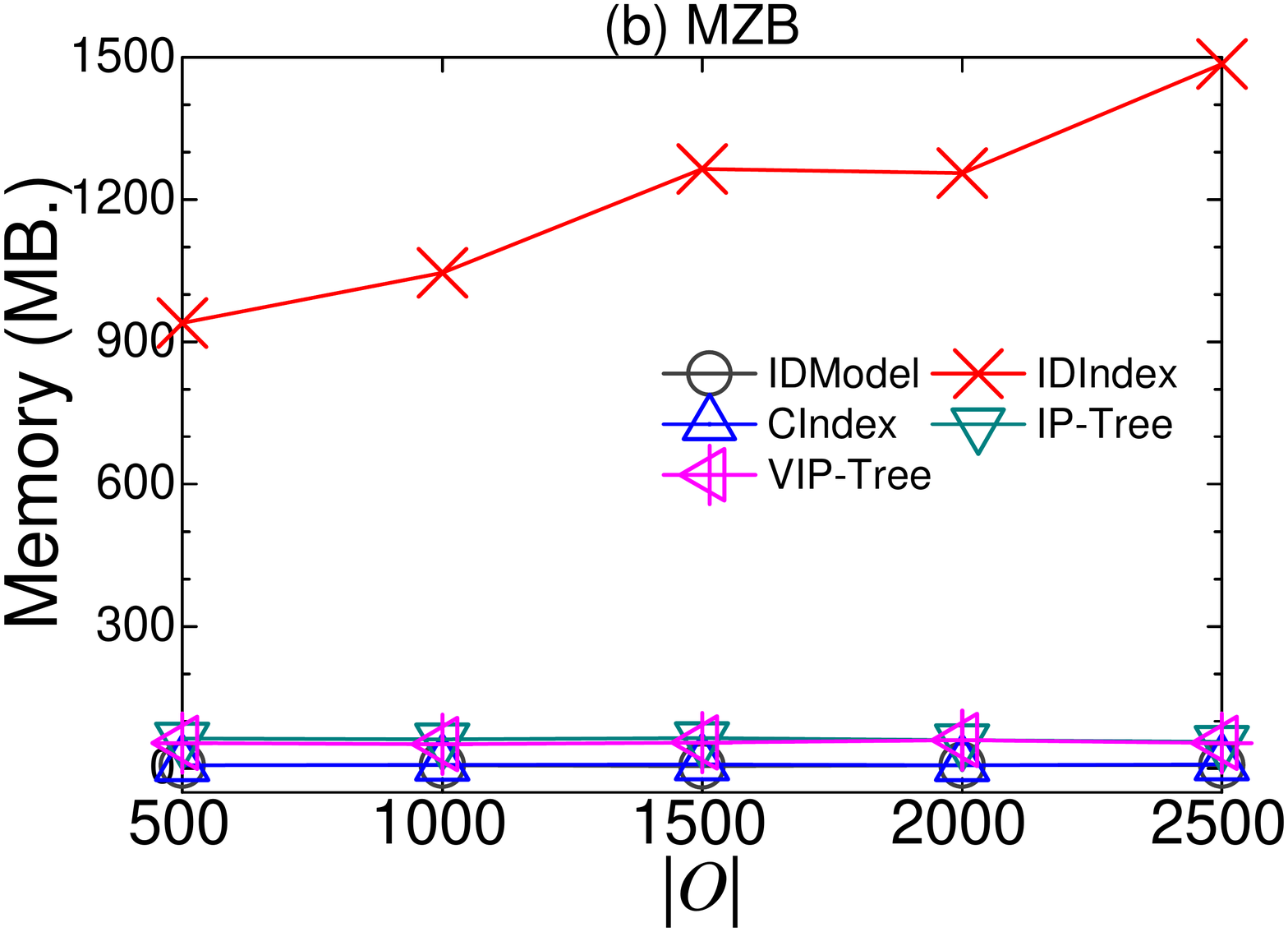}
\end{minipage}
\begin{minipage}[t]{0.245\textwidth}
\centering
\includegraphics[width=\textwidth, height = 3cm]{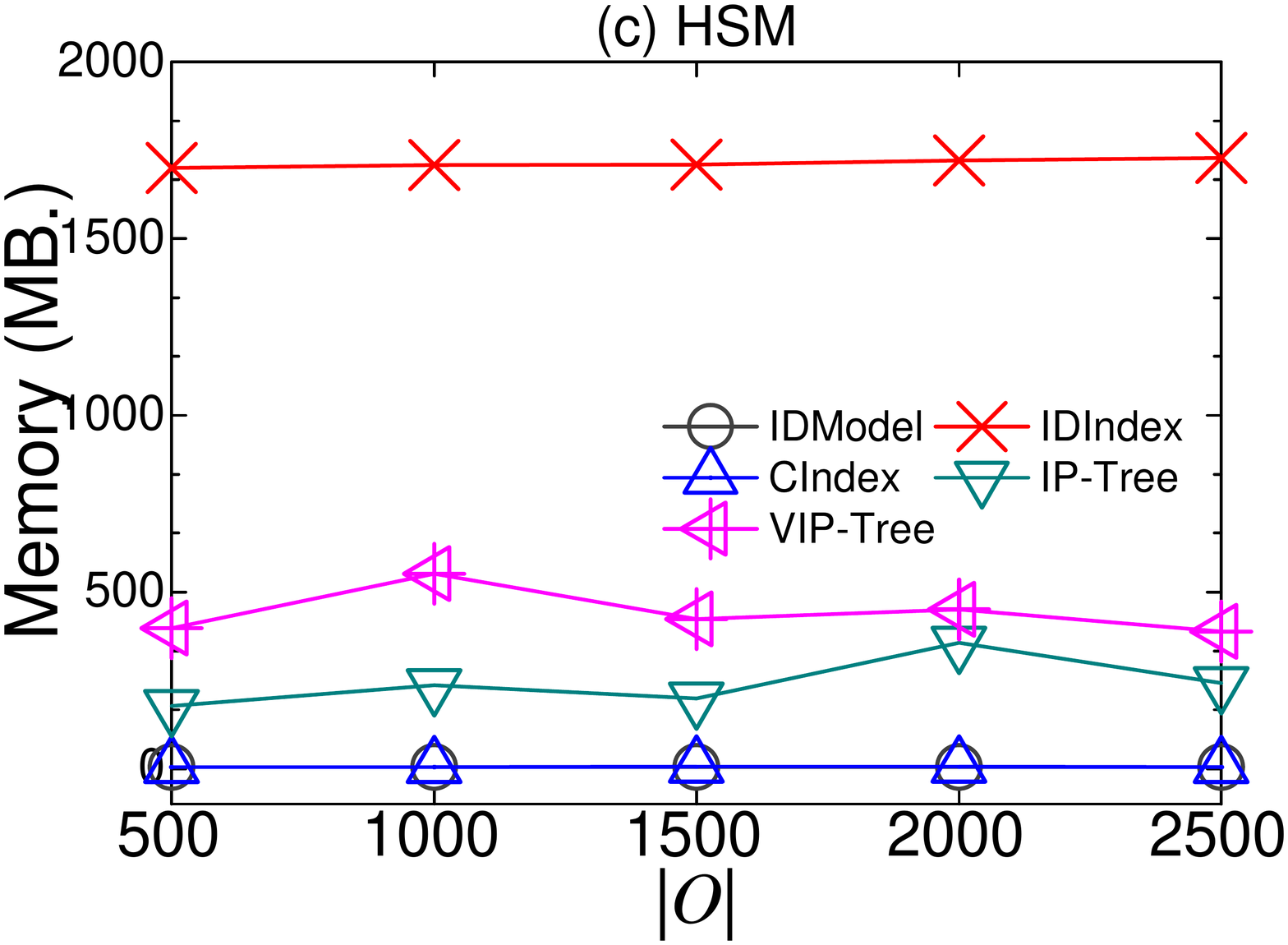}
\end{minipage}
\begin{minipage}[t]{0.245\textwidth}
\centering
\includegraphics[width=\textwidth, height = 3cm]{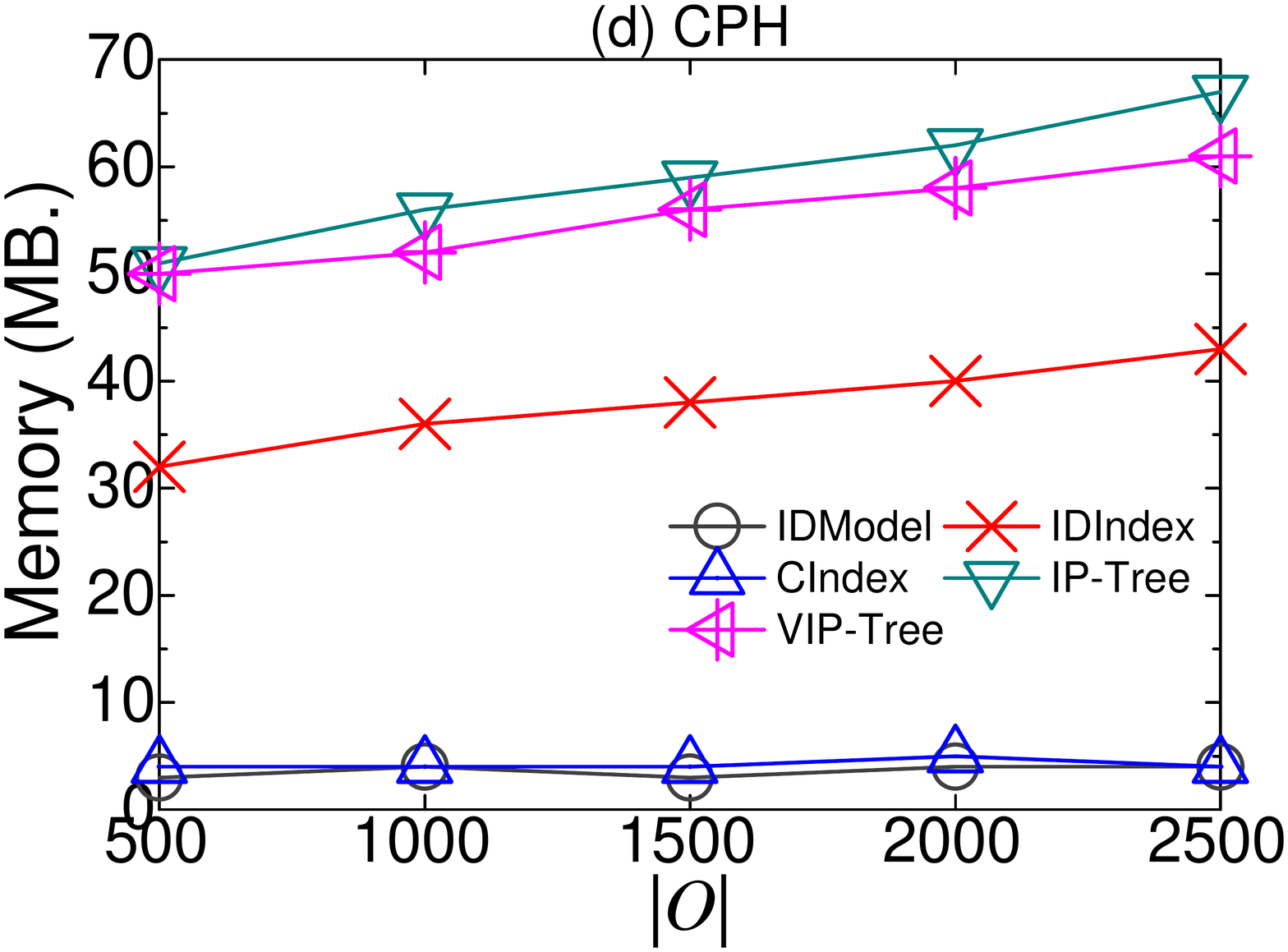}
\end{minipage}
\ExpCaption{\textsf{RQ} Memory vs. $|O|$}\label{fig:B2_O_RQ_mem}
\end{figure*}

\noindent\textsf{RQ}: With different sizes of $O$, the running time and memory use are reported in Figures~\ref{fig:B2_O_RQ_time} and~\ref{fig:B2_O_RQ_mem}, respectively.
\begin{itemize}[leftmargin=*]
\item Algorithms based on different model/indexes are almost insensitive to $|O|$ in running time, implying that each is able to prune irrelevant objects effectively and stop searching early. \change{A larger $|O|$ results in higher object density in an indoor space. This tends to increase the query processing time in general, as the query algorithms need to process larger object buckets. However, this impact is negligible according to the results in Figure~\ref{fig:B2_O_RQ_time}. This implies that all model/indexes are good at pruning indoor partitions and thus object buckets when processing \textsf{RQ}.}
\item \change{Referring to Figure~\ref{fig:B2_O_RQ_time}}, \textsc{IDIndex} runs faster than others by several orders of magnitude in all datasets, thanks to its precomputed global door-to-door distances. However, it also requires memory an order of magnitude higher to store the distance matrix \change{(Figure~\ref{fig:B2_O_RQ_mem})}.
A special case occurs on CPH \change{(Figure~\ref{fig:B2_O_RQ_mem}(d))} that \textsc{IP-Tree} and \textsc{VIP-Tree} consume more memory than others.
First, the door number of CPH is quite small such that the matrices of \textsc{IDIndex} are not large. Second, as there are fewer access doors, \textsc{IP-Tree}/\textsc{VIP-Tree} involves heavy on-the-fly computations on distances between doors and non-leaf nodes and thus needs more memory for the intermediate results.
\item On each dataset, \textsc{IDModel} and \textsc{CIndex} incur almost the same execution time \change{(see Figure~\ref{fig:B2_O_RQ_time})}, as they both use graph traversal to search for objects.
Under complex indoor topology, \textsc{CIndex} using R-tree does not have much advantage in spatial pruning.
\item \textsc{IP-Tree} and \textsc{VIP-Tree} perform differently on different datasets.
They outperform \textsc{IDModel} and \textsc{CIndex} on MZB but are worse on the others \change{(see Figure~\ref{fig:B2_O_RQ_time})}. Recall that MZB features some crucial partitions having up to 56 doors.
In such a case, the efficiency of graph traversal is much lower than searching on the tree structure.
On the contrary, when the number of candidate doors for the next hop is relatively small, the graph-based search algorithms are advantaged in range queries.
Therefore, we recommend using \textsc{IP-Tree}/\textsc{VIP-Tree} to perform \textsf{RQ} in spaces with very large main corridors.
\item \change{Referring to Figure~\ref{fig:B2_O_RQ_time}}, \textsc{VIP-Tree} is generally faster than \textsc{IP-Tree} because of more cached distances. \textsc{IP-Tree} needs to compute more intermediate results on the fly. However, memory use is close between the two \change{(see Figure~\ref{fig:B2_O_RQ_mem})}.
\end{itemize}

\begin{figure*}[!htbp]
\centering
\begin{minipage}[t]{0.245\textwidth}
\centering
\includegraphics[width=\textwidth, height = 3cm]{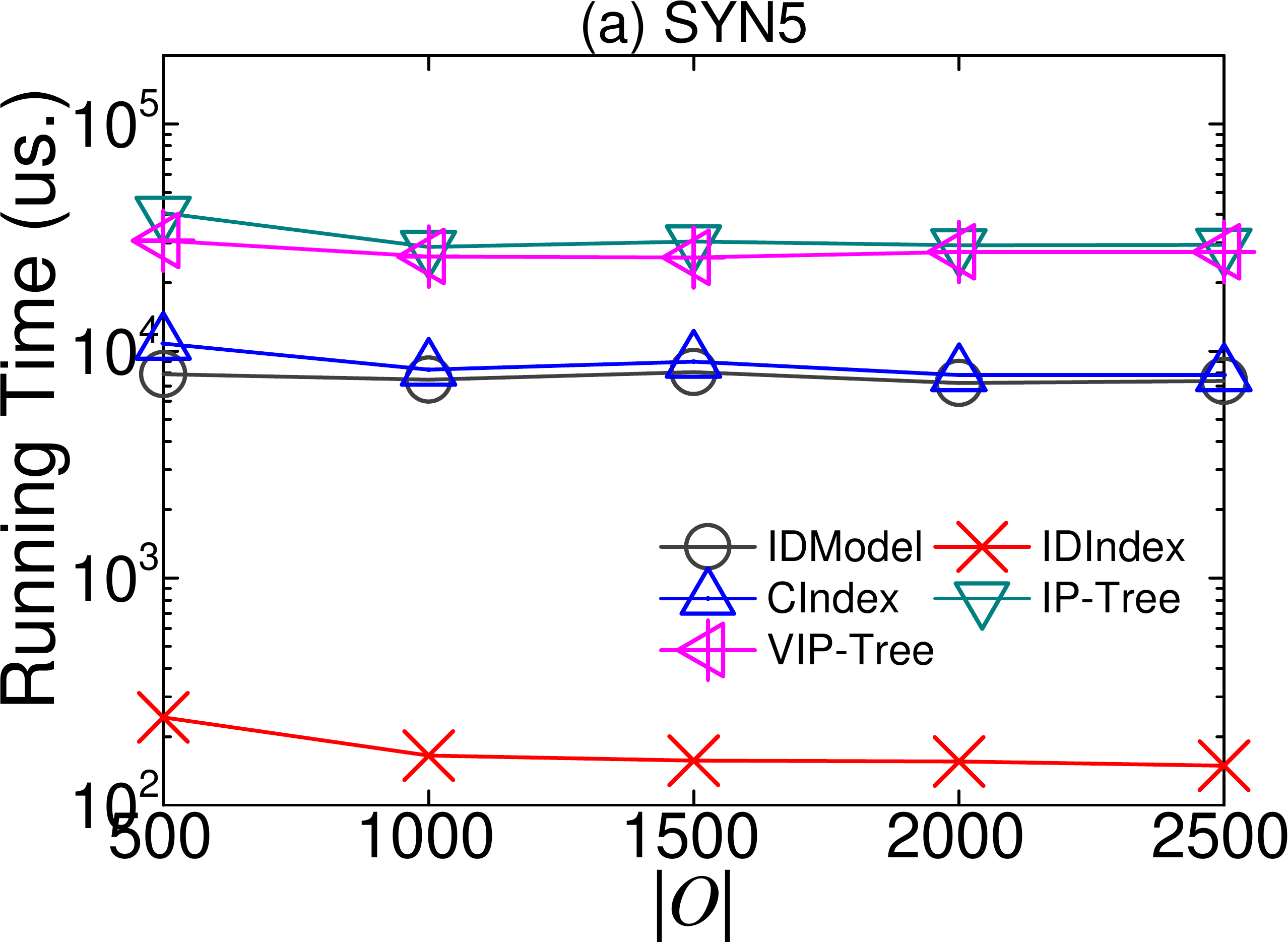}
\end{minipage}
\begin{minipage}[t]{0.245\textwidth}
\centering
\includegraphics[width=\textwidth, height = 3cm]{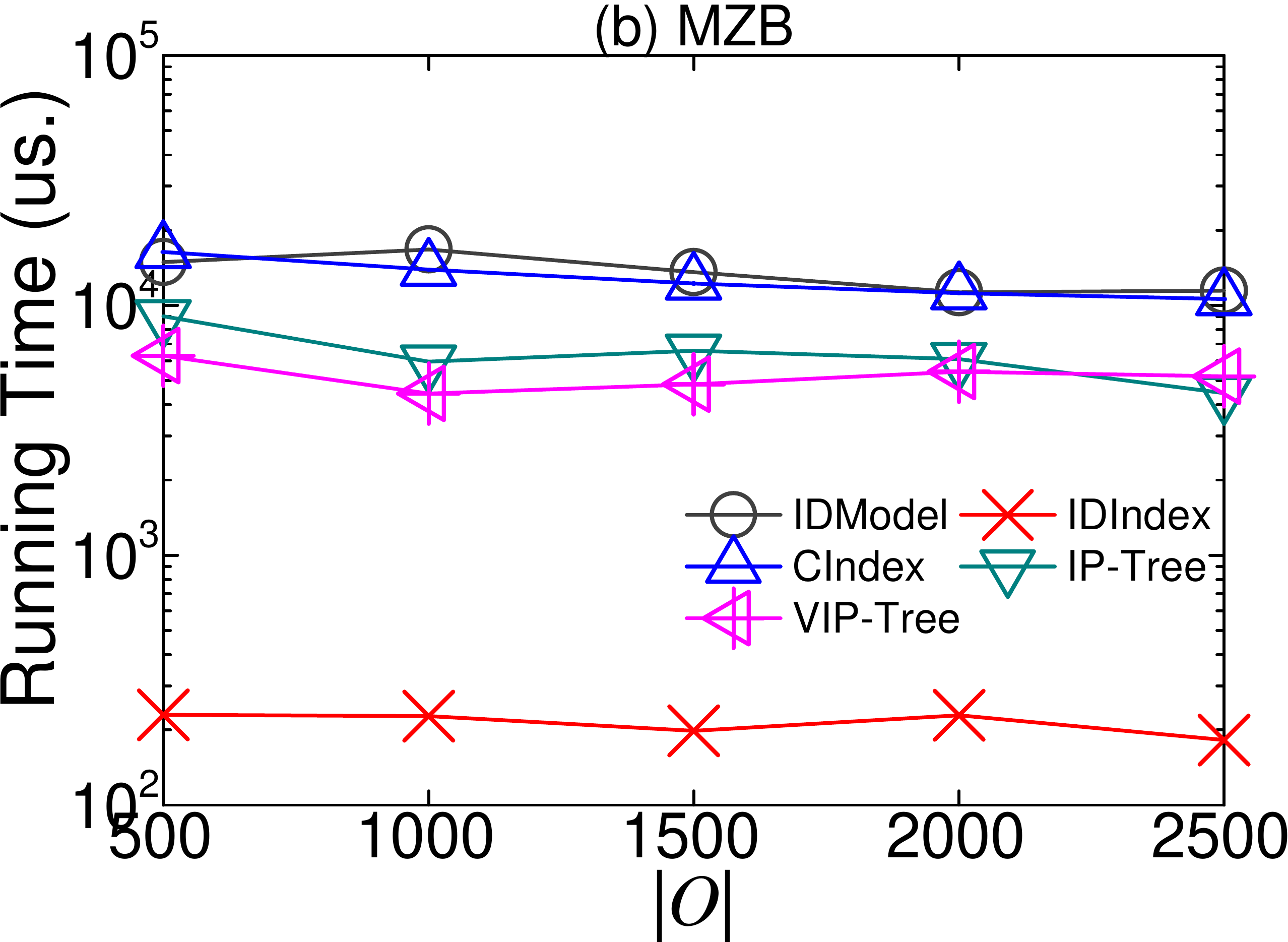}
\end{minipage}
\begin{minipage}[t]{0.245\textwidth}
\centering
\includegraphics[width=\textwidth, height = 3cm]{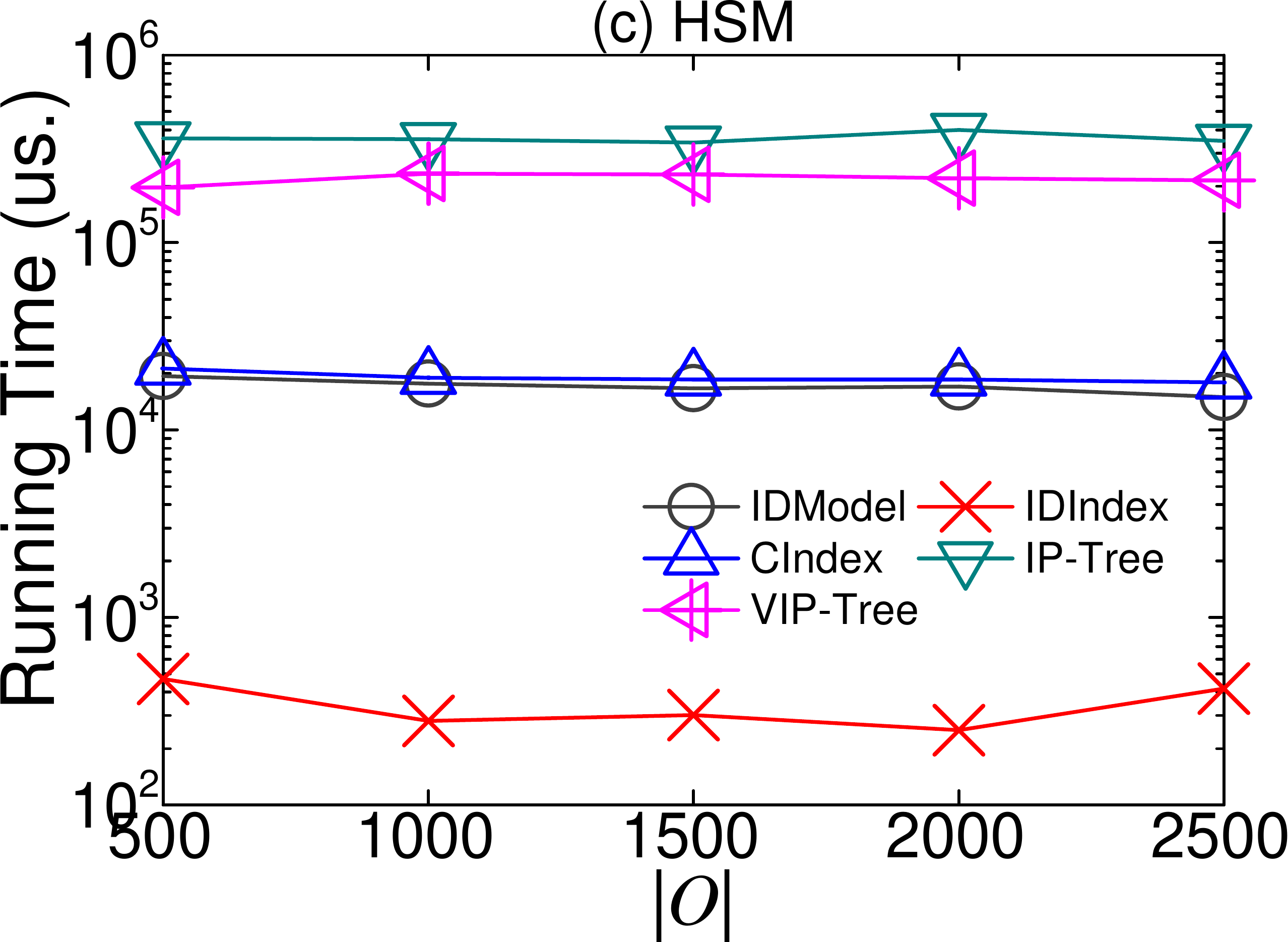}
\end{minipage}
\begin{minipage}[t]{0.245\textwidth}
\centering
\includegraphics[width=\textwidth, height = 3cm]{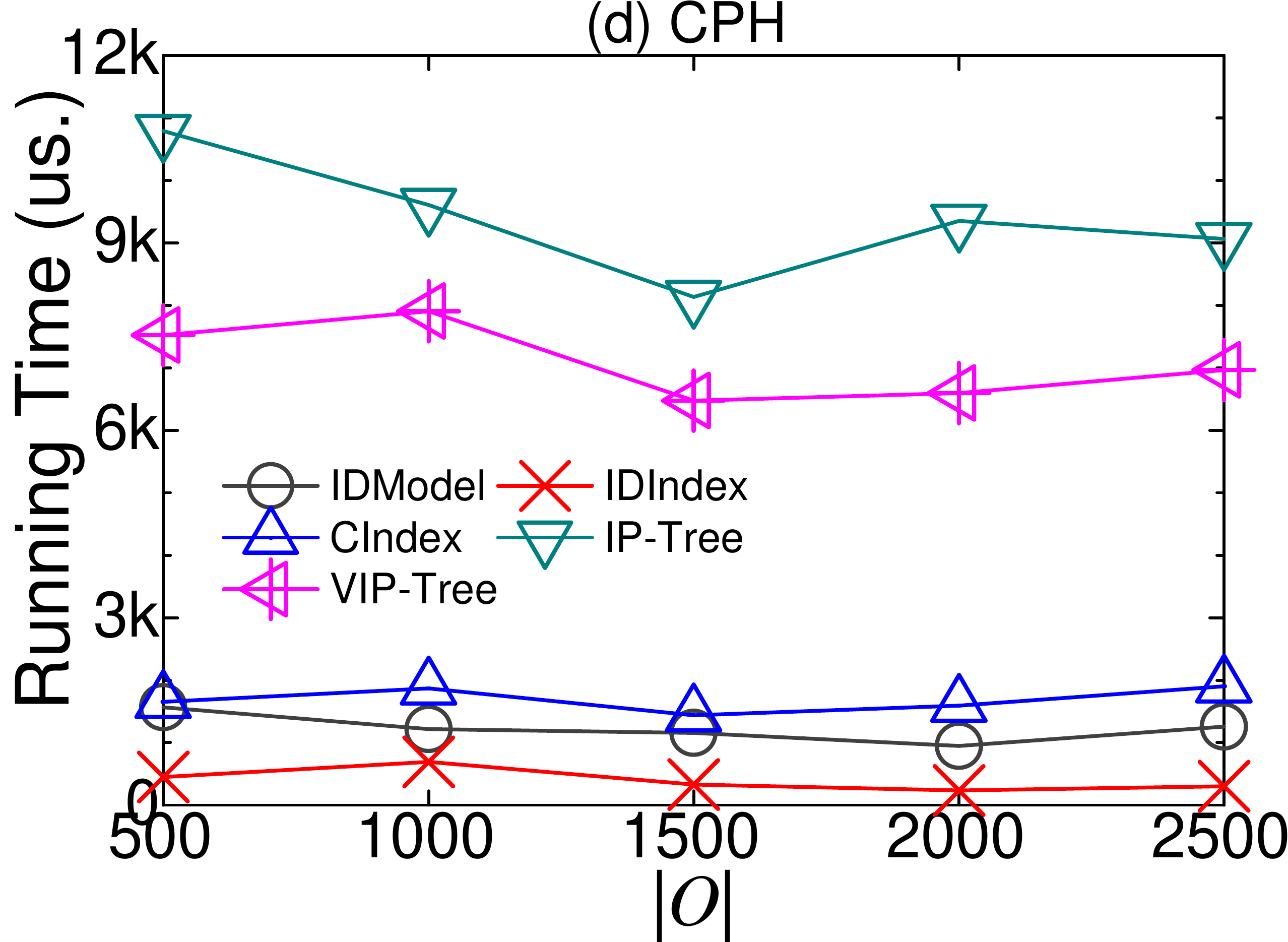}
\end{minipage}
\ExpCaption{$k$\textsf{NNQ} Time vs. $|O|$}\label{fig:B2_O_kNNQ_time}
\end{figure*}

\begin{figure*}[!htbp]
\centering
\begin{minipage}[t]{0.245\textwidth}
\centering
\includegraphics[width=\textwidth, height = 3cm]{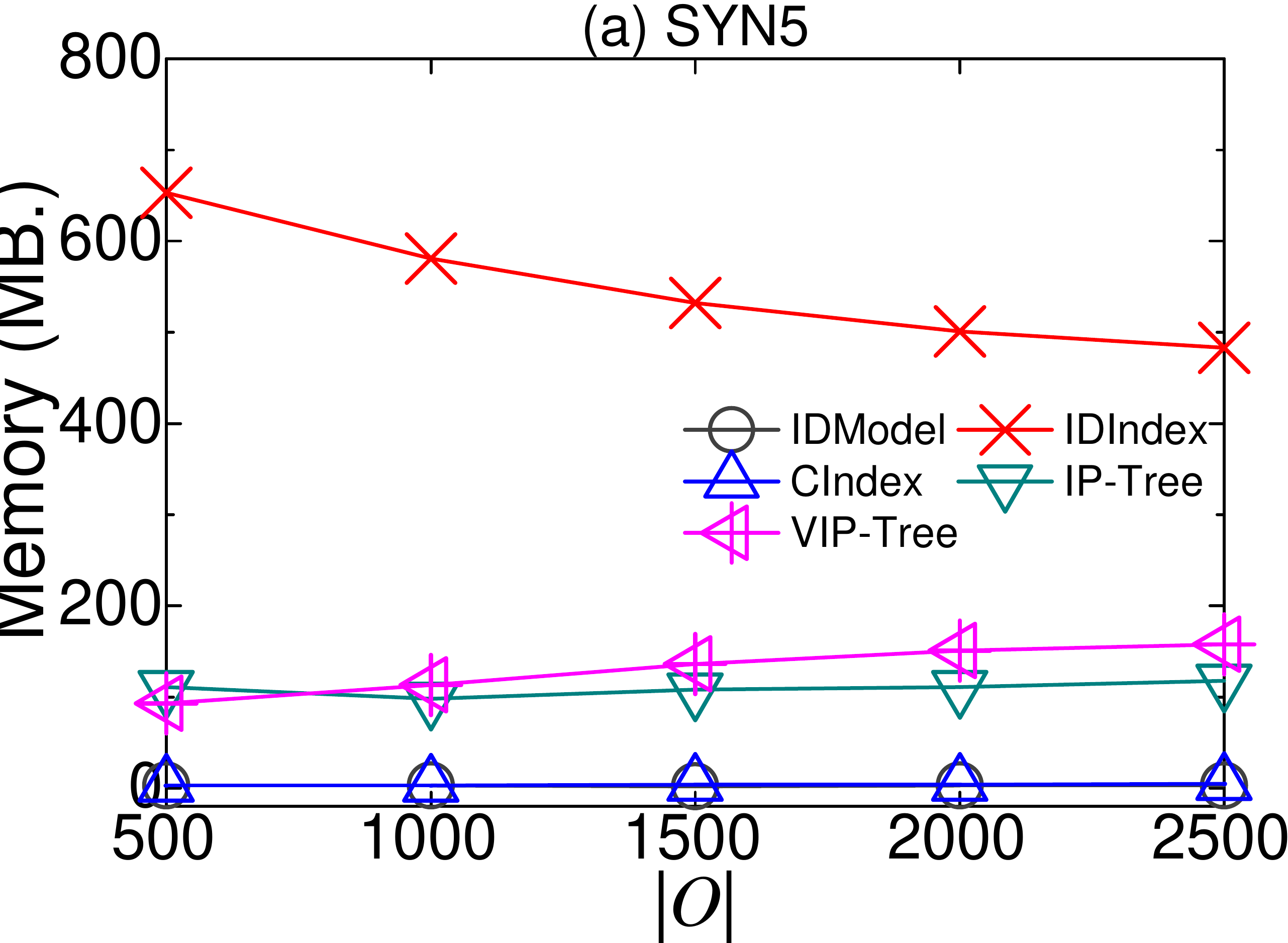}
\end{minipage}
\begin{minipage}[t]{0.245\textwidth}
\centering
\includegraphics[width=\textwidth, height = 3cm]{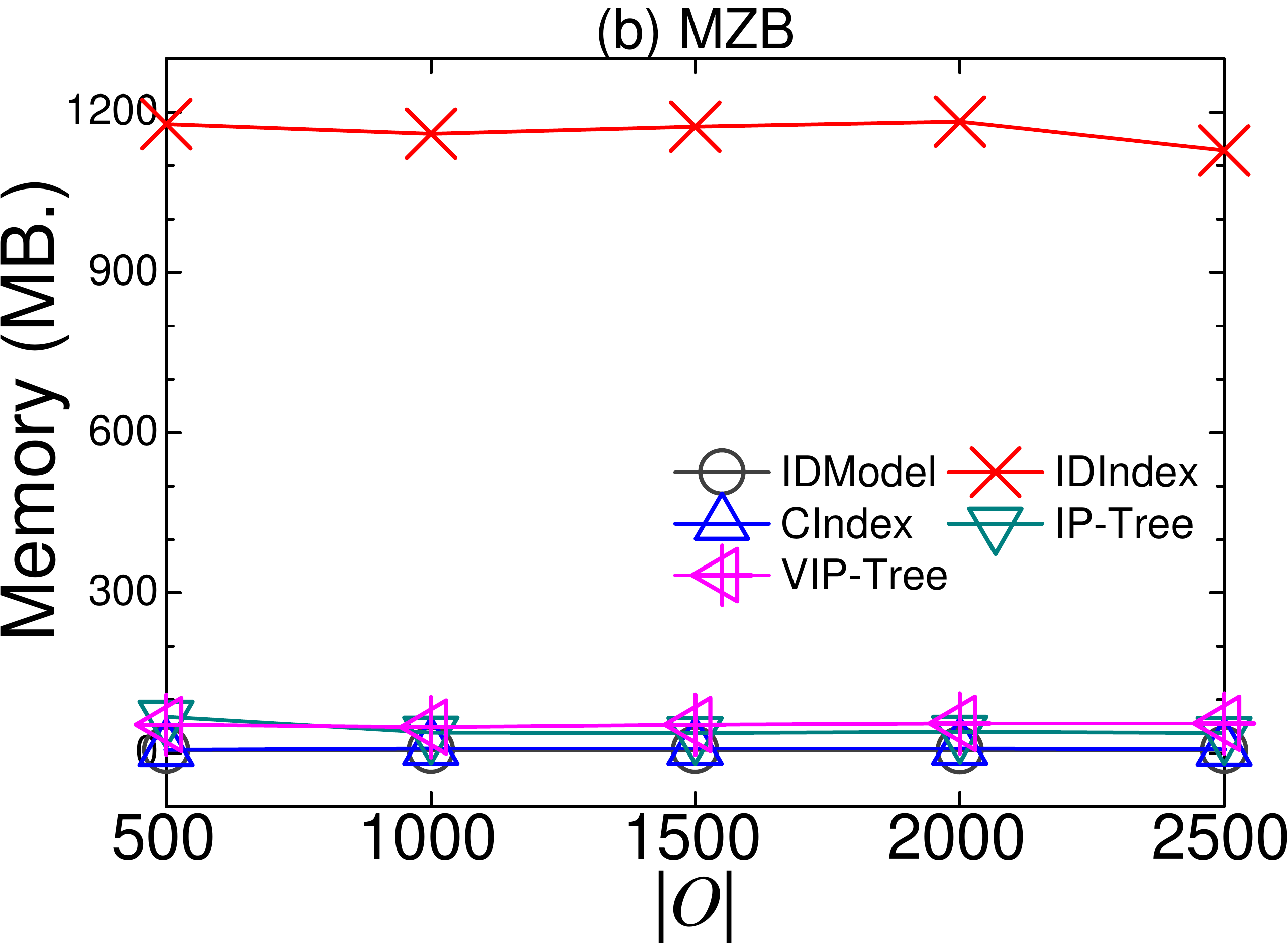}
\end{minipage}
\begin{minipage}[t]{0.245\textwidth}
\centering
\includegraphics[width=\textwidth, height = 3cm]{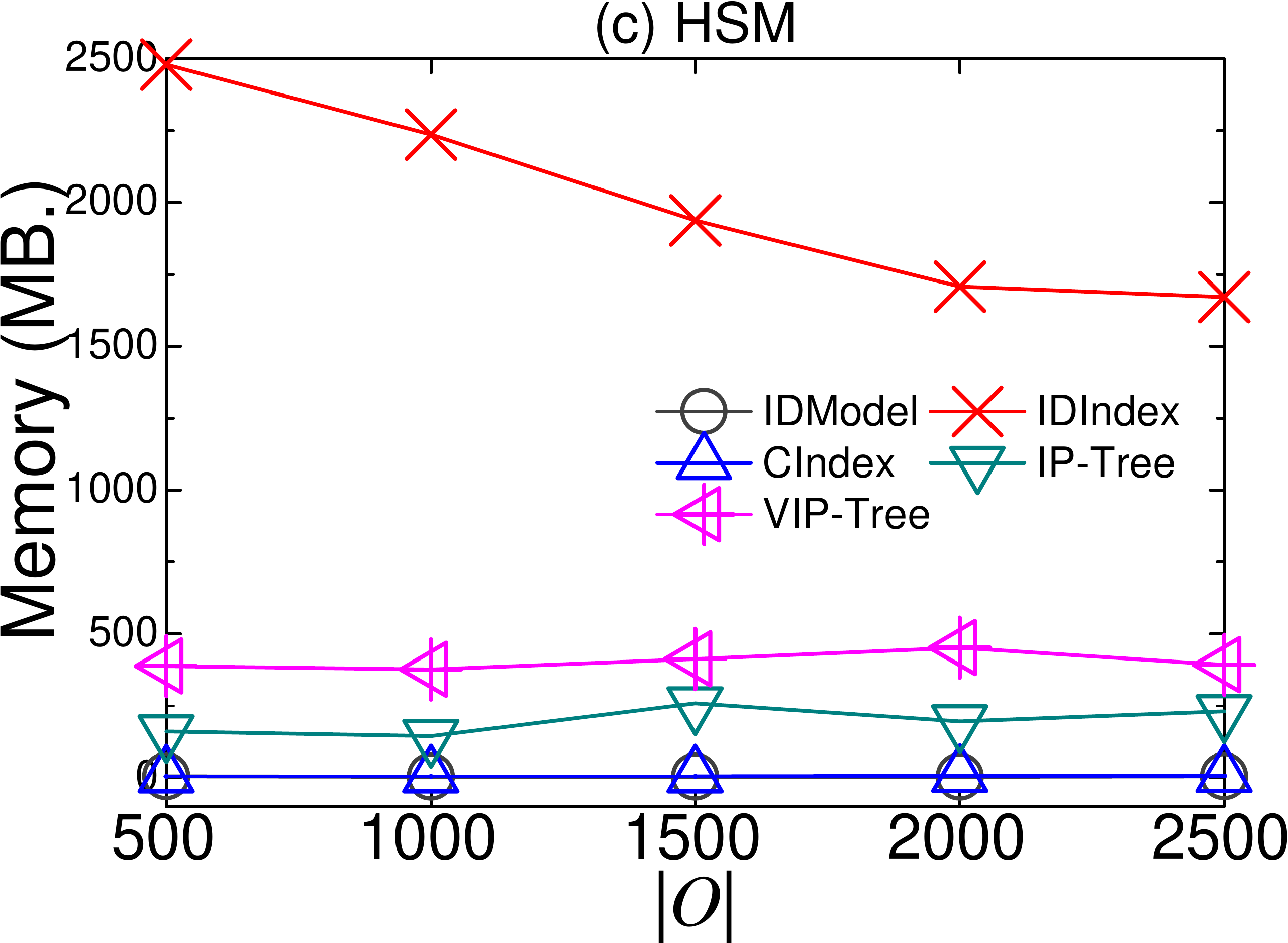}
\end{minipage}
\begin{minipage}[t]{0.245\textwidth}
\centering
\includegraphics[width=\textwidth, height = 3cm]{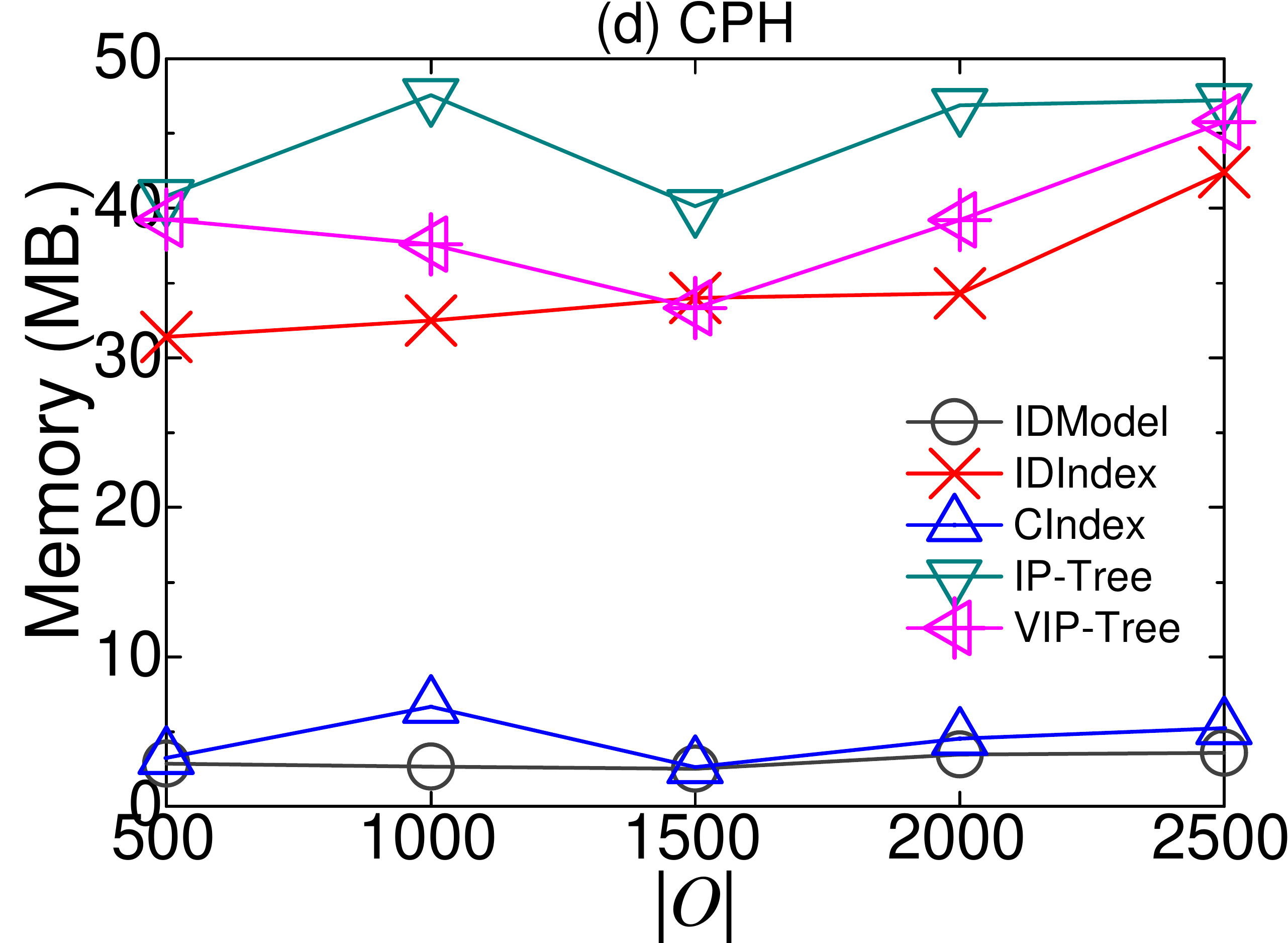}
\end{minipage}
\ExpCaption{$k$\textsf{NNQ} Memory vs. $|O|$}\label{fig:B2_O_kNNQ_mem}
\end{figure*}

\noindent$k$\textsf{NNQ}: Figures~\ref{fig:B2_O_kNNQ_time} and~\ref{fig:B2_O_kNNQ_mem} report $|O|$'s impact on the time and memory costs, respectively.
In general, each model/index's performance on $k$\textsf{NNQ} exhibits similar trend as that on \textsf{RQ}.
\begin{itemize}[leftmargin=*]
\item \change{Referring to Figure~\ref{fig:B2_O_kNNQ_time}, the time cost of each algorithm on each dataset remains stable, showing that large object workloads (and high object density) have little effect on all models.}
\item On datasets with relatively large numbers of doors and partitions (i.e., SYN5, MZB, and HSM), \textsc{IDIndex} runs faster by orders of magnitude. However, its memory use is clearly larger.
\item On one-floor CPH with small numbers of doors and partitions, \textsc{IP-Tree} and \textsc{VIP-Tree} incur more running time as well as higher memory use \change{(Figures~\ref{fig:B2_O_kNNQ_time}(d) and~\ref{fig:B2_O_kNNQ_mem}(d))}. However, they run faster on MZB \change{(Figure~\ref{fig:B2_O_kNNQ_time}(b))} in which many access doors exist due to many crucial partitions (see Table~\ref{tab:dataset}).
\item \textsc{IDModel} and \textsc{CIndex} perform comparably \change{as shown in Figures~\ref{fig:B2_O_kNNQ_time} and~\ref{fig:B2_O_kNNQ_mem}}. Without a specially designed partition R-tree, \textsc{IDModel} achieves quite good object pruning due to the efficient distance mapping maintained in its edges and vertexes.
\end{itemize}

\ExpHead{B3 Effect of Range Distance $r$}

\begin{figure*}[!htbp]
\centering
\begin{minipage}[t]{0.245\textwidth}
\centering
\includegraphics[width=\textwidth, height = 3cm]{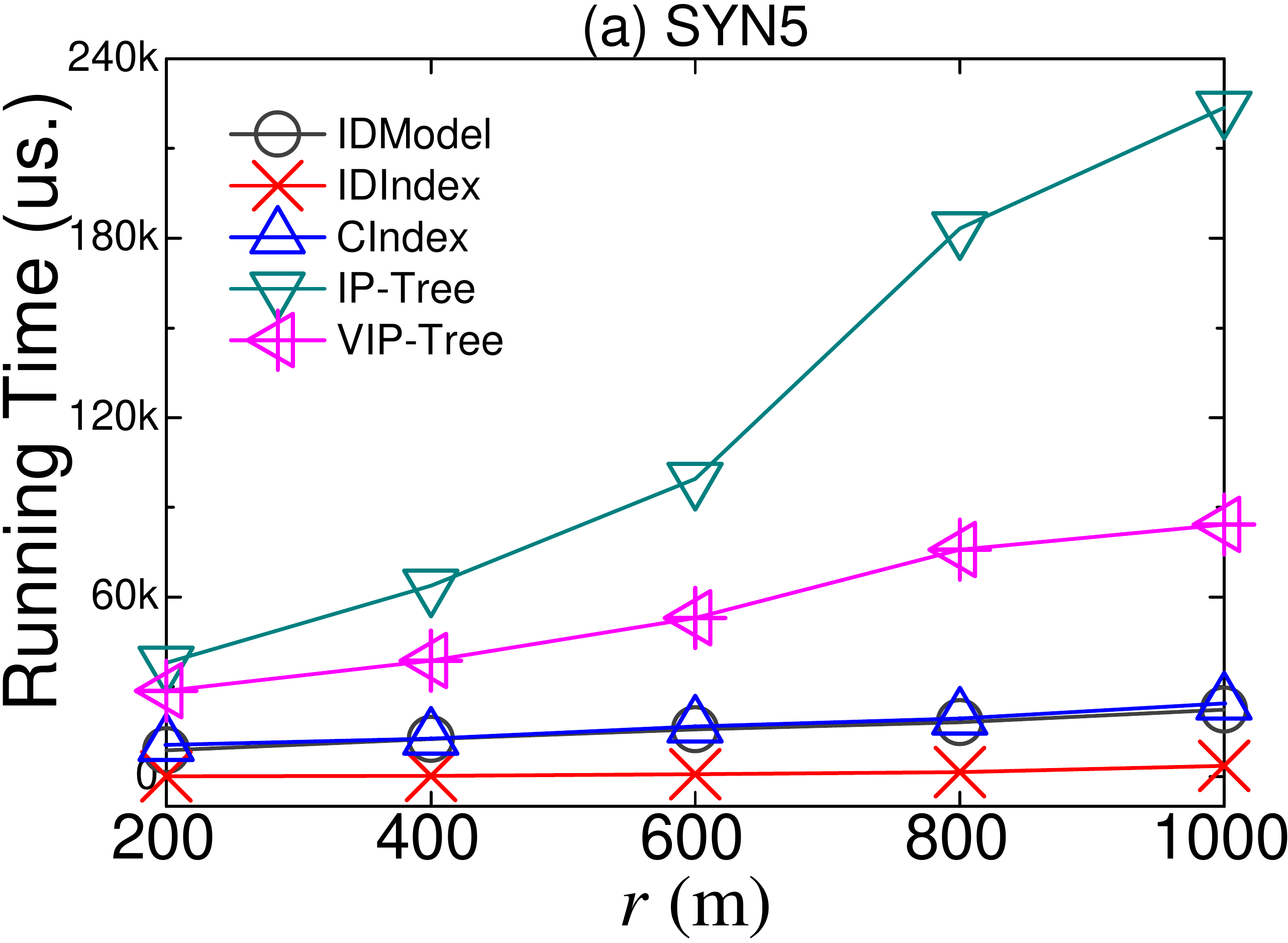}
\end{minipage}
\begin{minipage}[t]{0.245\textwidth}
\centering
\includegraphics[width=\textwidth, height = 3cm]{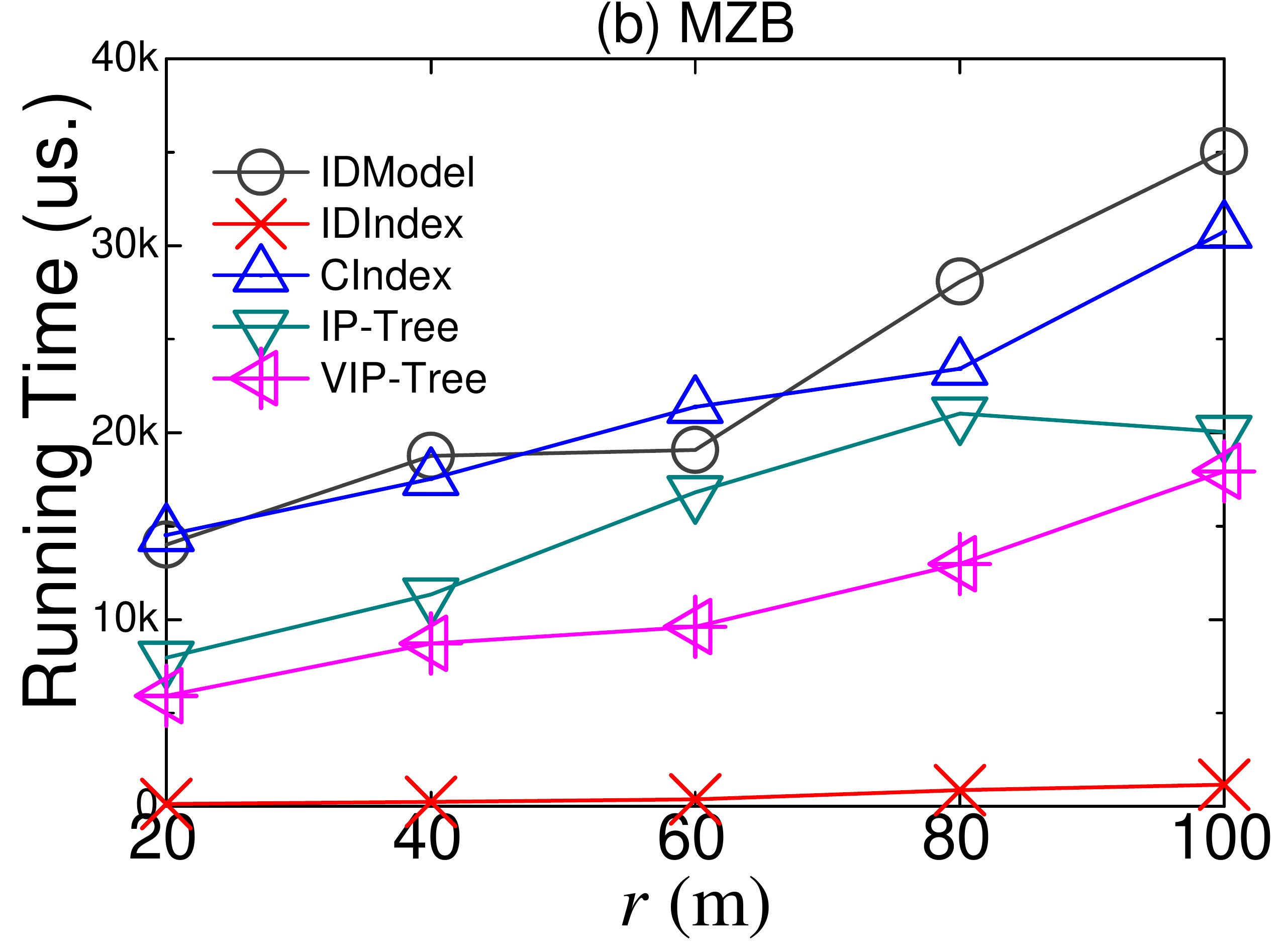}
\end{minipage}
\begin{minipage}[t]{0.245\textwidth}
\centering
\includegraphics[width=\textwidth, height = 3cm]{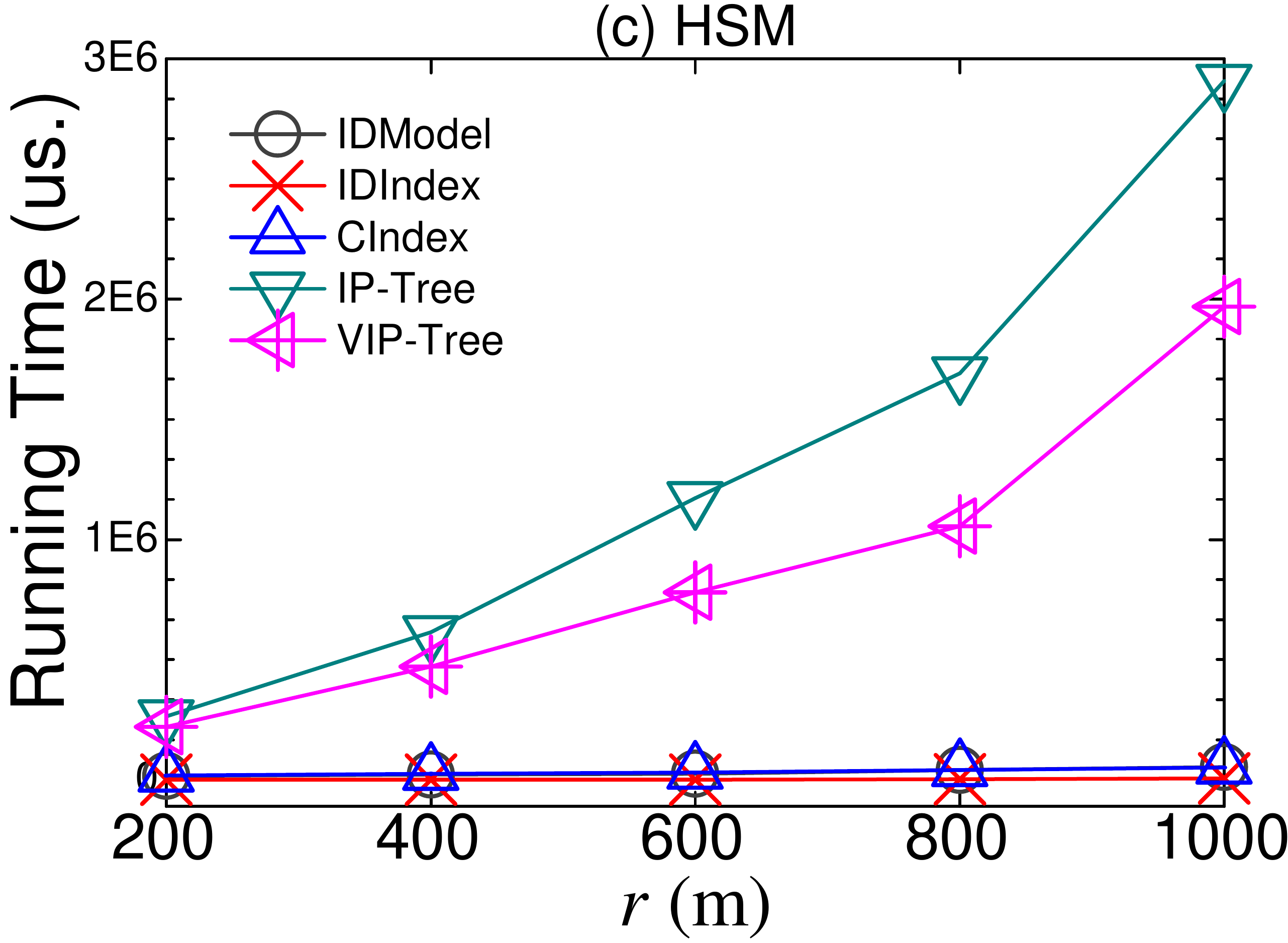}
\end{minipage}
\begin{minipage}[t]{0.245\textwidth}
\centering
\includegraphics[width=\textwidth, height = 3cm]{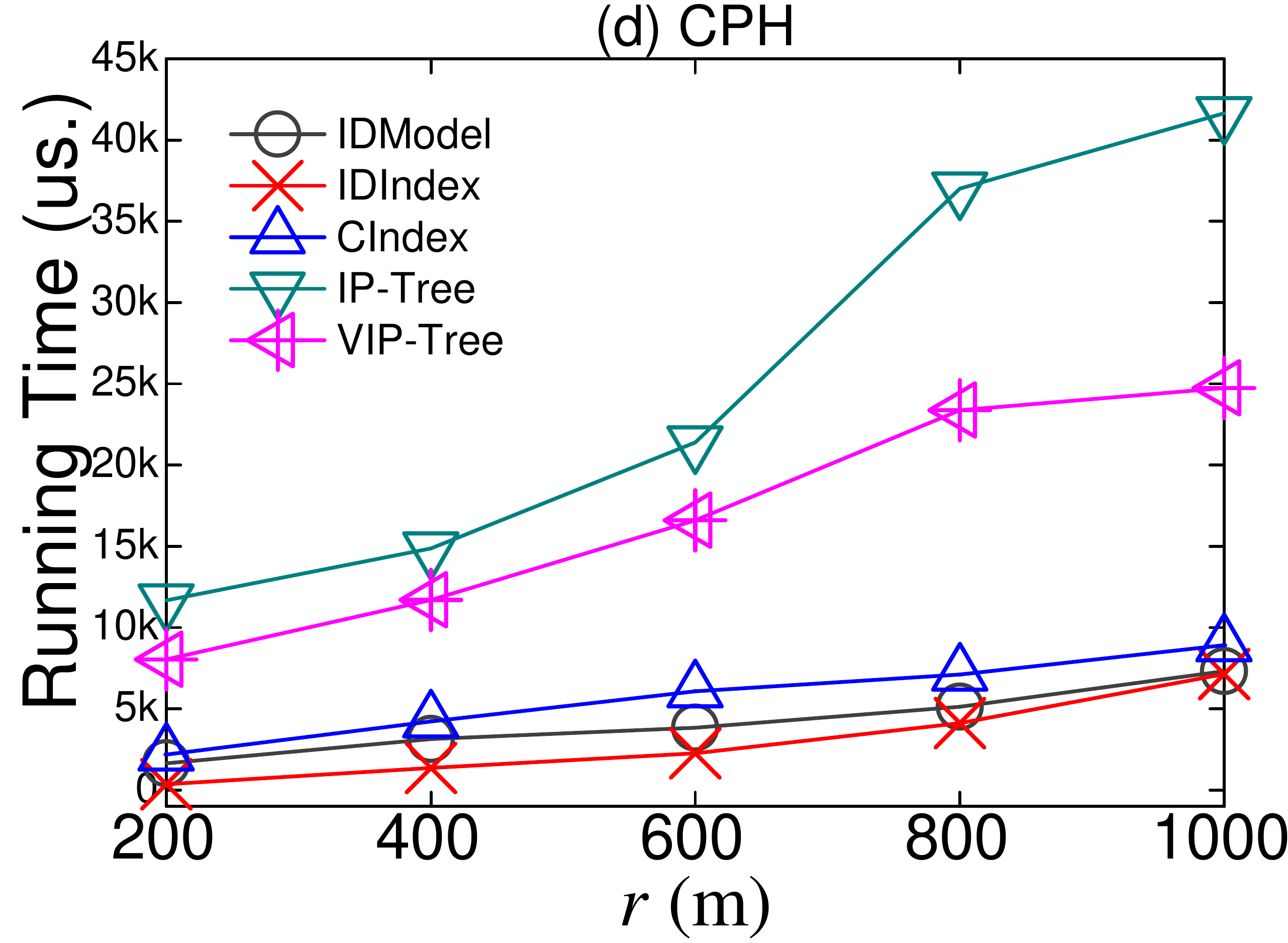}
\end{minipage}
\ExpCaption{\textsf{RQ} Time vs. $r$}\label{fig:B3_r_time}
\end{figure*}

\begin{figure*}[!htbp]
\centering
\begin{minipage}[t]{0.245\textwidth}
\centering
\includegraphics[width=\textwidth, height = 3cm]{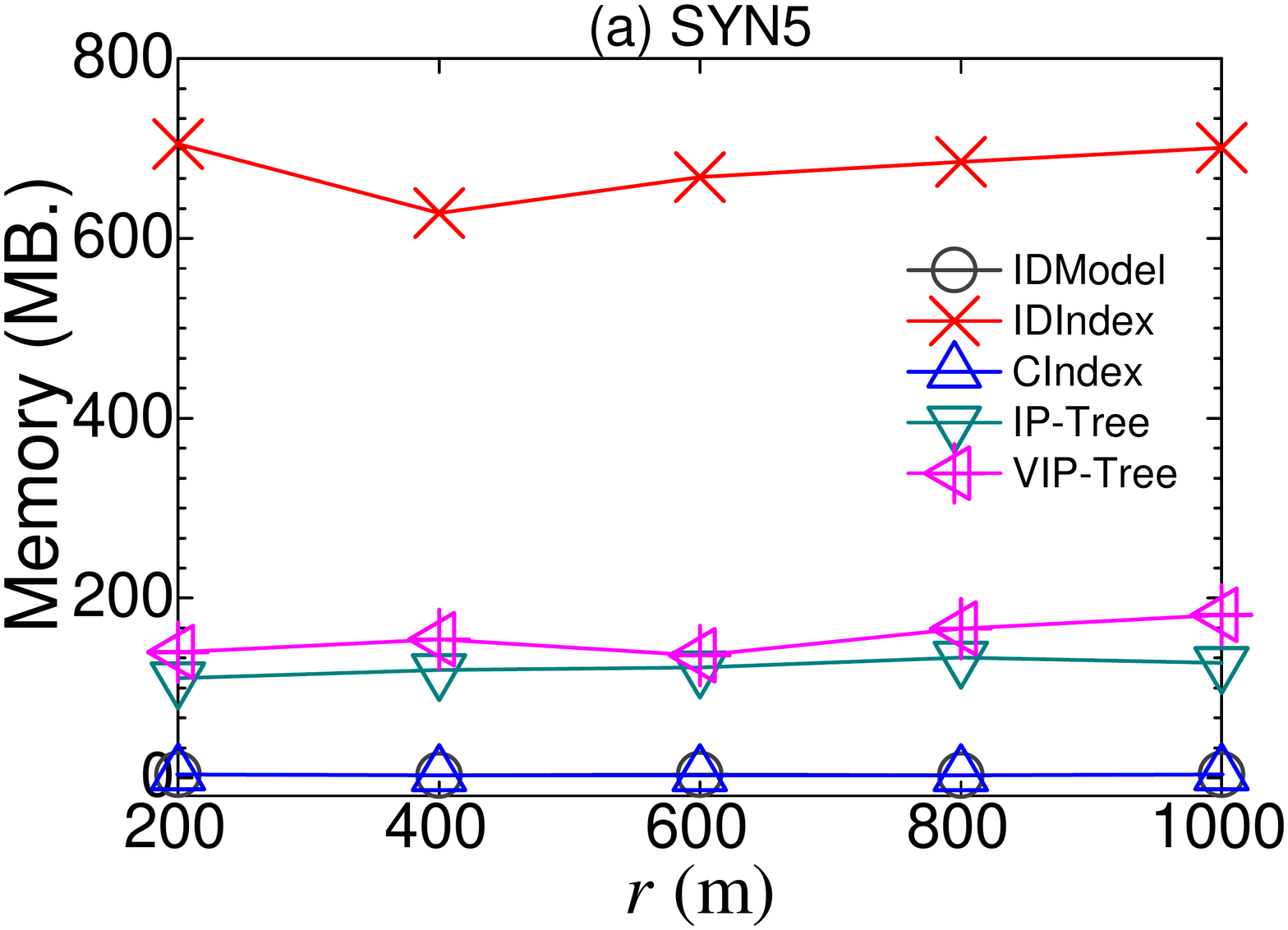}
\end{minipage}
\begin{minipage}[t]{0.245\textwidth}
\centering
\includegraphics[width=\textwidth, height = 3cm]{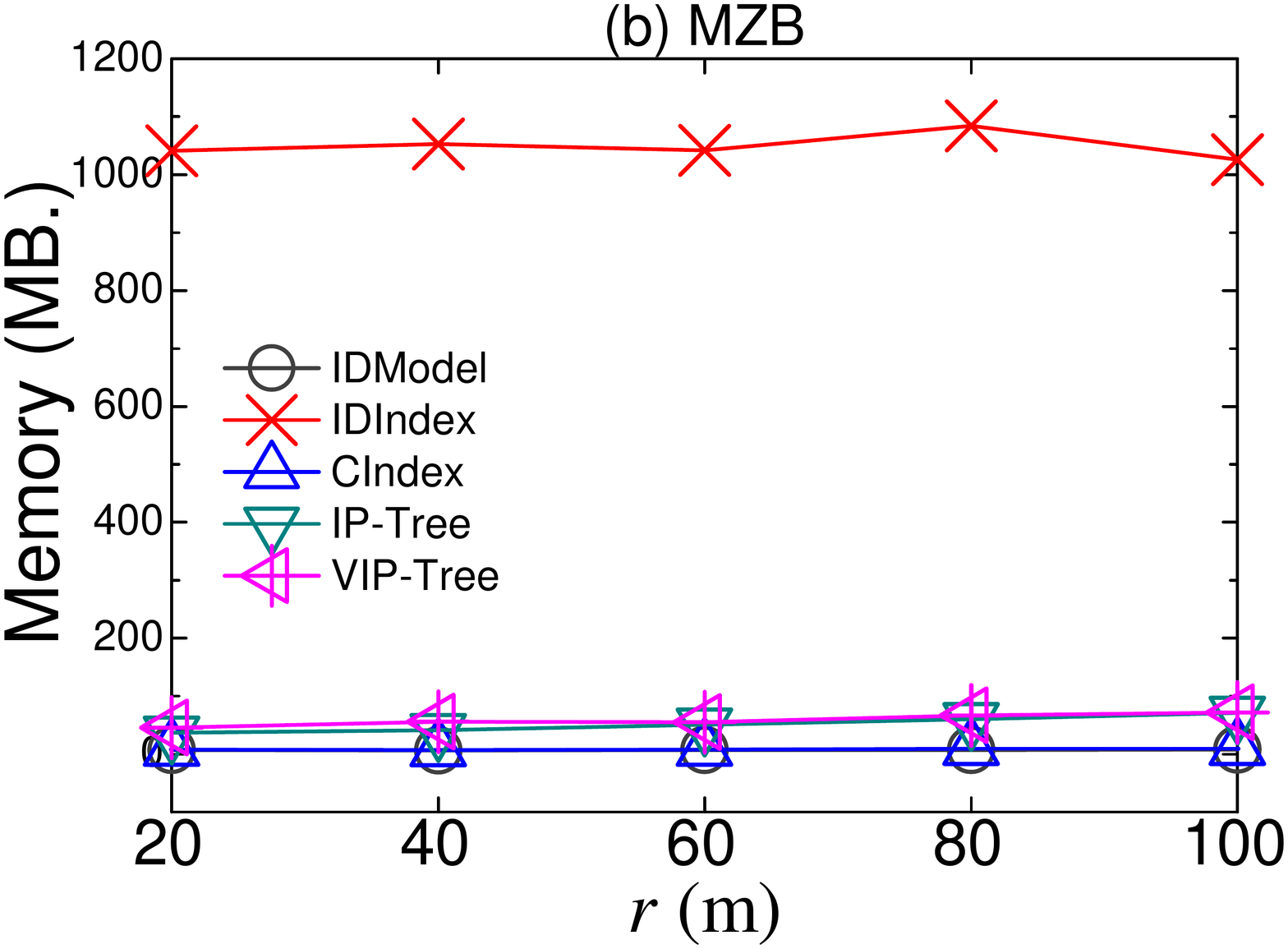}
\end{minipage}
\begin{minipage}[t]{0.245\textwidth}
\centering
\includegraphics[width=\textwidth, height = 3cm]{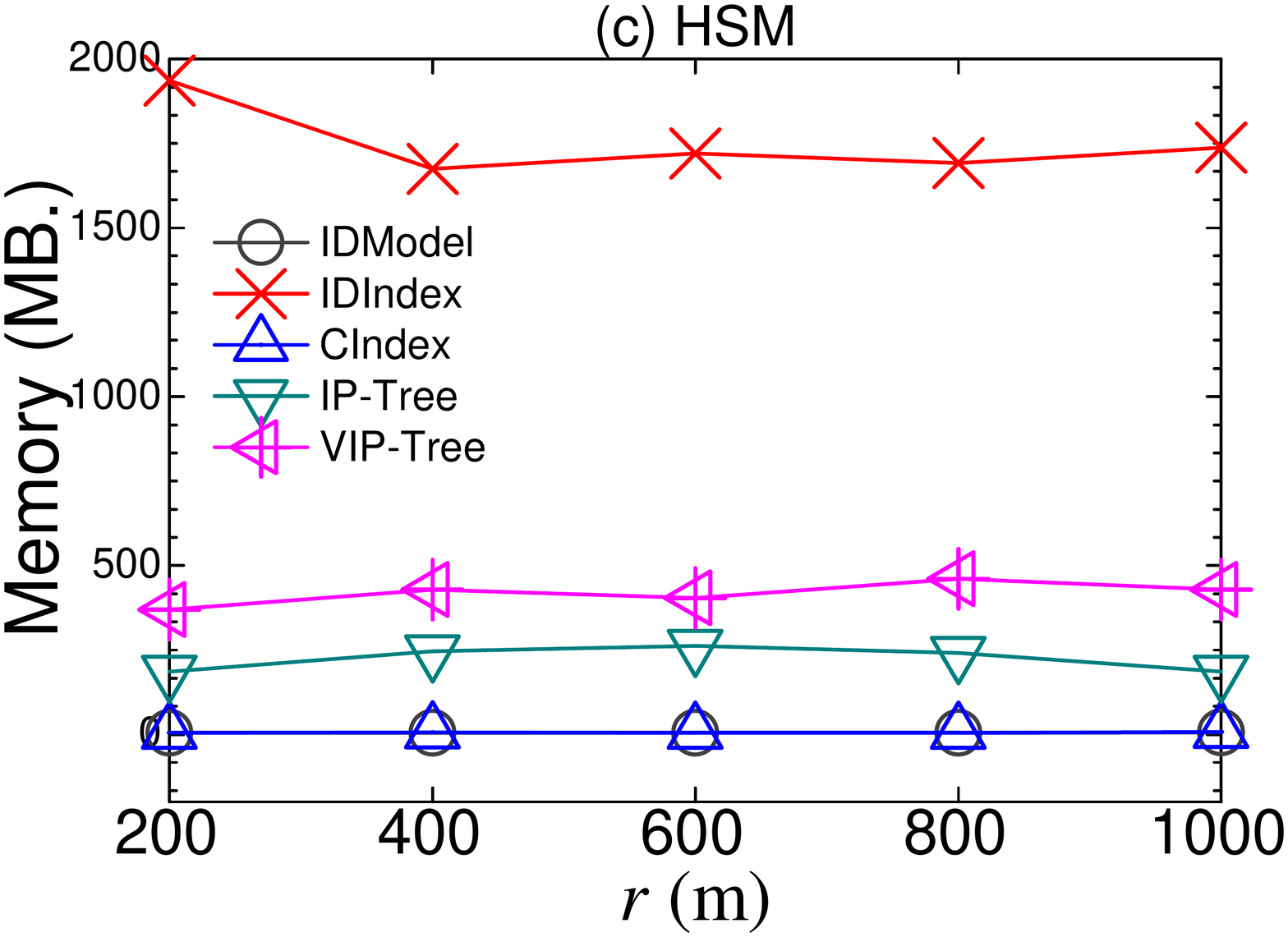}
\end{minipage}
\begin{minipage}[t]{0.245\textwidth}
\centering
\includegraphics[width=\textwidth, height = 3cm]{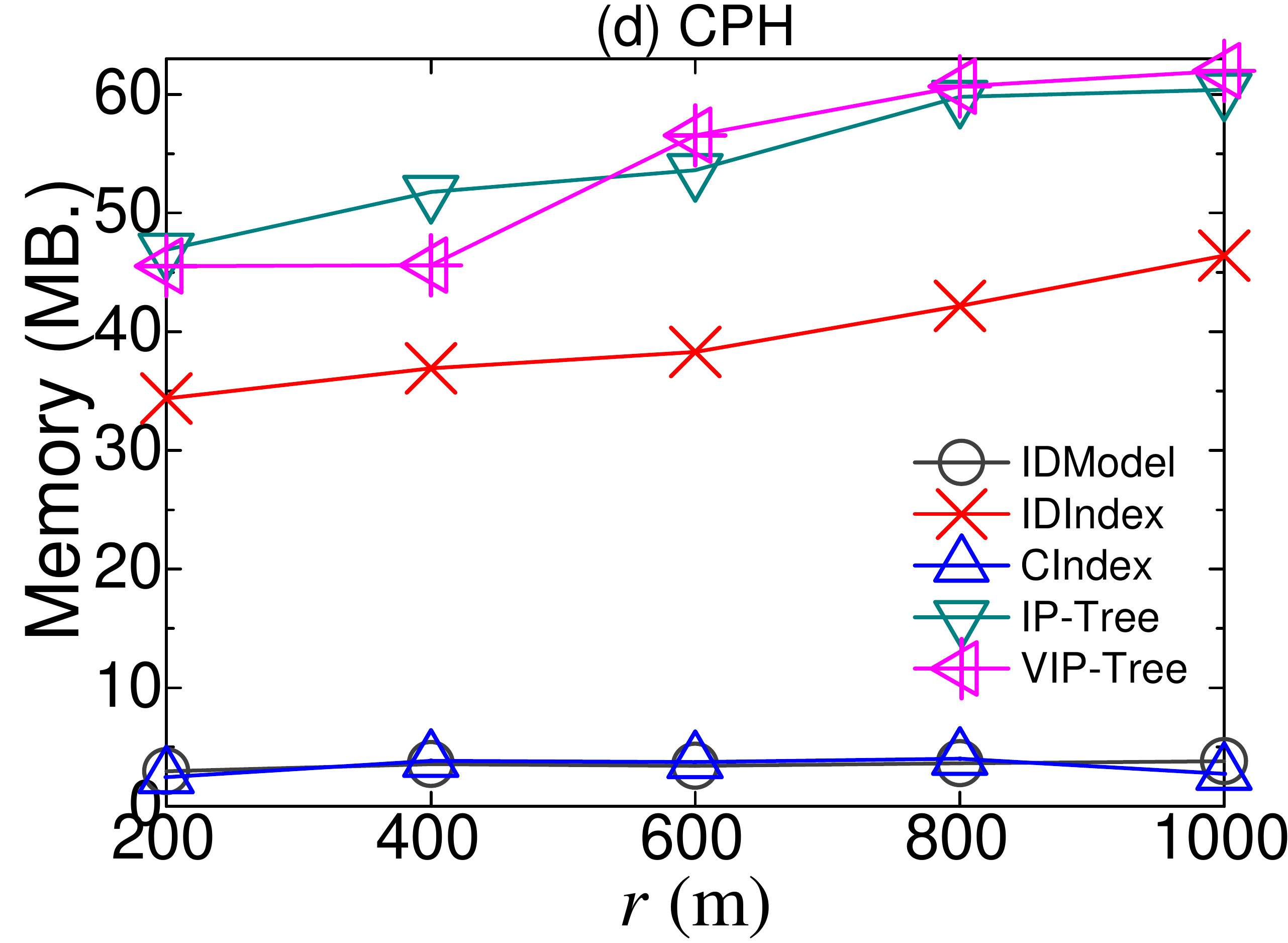}
\end{minipage}
\ExpCaption{\textsf{RQ} Memory vs. $r$}\label{fig:B3_r_mem}
\end{figure*}

\noindent\textsf{RQ}: The time and memory costs with respect to varied $r$ are reported in Figures~\ref{fig:B3_r_time} and~\ref{fig:B3_r_mem}, respectively.
\begin{itemize}[leftmargin=*]
\item On SYN5, MZB, and HSM with complex indoor topology, \textsc{IDIndex}'s running time \change{reported in Figure~\ref{fig:B3_r_time}} increases slowly with a growing $r$. In contrast, on the simple-topology CPH, the advantage of \textsc{IDIndex} over others is not marked.
\item \textsc{IDModel} and \textsc{CIndex} perform well on all datasets, except on MZB \change{(Figure~\ref{fig:B3_r_time}(b))} that has a large number of crucial partitions. This again reflects the disadvantages of the graph-based traversal algorithms when dealing with this particular topology type. Nevertheless, through efficient node search and on-the-fly distance computation, these two model/indexes always have the smallest memory overhead.
\item When increasing $r$, the running time of \textsc{IP-Tree} and \textsc{VIP-Tree} \change{in Figure~\ref{fig:B3_r_time}} increase steadily on all datasets. A larger $r$ needs to consider a tree node farther from the node where the query point is located, and thus introduces more computations on the distance from a door to some non-leaf nodes. As the distance to the access door of each ancestor node is materialized at the leaf node, \textsc{VIP-Tree} runs faster than \textsc{IP-Tree}.
\end{itemize}

\ExpHead{B4 Effect of $k$}

\begin{figure*}[!htbp]
\centering
\begin{minipage}[t]{0.245\textwidth}
\centering
\includegraphics[width=\textwidth, height = 3cm]{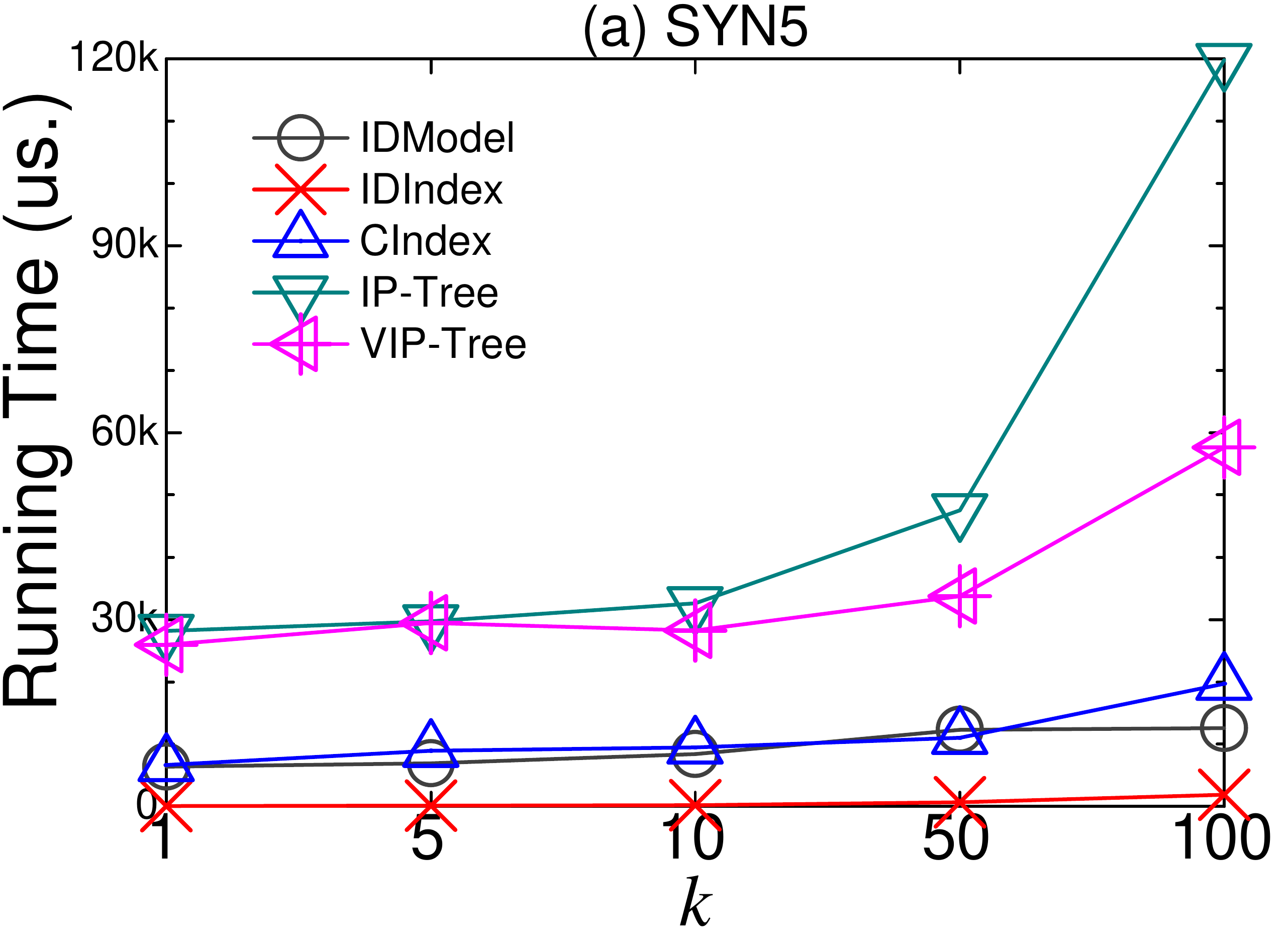}
\end{minipage}
\begin{minipage}[t]{0.245\textwidth}
\centering
\includegraphics[width=\textwidth, height = 3cm]{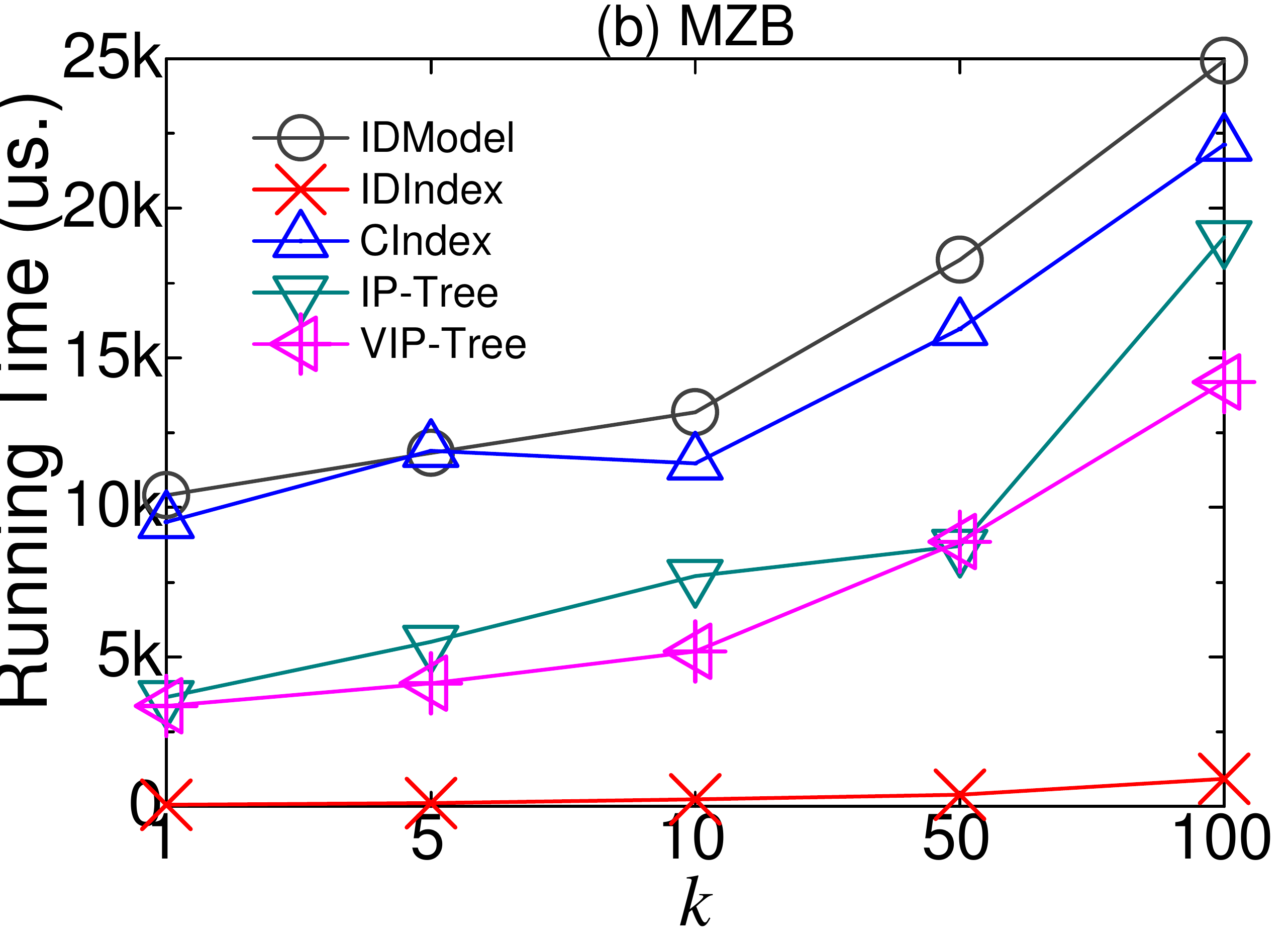}
\end{minipage}
\begin{minipage}[t]{0.245\textwidth}
\centering
\includegraphics[width=\textwidth, height = 3cm]{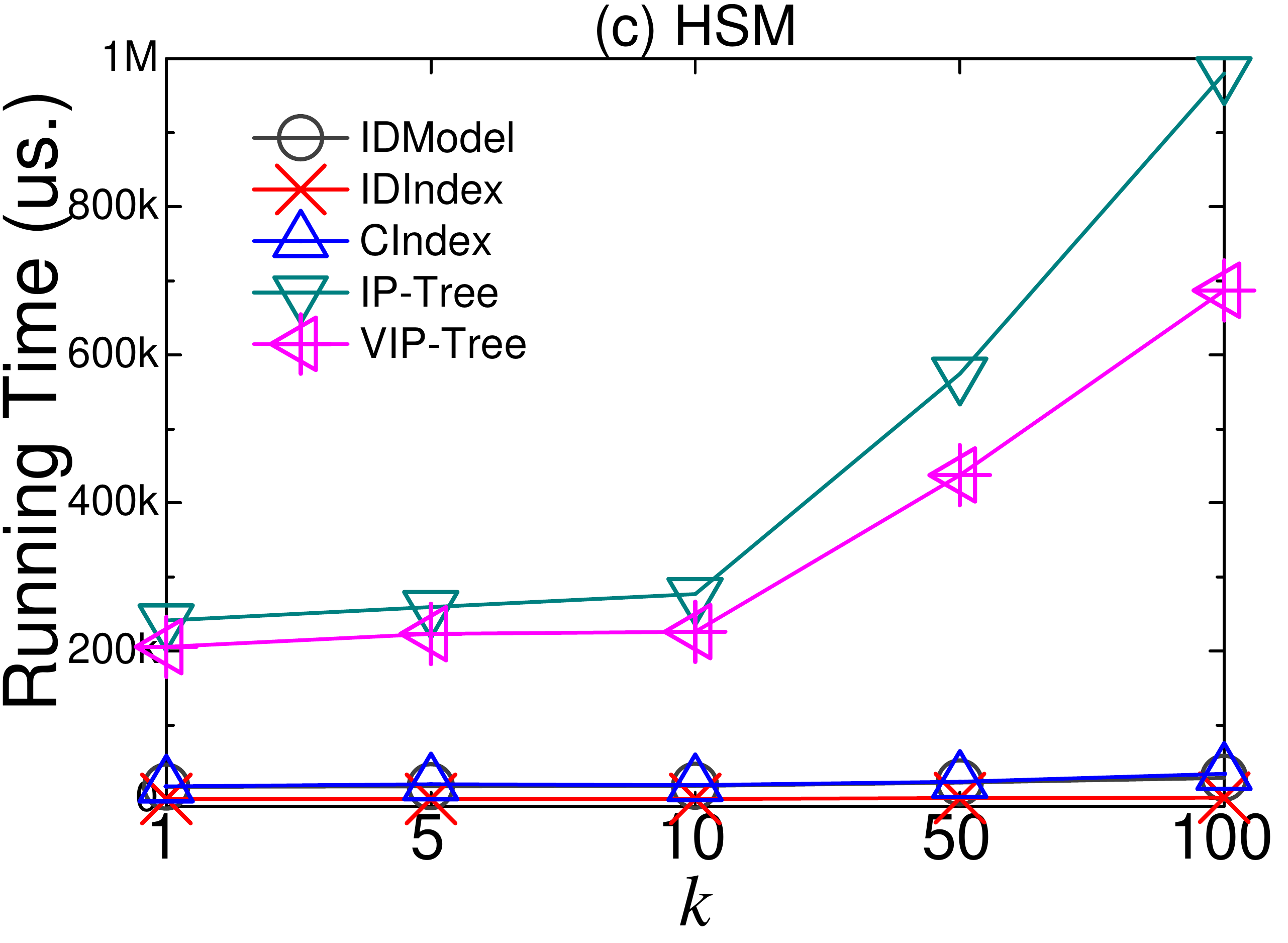}
\end{minipage}
\begin{minipage}[t]{0.245\textwidth}
\centering
\includegraphics[width=\textwidth, height = 3cm]{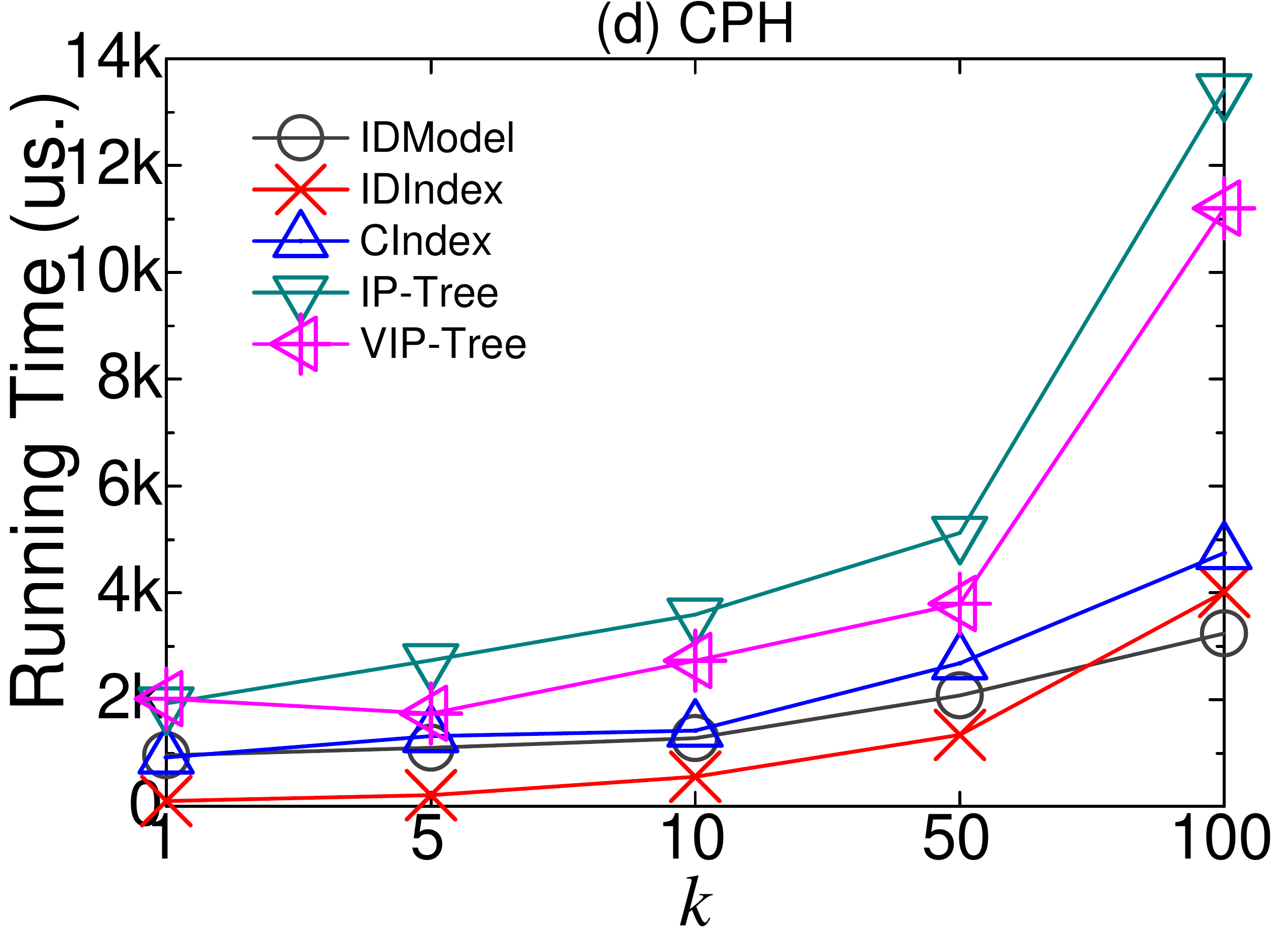}
\end{minipage}
\ExpCaption{$k$\textsf{NNQ} Time vs. $k$}\label{fig:B4_k_time}
\end{figure*}

\begin{figure*}[!htbp]
\centering
\begin{minipage}[t]{0.245\textwidth}
\centering
\includegraphics[width=\textwidth, height = 3cm]{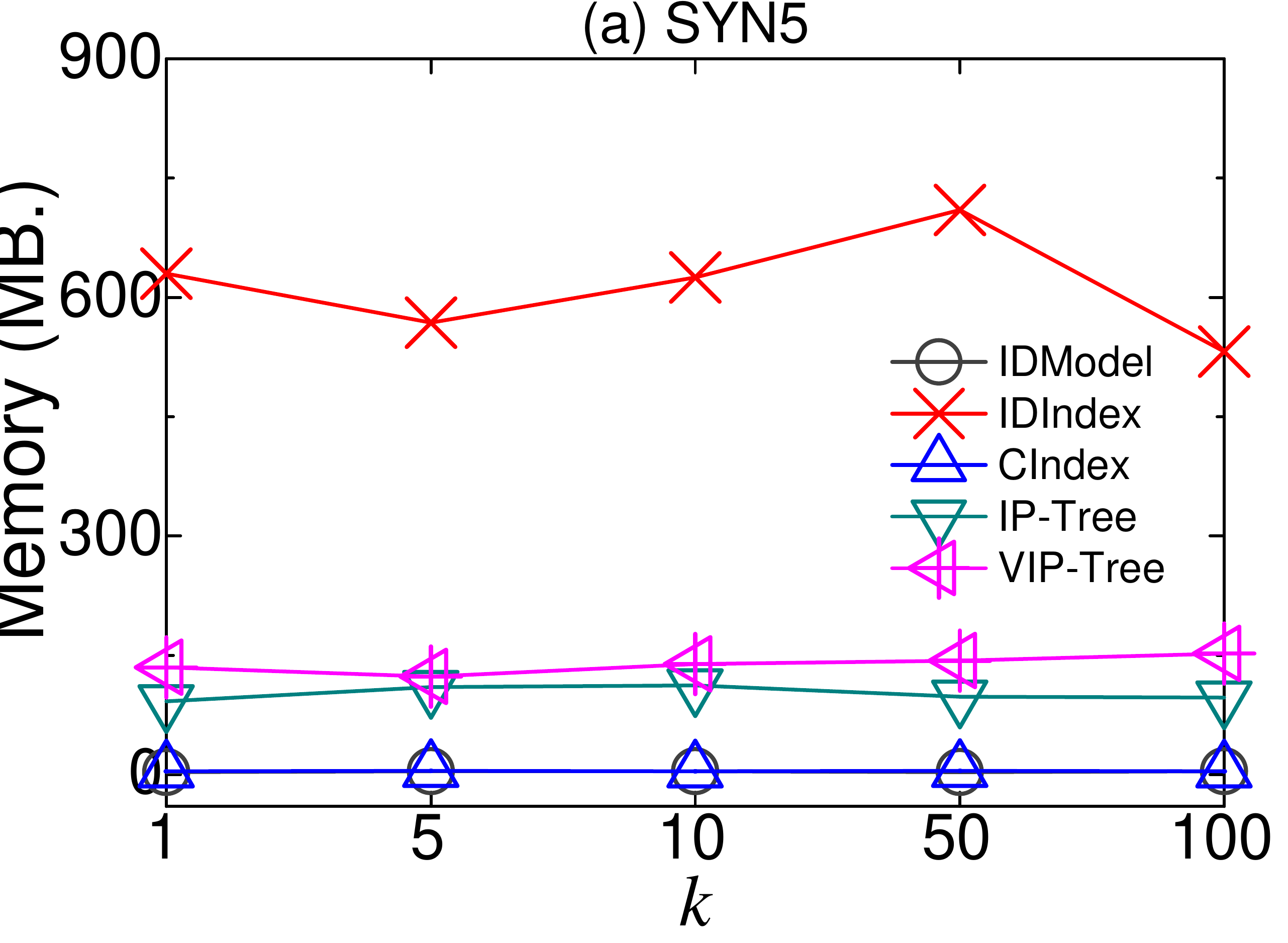}
\end{minipage}
\begin{minipage}[t]{0.245\textwidth}
\centering
\includegraphics[width=\textwidth, height = 3cm]{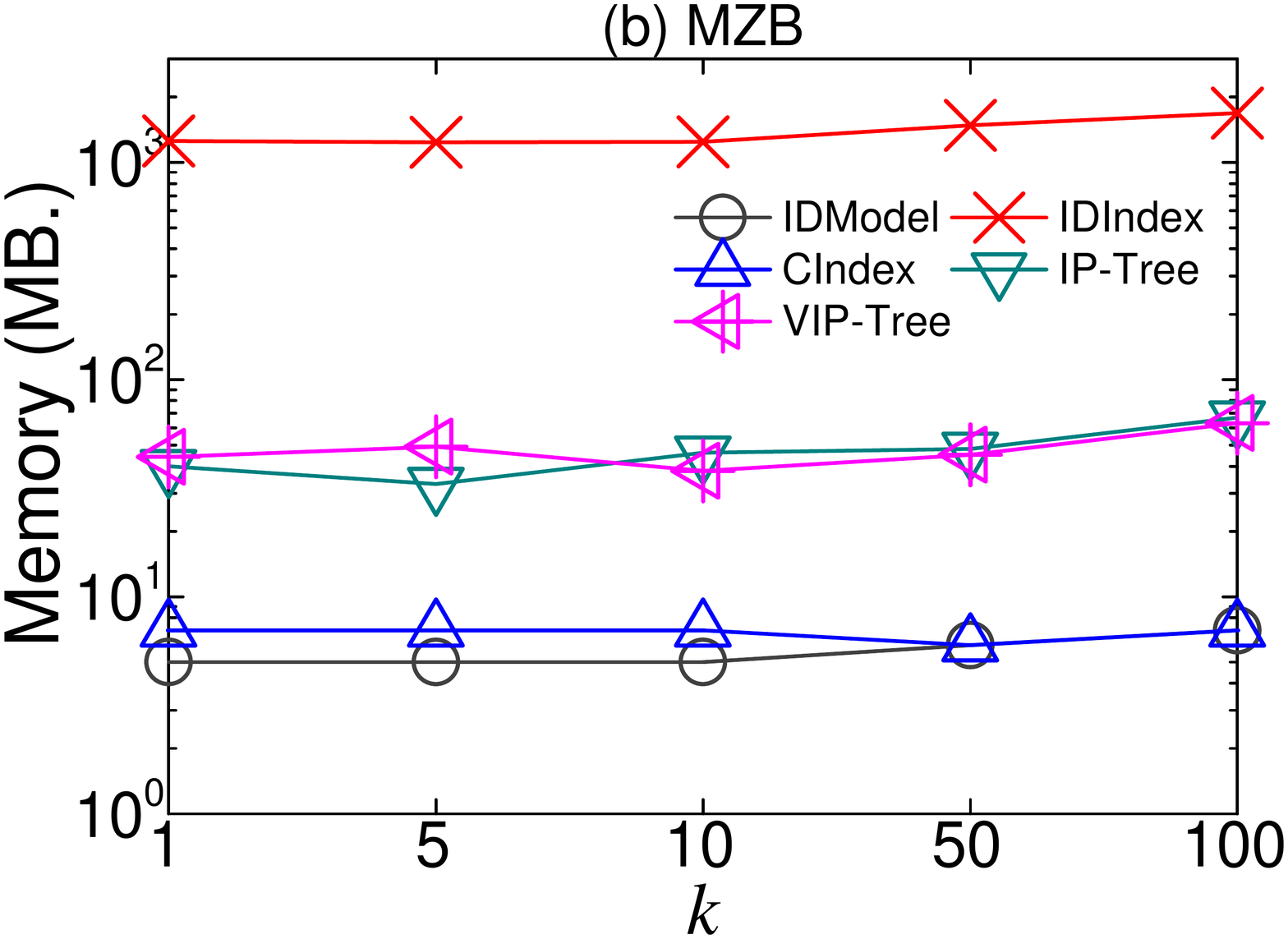}
\end{minipage}
\begin{minipage}[t]{0.245\textwidth}
\centering
\includegraphics[width=\textwidth, height = 3cm]{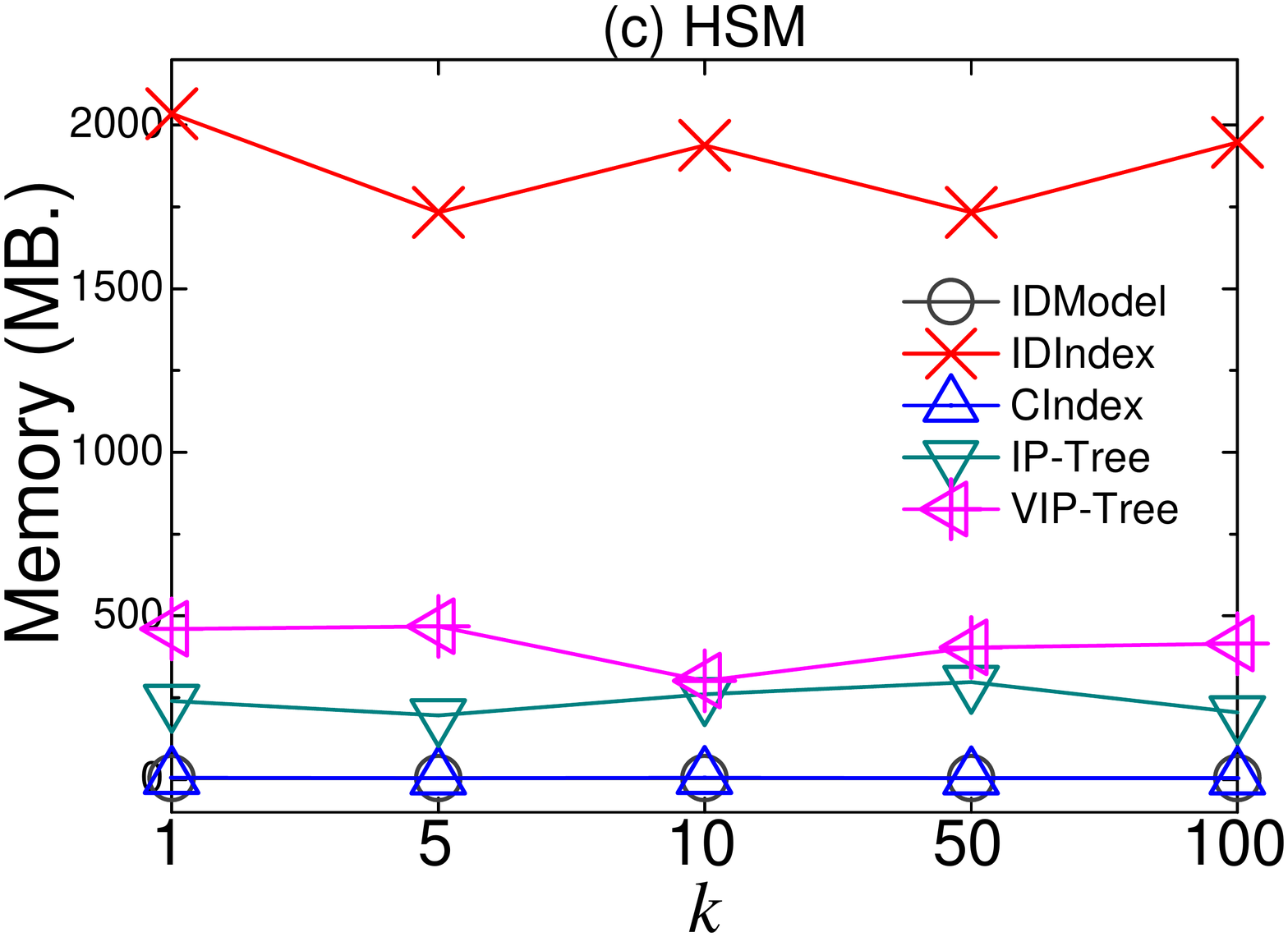}
\end{minipage}
\begin{minipage}[t]{0.245\textwidth}
\centering
\includegraphics[width=\textwidth, height = 3cm]{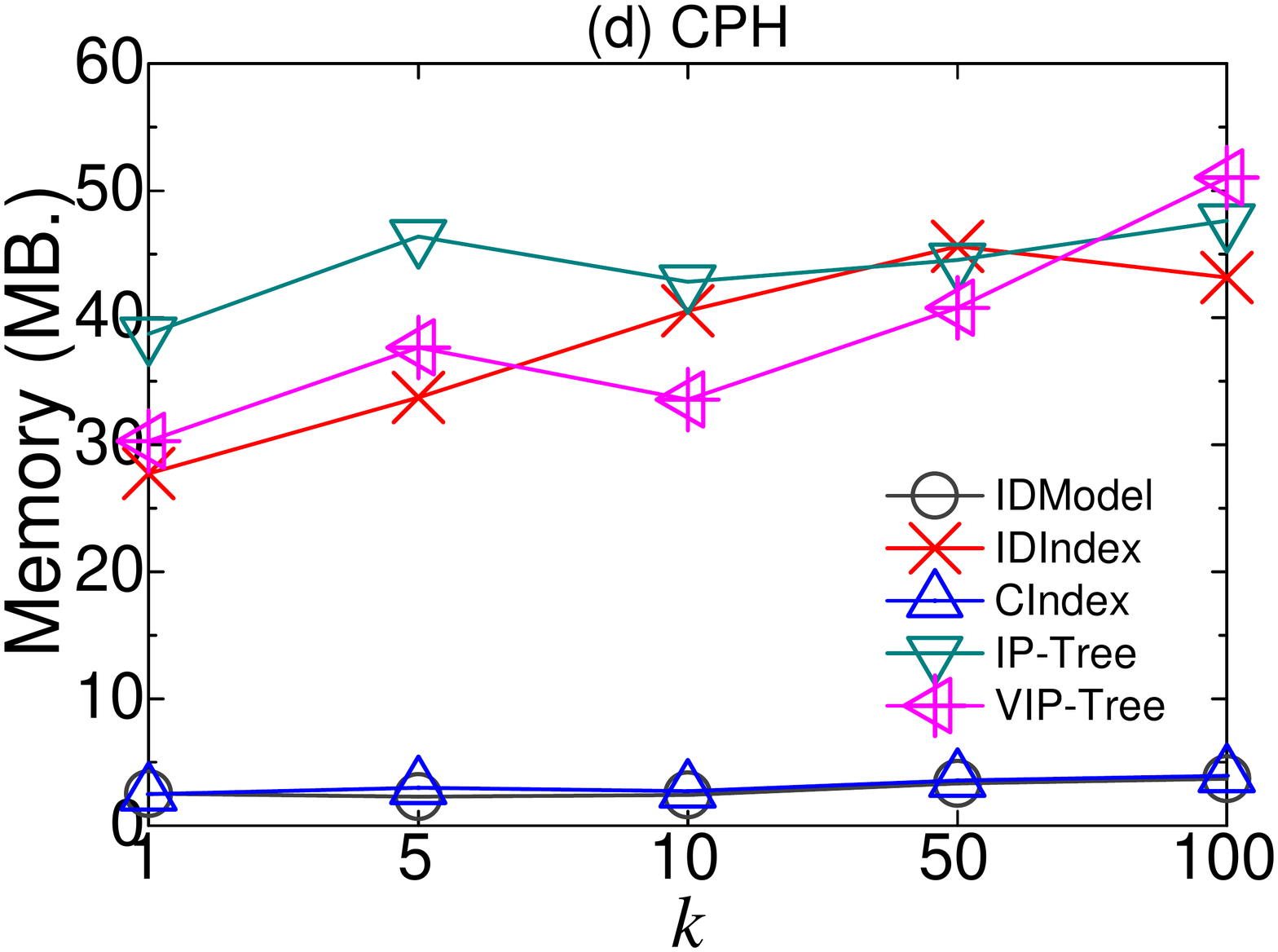}
\end{minipage}
\ExpCaption{$k$\textsf{NNQ} Memory vs. $k$}\label{fig:B4_k_mem}
\end{figure*}

\noindent$k$\textsf{NNQ}: The time and memory costs with respect to different $k$ values are reported in Figures~\ref{fig:B4_k_time} and~\ref{fig:B4_k_mem}, respectively.
\begin{itemize}[leftmargin=*]
\item Similar to increasing $r$ value in \textsf{RQ}, increasing $k$ leads to more search time by each model/index \change{according to the results reported in Figure~\ref{fig:B4_k_time}}. Among them, \textsc{IDIndex}'s running time increases slowest. In addition, \textsc{IP-Tree}/\textsc{VIP-Tree} show exponential growth on SYN, HSM, and CPH. This is because the two indexes need to access the topologically far-away partitions and compute the distances to them on the fly when $k$ is large.
\item Considering both time and memory costs, \textsc{IDModel} and \textsc{CIndex} achieve a good balance when searching for nearest neighbor objects \change{(see Figures~\ref{fig:B4_k_time} and~\ref{fig:B4_k_mem})}.
\end{itemize}

\begin{figure*}[!htbp]
\centering
\begin{minipage}[t]{0.245\textwidth}
\centering
\includegraphics[width=\textwidth, height = 3cm]{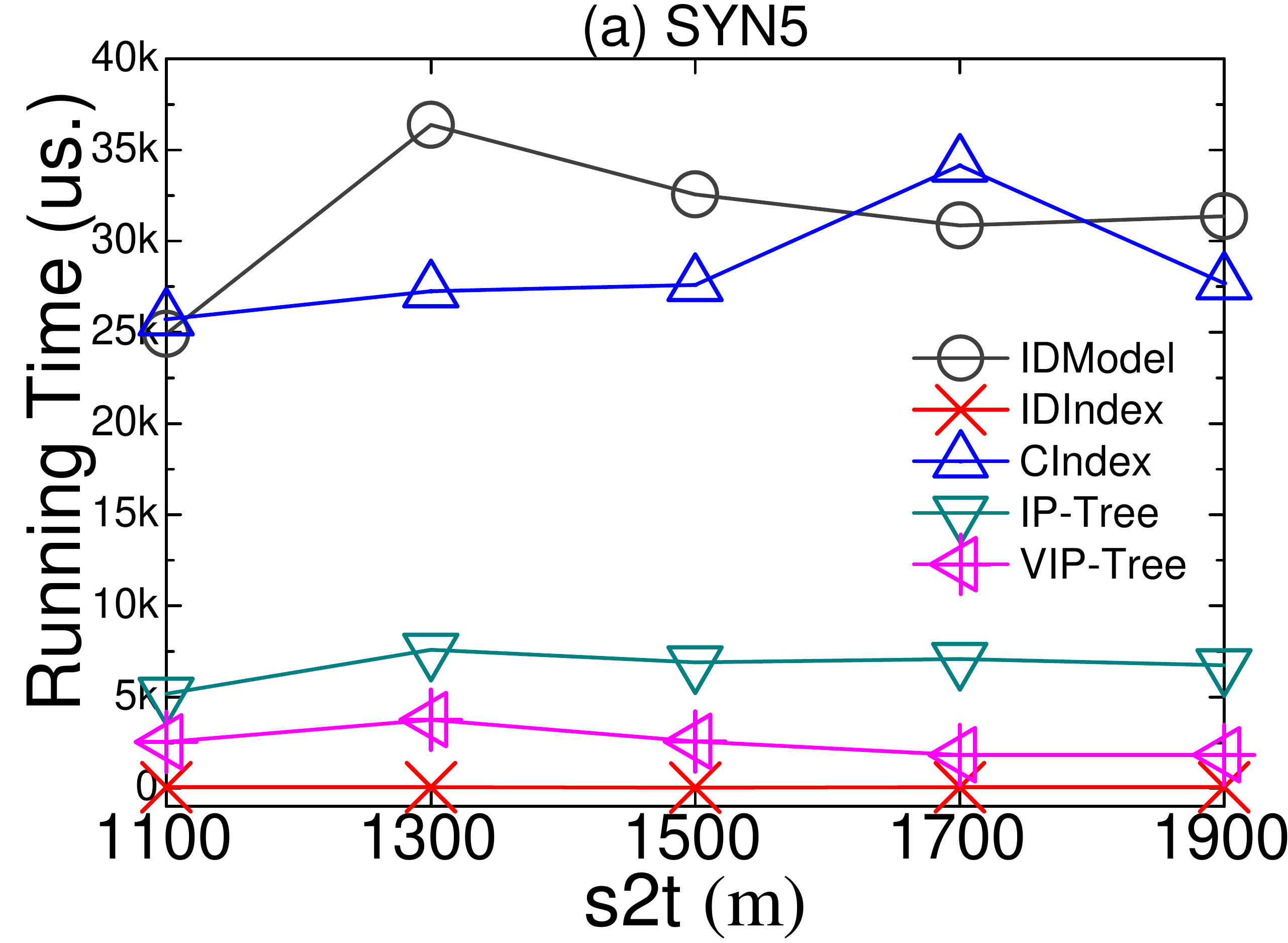}
\end{minipage}
\begin{minipage}[t]{0.245\textwidth}
\centering
\includegraphics[width=\textwidth, height = 3cm]{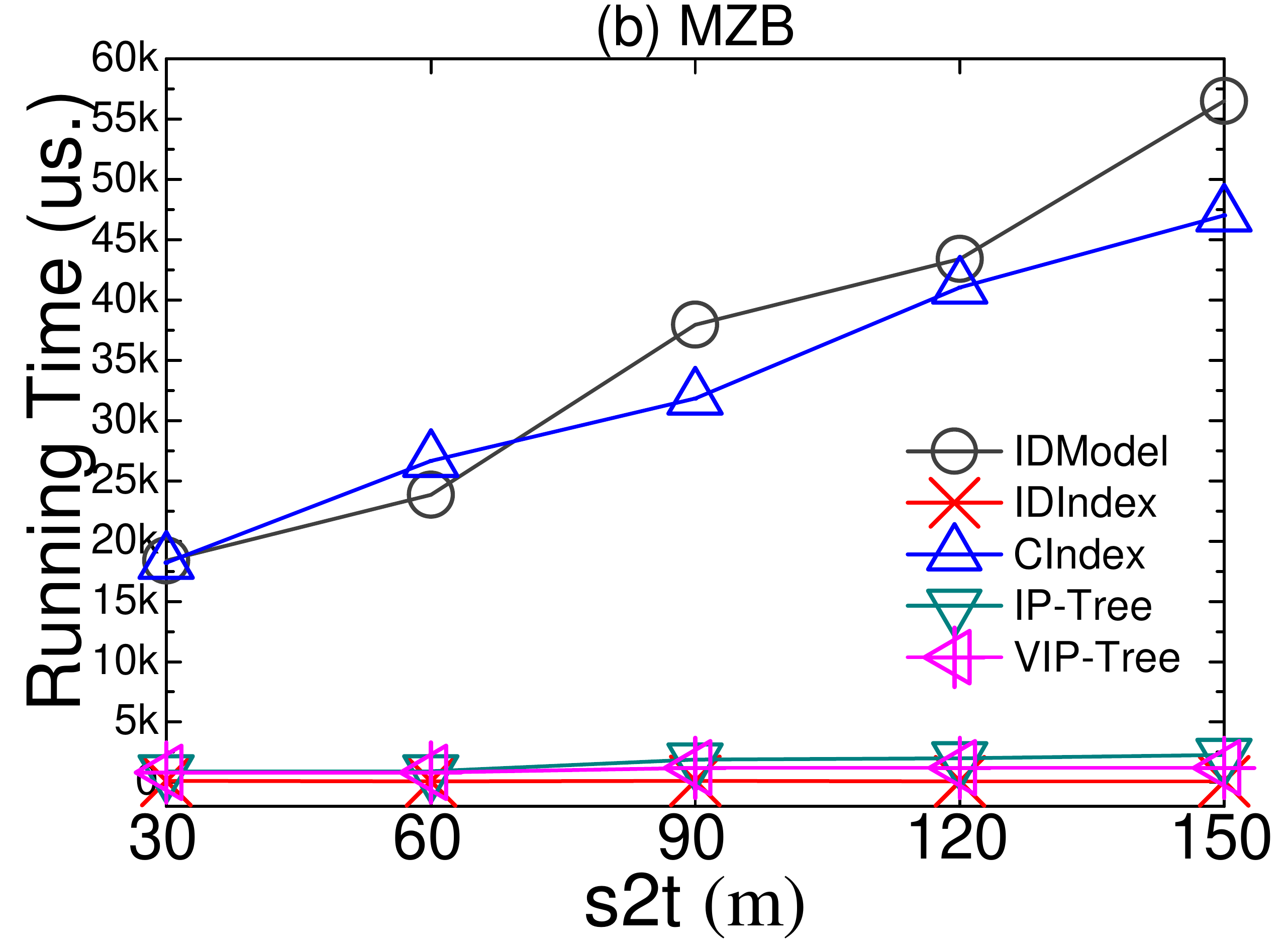}
\end{minipage}
\begin{minipage}[t]{0.245\textwidth}
\centering
\includegraphics[width=\textwidth, height = 3cm]{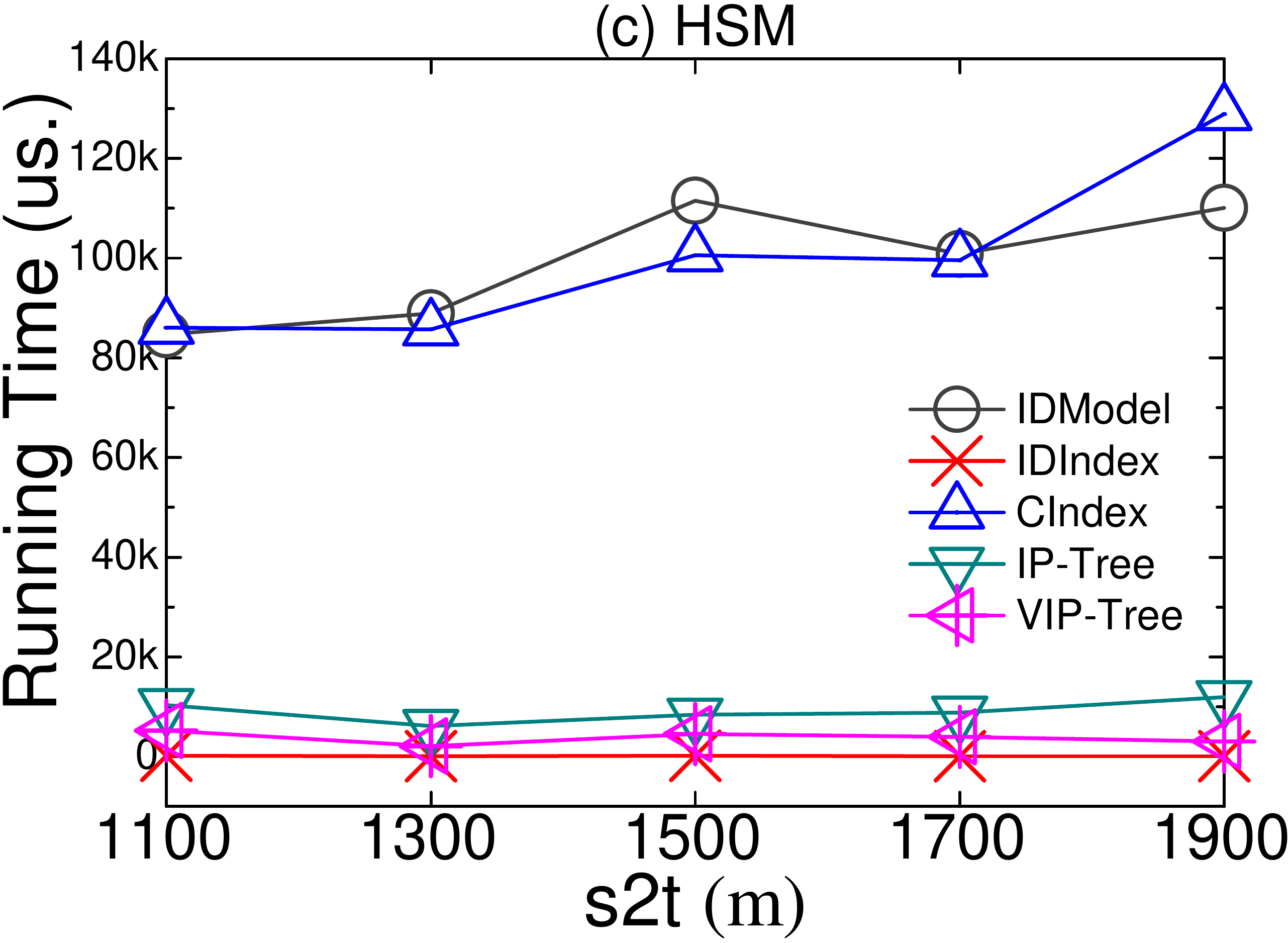}
\end{minipage}
\begin{minipage}[t]{0.245\textwidth}
\centering
\includegraphics[width=\textwidth, height = 3cm]{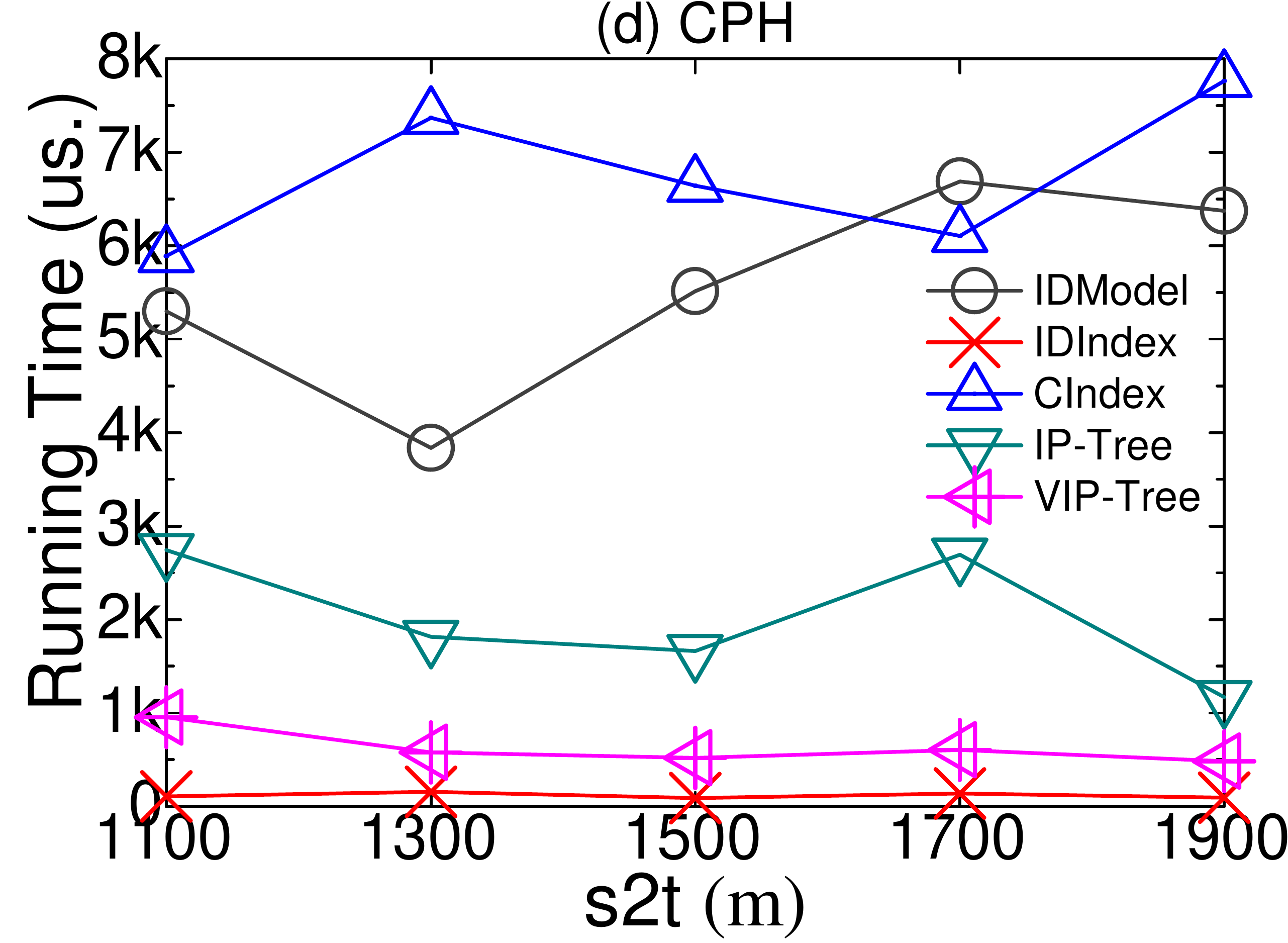}
\end{minipage}
\ExpCaption{\textsf{SPDQ} Time vs. s2t}\label{fig:B5_ds2t_time}
\end{figure*}

\begin{figure*}[!htbp]
\centering
\begin{minipage}[t]{0.245\textwidth}
\centering
\includegraphics[width=\textwidth, height = 3cm]{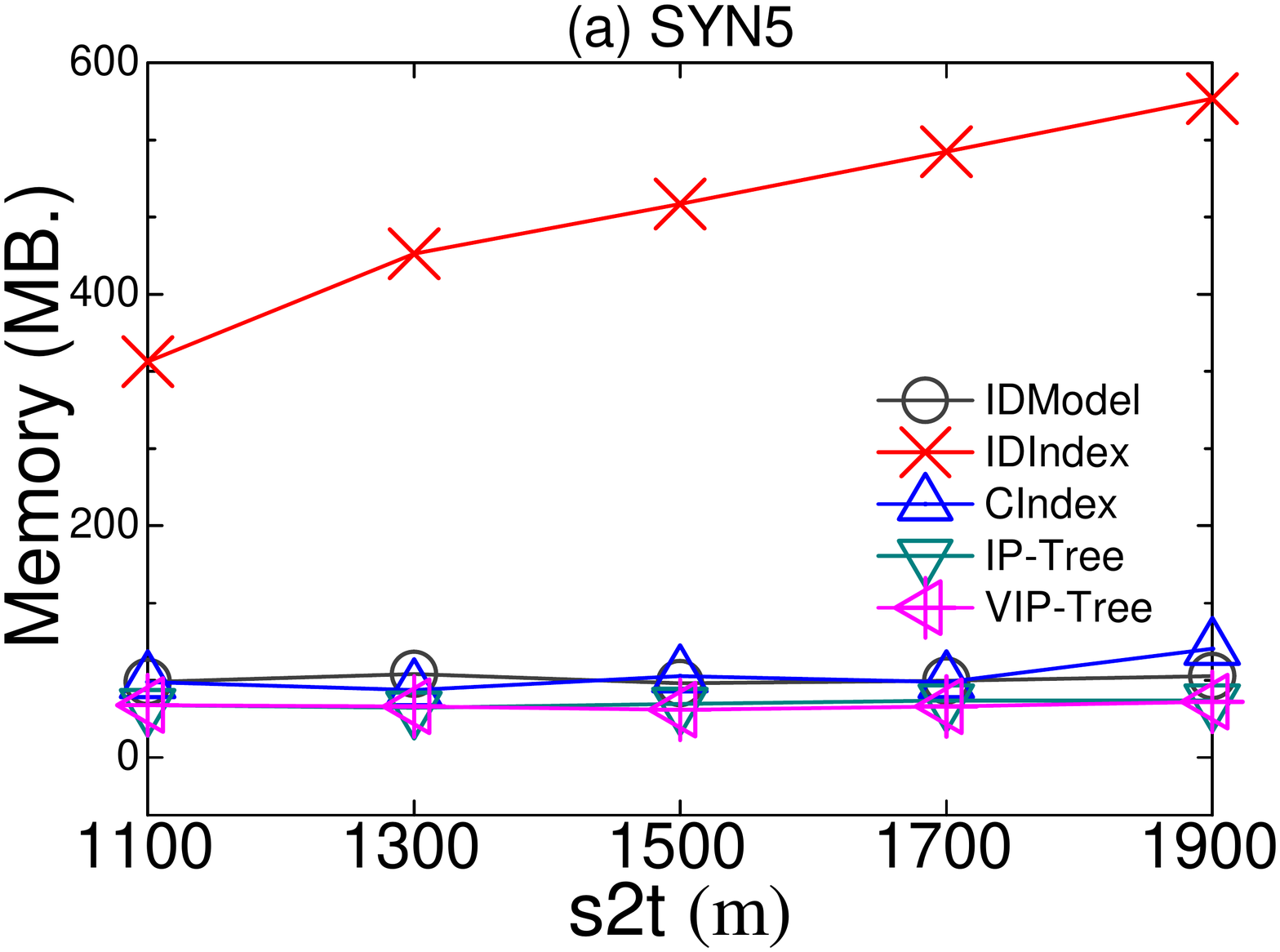}
\end{minipage}
\begin{minipage}[t]{0.245\textwidth}
\centering
\includegraphics[width=\textwidth, height = 3cm]{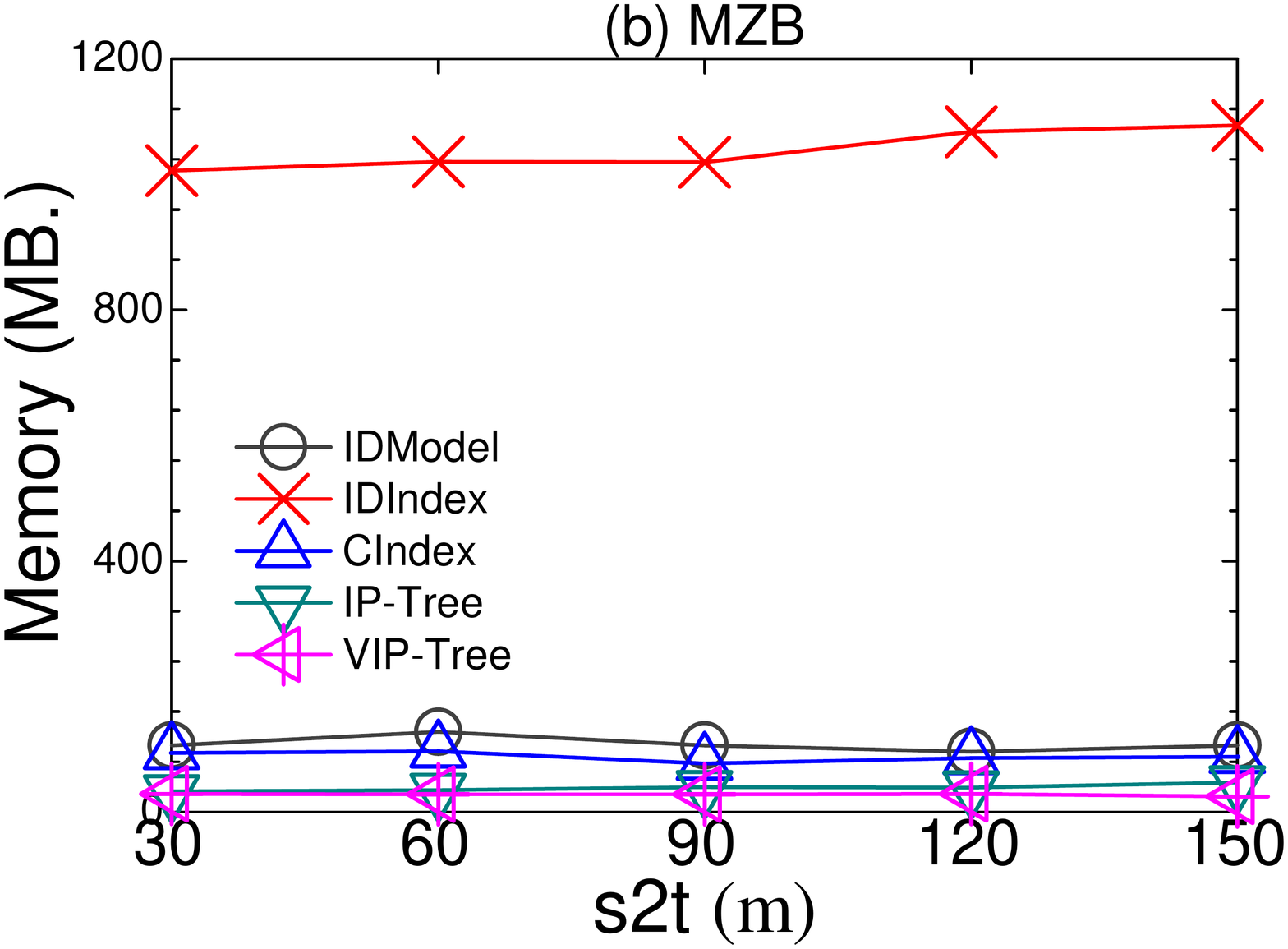}
\end{minipage}
\begin{minipage}[t]{0.245\textwidth}
\centering
\includegraphics[width=\textwidth, height = 3cm]{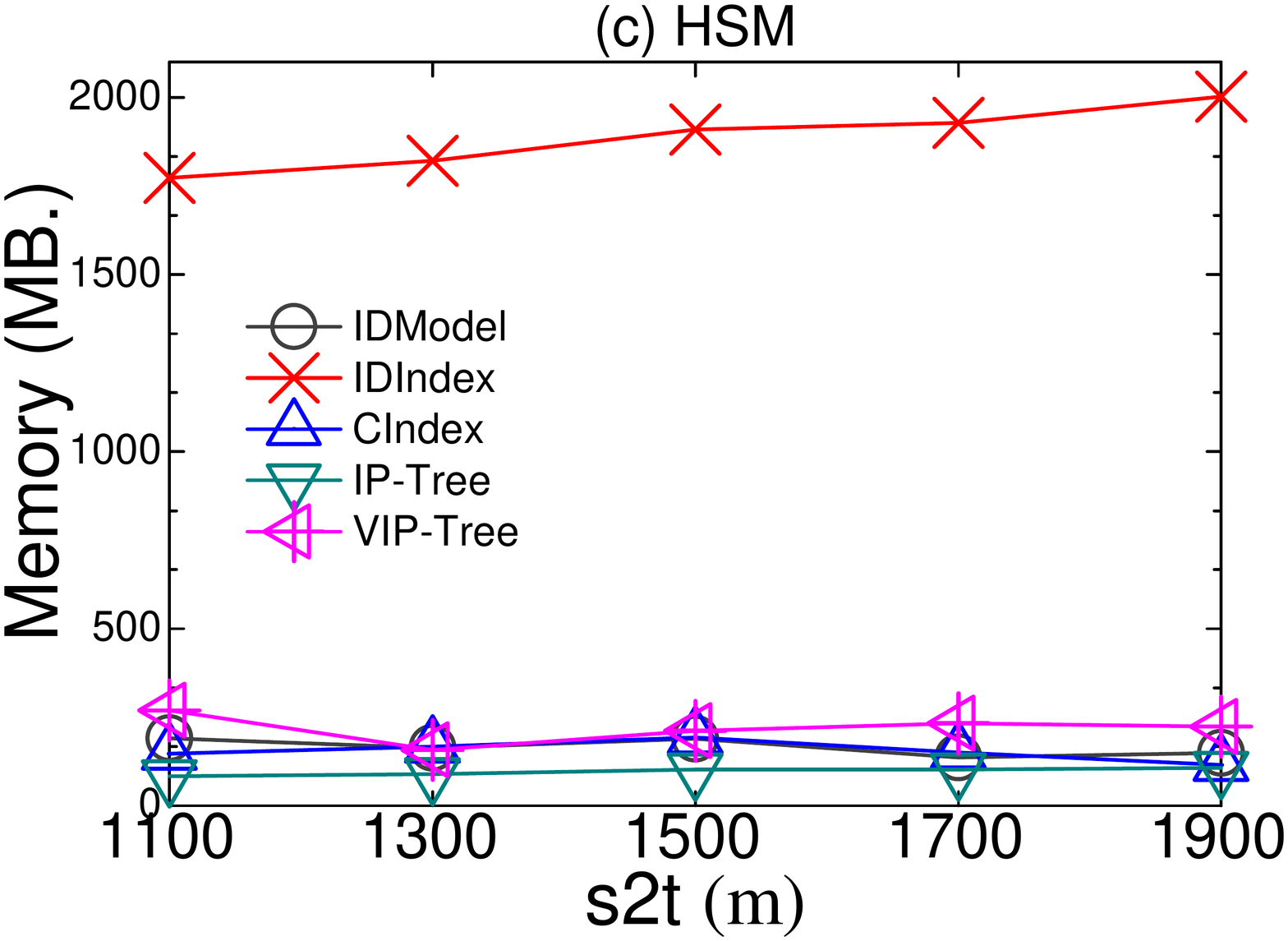}
\end{minipage}
\begin{minipage}[t]{0.245\textwidth}
\centering
\includegraphics[width=\textwidth, height = 3cm]{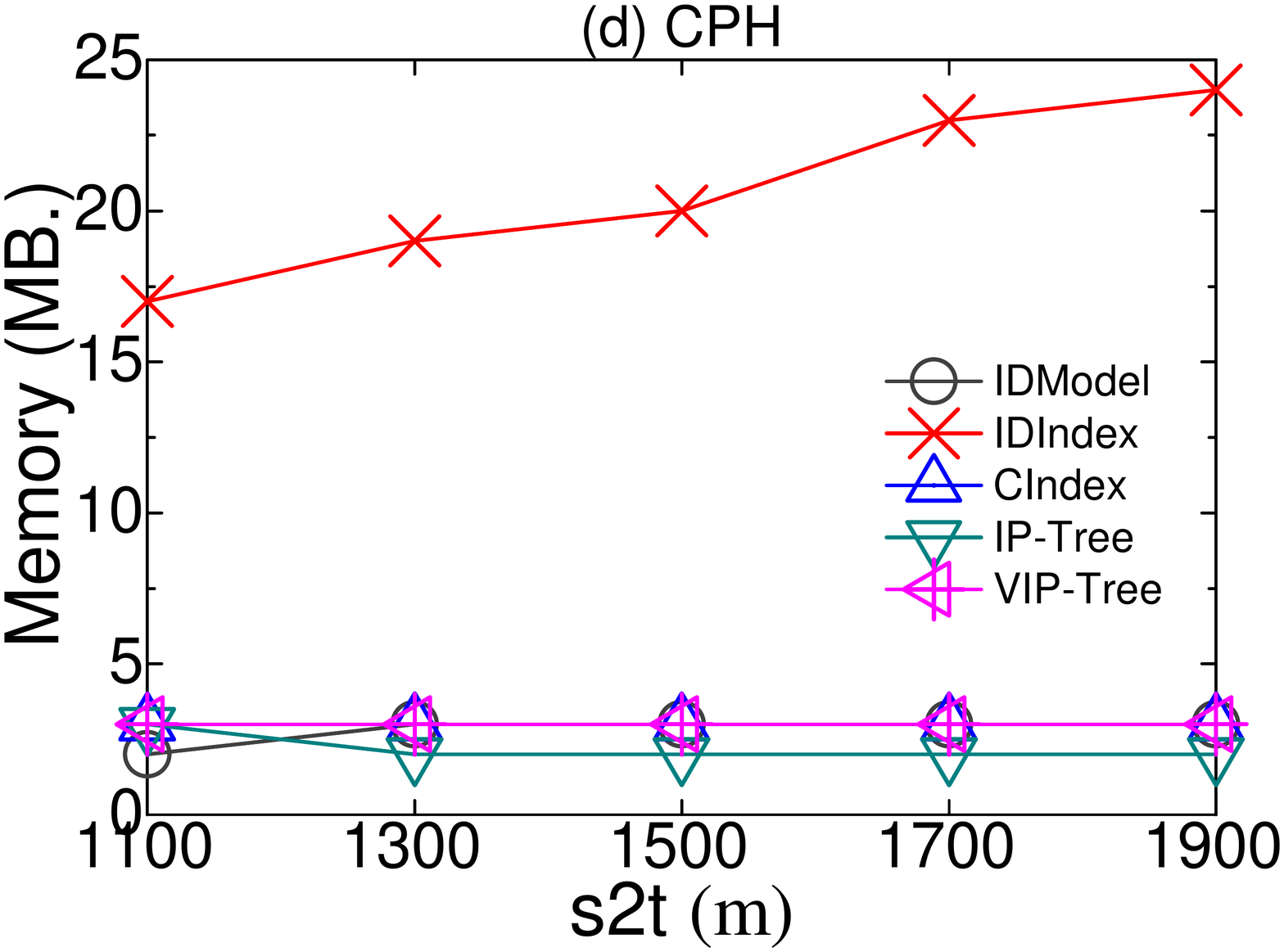}
\end{minipage}
\ExpCaption{\textsf{SPDQ} Memory vs. s2t}\label{fig:B5_ds2t_mem}
\end{figure*}

\begin{figure*}[!htbp]
\centering
\begin{minipage}[t]{0.245\textwidth}
\centering
\includegraphics[width=\textwidth, height = 3cm]{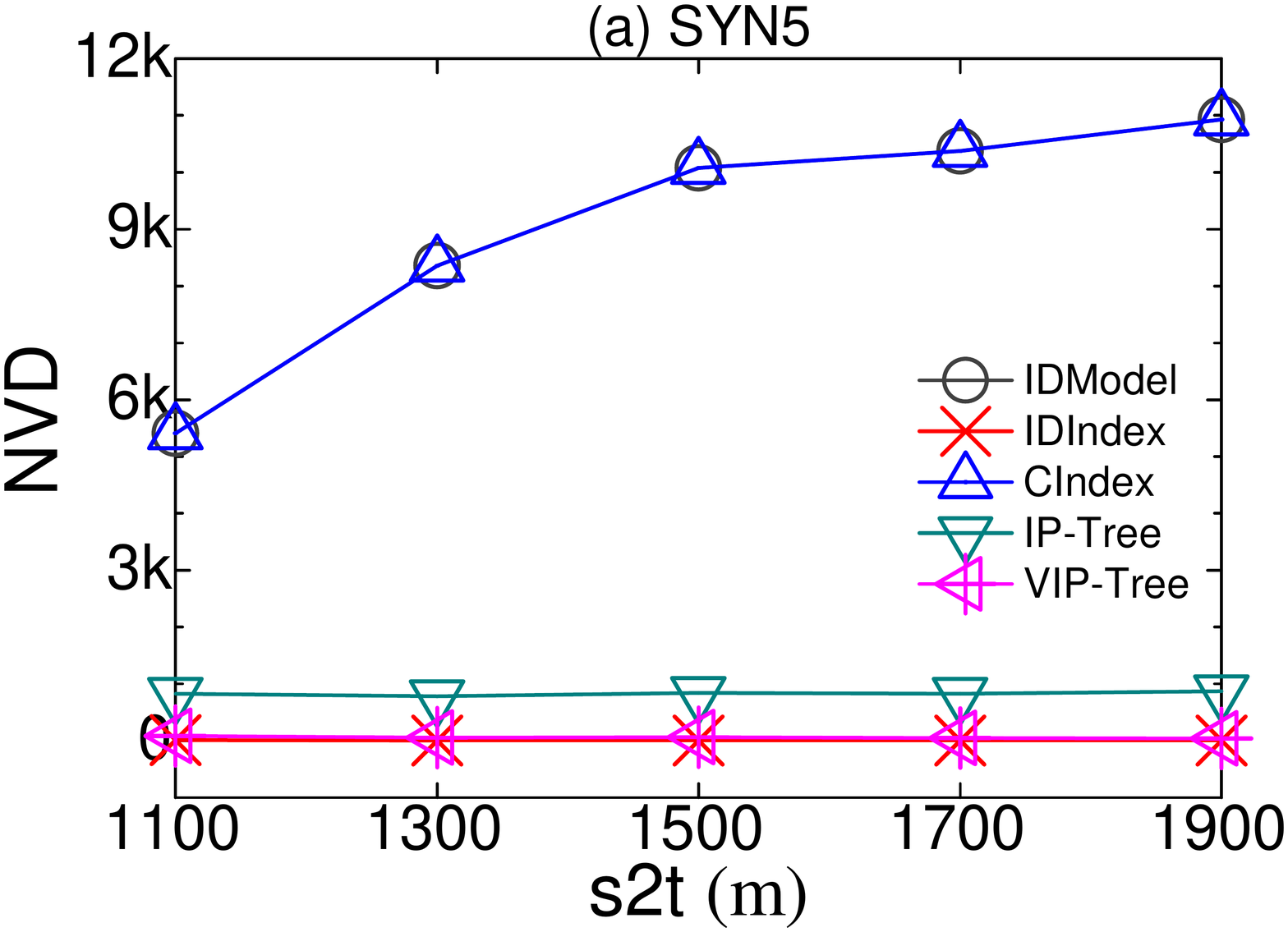}
\end{minipage}
\begin{minipage}[t]{0.245\textwidth}
\centering
\includegraphics[width=\textwidth, height = 3cm]{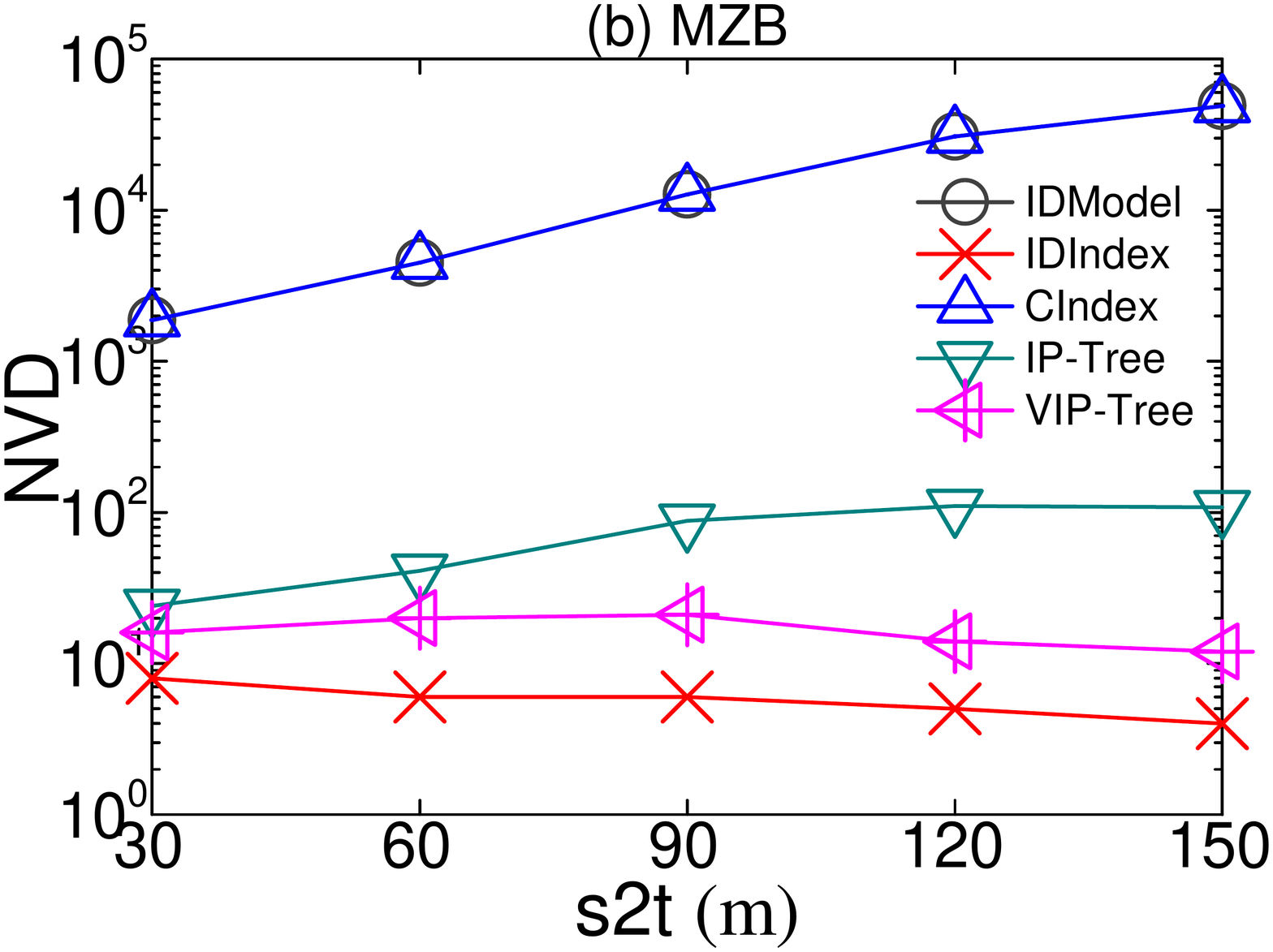}
\end{minipage}
\begin{minipage}[t]{0.245\textwidth}
\centering
\includegraphics[width=\textwidth, height = 3cm]{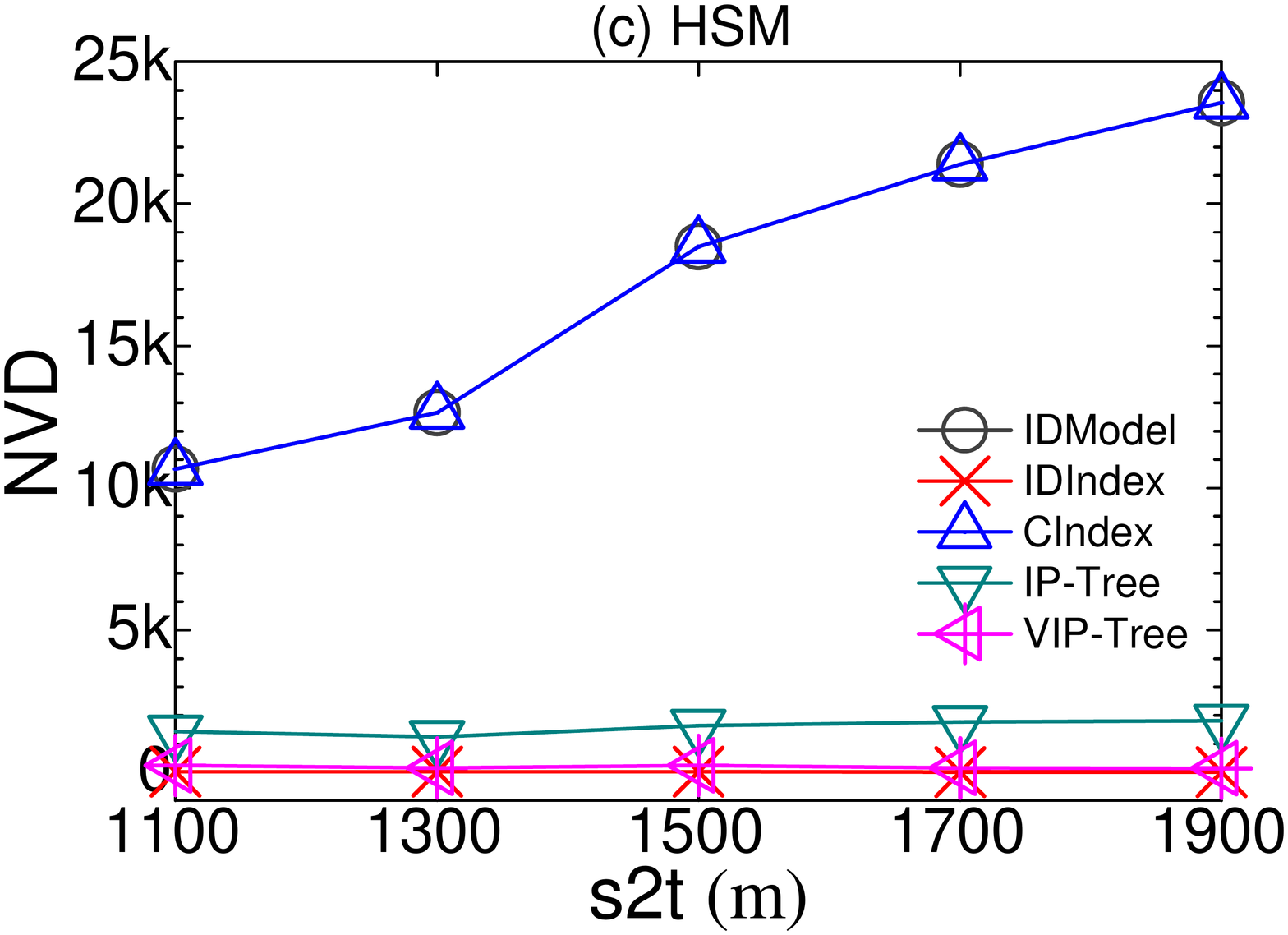}
\end{minipage}
\begin{minipage}[t]{0.245\textwidth}
\centering
\includegraphics[width=\textwidth, height = 3cm]{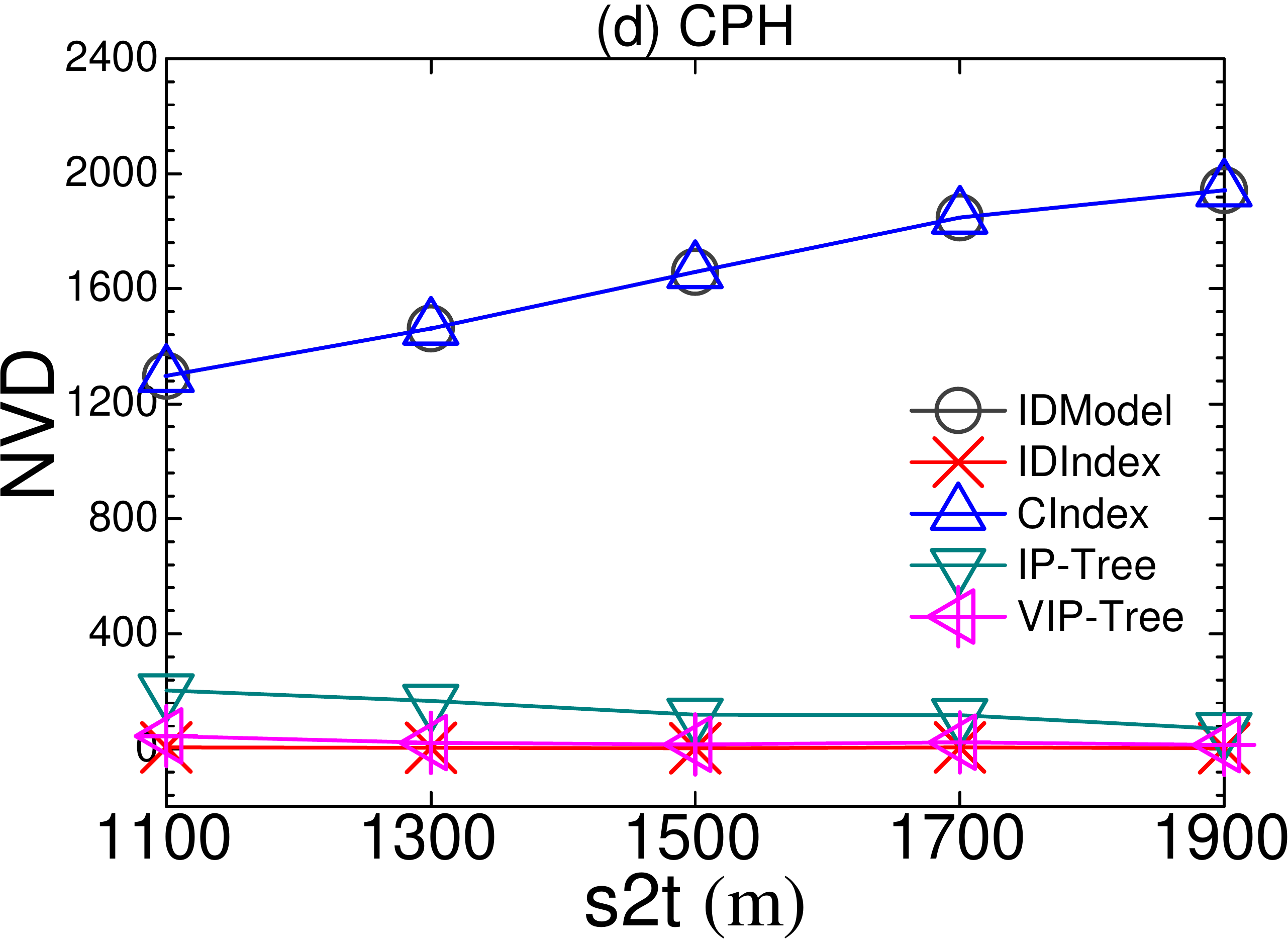}
\end{minipage}
\ExpCaption{NVD in \textsf{SPDQ} vs. s2t}\label{fig:B5_ds2t_nvd}
\end{figure*}

\ExpHead{B5 Effect of Source-Target Distance s2t}

\noindent\textsf{SPDQ}: The time cost, memory use, and NVD for different s2t values are reported in Figures~\ref{fig:B5_ds2t_time},~\ref{fig:B5_ds2t_mem}, and~\ref{fig:B5_ds2t_nvd}, respectively.
\begin{itemize}[leftmargin=*]
\item \textsc{IDIndex} runs the fastest and is not affected by s2t \change{as reported in Figure~\ref{fig:B5_ds2t_time}}. As only a small number of doors are required to process after the source point and before the target point, its NVD is always small \change{(Figure~\ref{fig:B5_ds2t_nvd})}. Nevertheless, its global distance matrix takes up a lot of memory \change{(Figure~\ref{fig:B5_ds2t_mem})}.
\item \textsc{IDModel} and \textsc{CIndex} use the same graph search process. Note that because the Euclidean distance is no larger than the indoor distance, using R-tree to prune space by Euclidean distance does not really reduce the number of doors to visit. Therefore, the two models' NVDs \change{in Figure~\ref{fig:B5_ds2t_nvd}} are almost the same. Also, as s2t increases, the candidate space becomes larger and the running time of the two becomes longer \change{(see Figure~\ref{fig:B5_ds2t_time})}.
\item On MZB and HSM \change{(Figure~\ref{fig:B5_ds2t_time}(b) and (c))}, \textsc{VIP-Tree} achieves query performance comparable to \textsc{IDIndex} that precomputes door-to-door distances. Both MZB and HSM are large-scale and have over 1000 doors. In the routing process based on \textsc{VIP-Tree}, the precomputed distances in non-leaf nodes greatly accelerate the expansion to the target point. Therefore, \textsc{VIP-Tree} is particularly suitable for the shortest path search in indoor spaces with complex structures.
\end{itemize}

\begin{figure*}[!htbp]
    \begin{minipage}[t]{0.24\textwidth}
    \centering
    \includegraphics[width=\textwidth, height = 3cm]{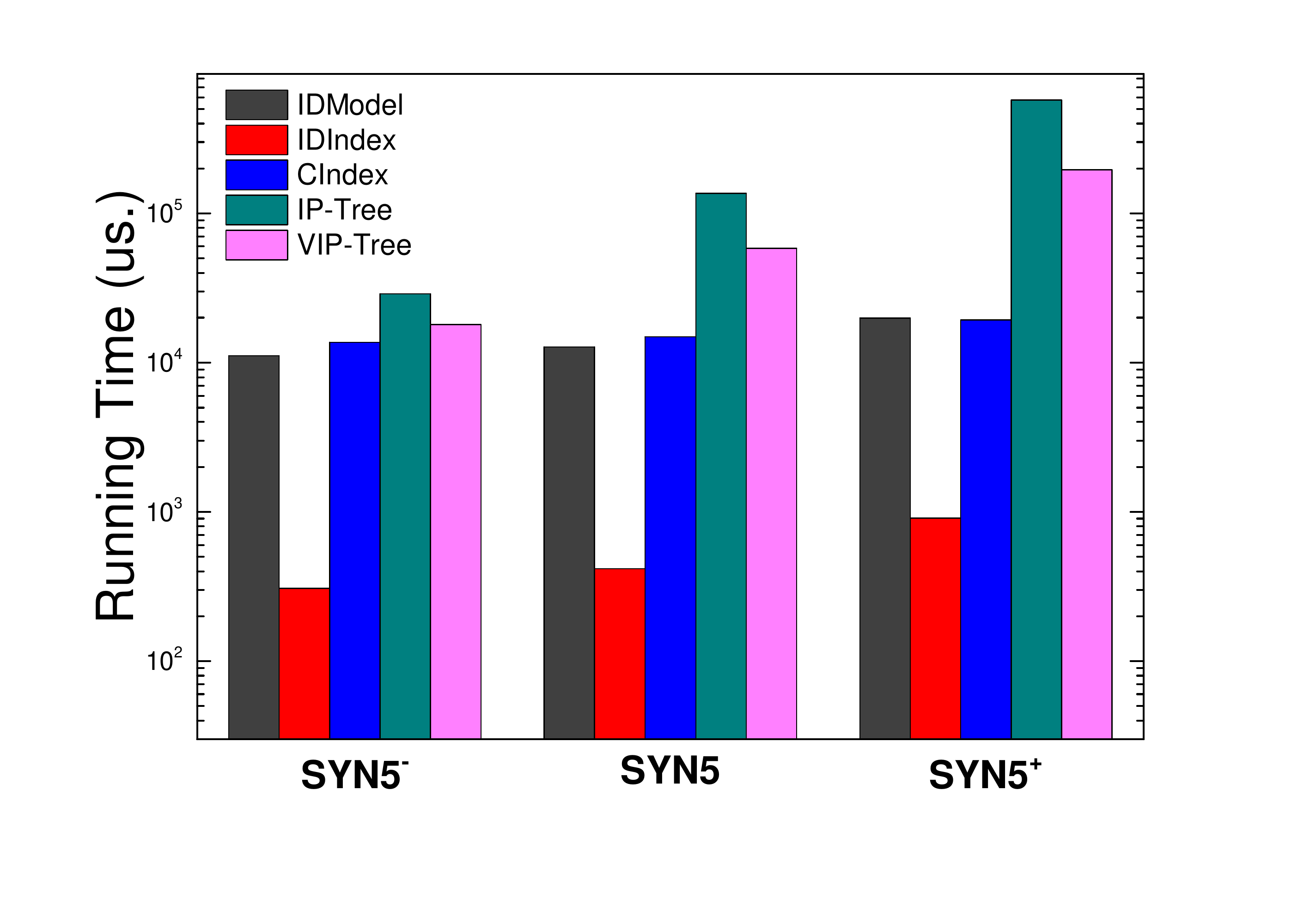}
    \ExpCaption{\textsf{RQ} Time vs. \\Topology}\label{fig:B6_RQ_time}
    \end{minipage}
    \begin{minipage}[t]{0.24\textwidth}
    \centering
    \includegraphics[width=\textwidth, height = 3cm]{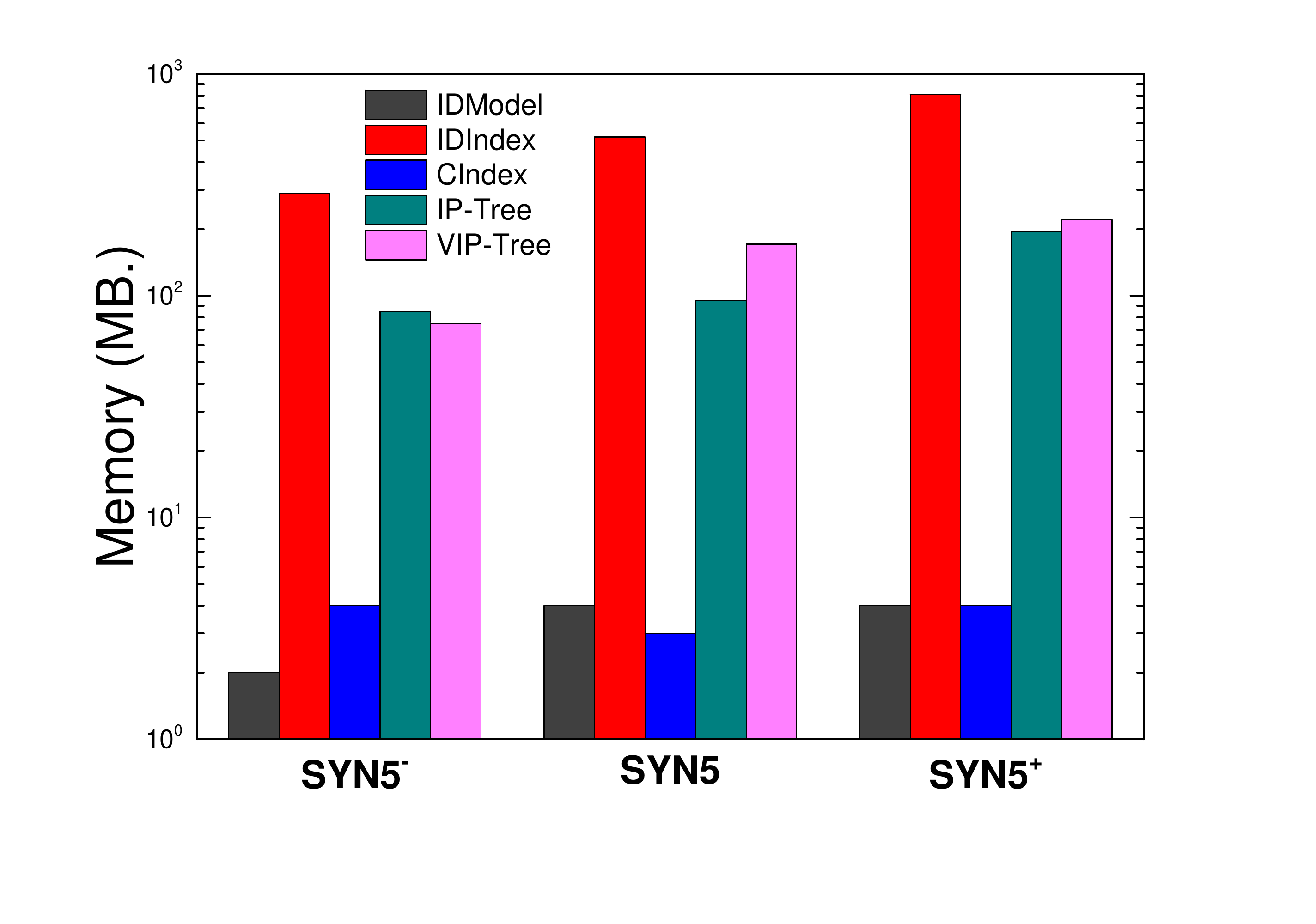}
    \ExpCaption{\textsf{RQ} Memory \\vs. Topology}\label{fig:B6_RQ_mem}
    \end{minipage}
    \begin{minipage}[t]{0.24\textwidth}
    \centering
    \includegraphics[width=\textwidth, height = 3cm]{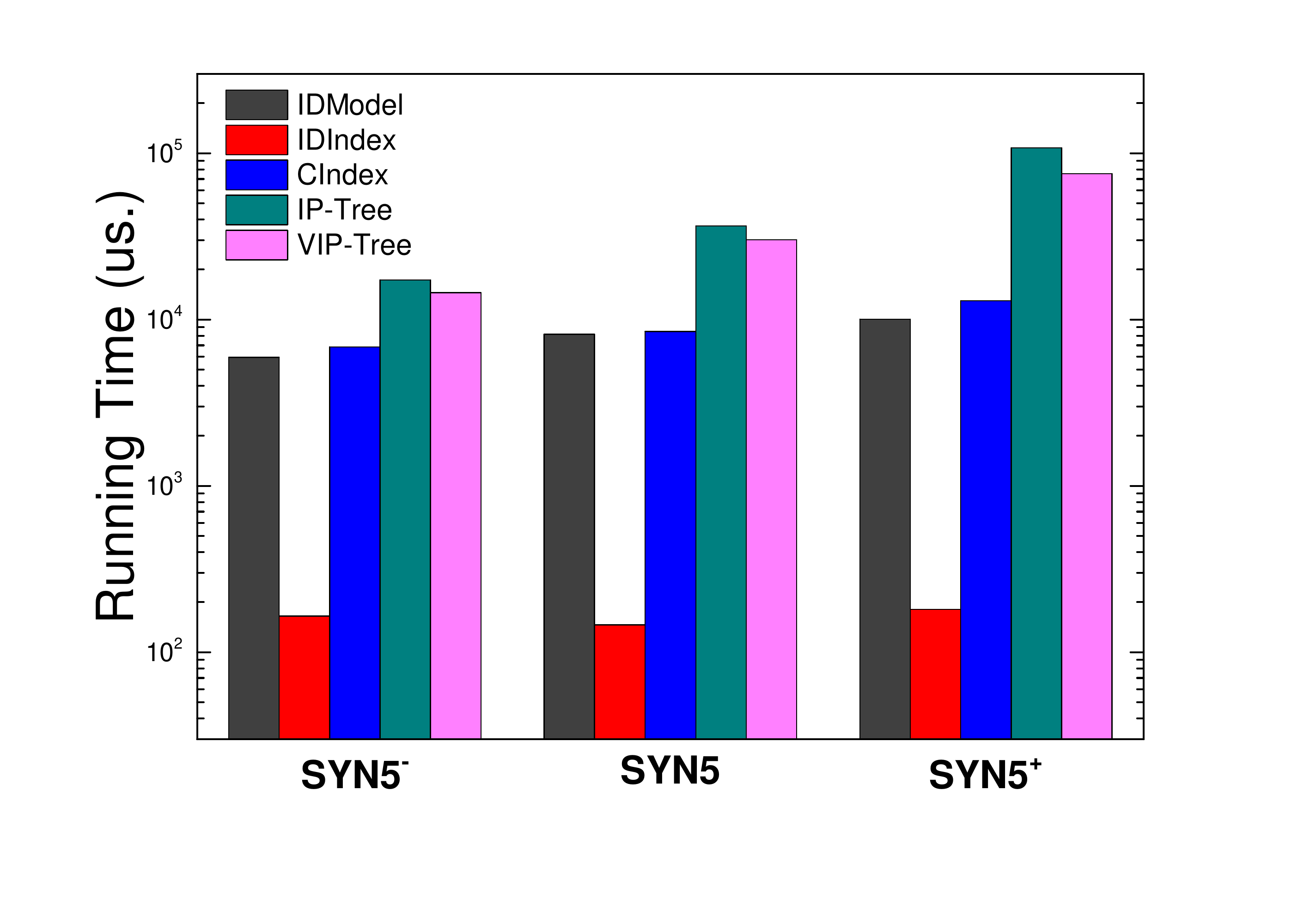}
    \ExpCaption{$k$\textsf{NNQ} Time \\vs. Topology}\label{fig:B6_KNNQ_time}
    \end{minipage}
    \begin{minipage}[t]{0.24\textwidth}
    \centering
    \includegraphics[width=\textwidth, height = 3cm]{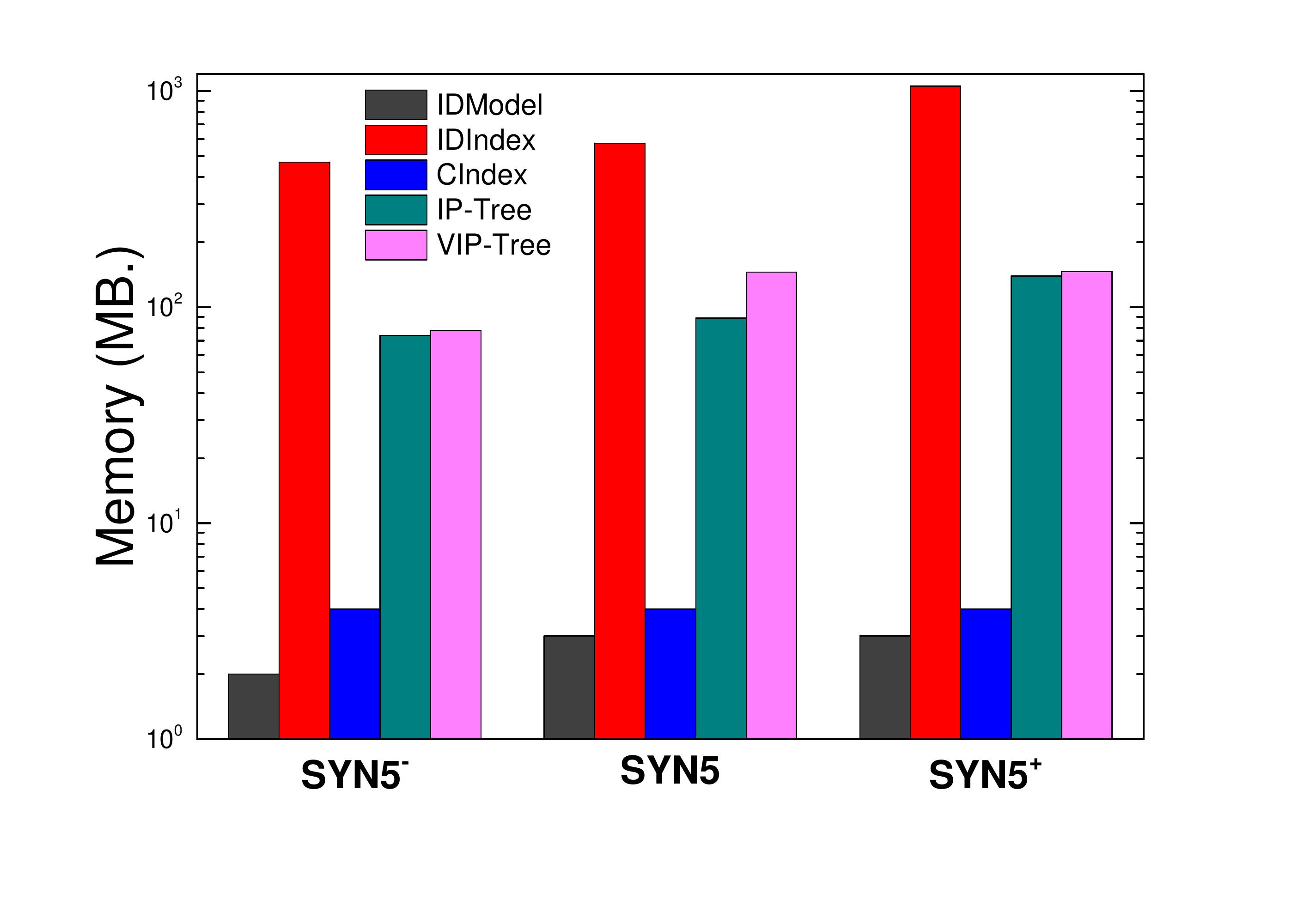}
    \ExpCaption{$k$\textsf{NNQ} Memory vs. Topology}\label{fig:B6_KNNQ_mem}
    \end{minipage}
\end{figure*}

\begin{figure*}[!htbp]
    \begin{minipage}[t]{0.328\textwidth}
    \centering
    \includegraphics[width=\textwidth, height = 3cm]{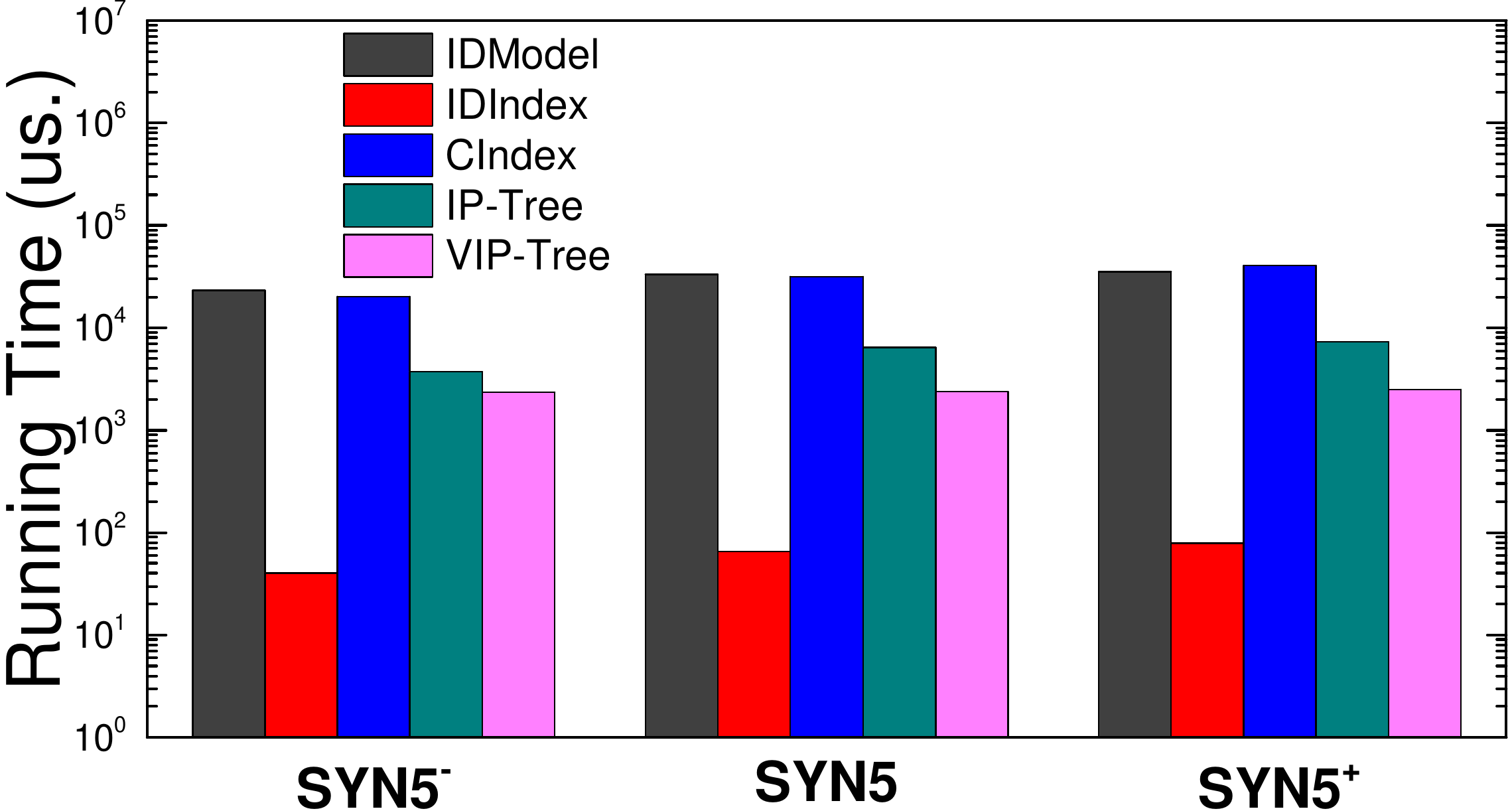}
    \ExpCaption{\textsf{SPDQ} Time \\vs. Topology}\label{fig:B6_SPDQ_time}
    \end{minipage}
    \begin{minipage}[t]{0.328\textwidth}
    \centering
    \includegraphics[width=\textwidth, height = 3cm]{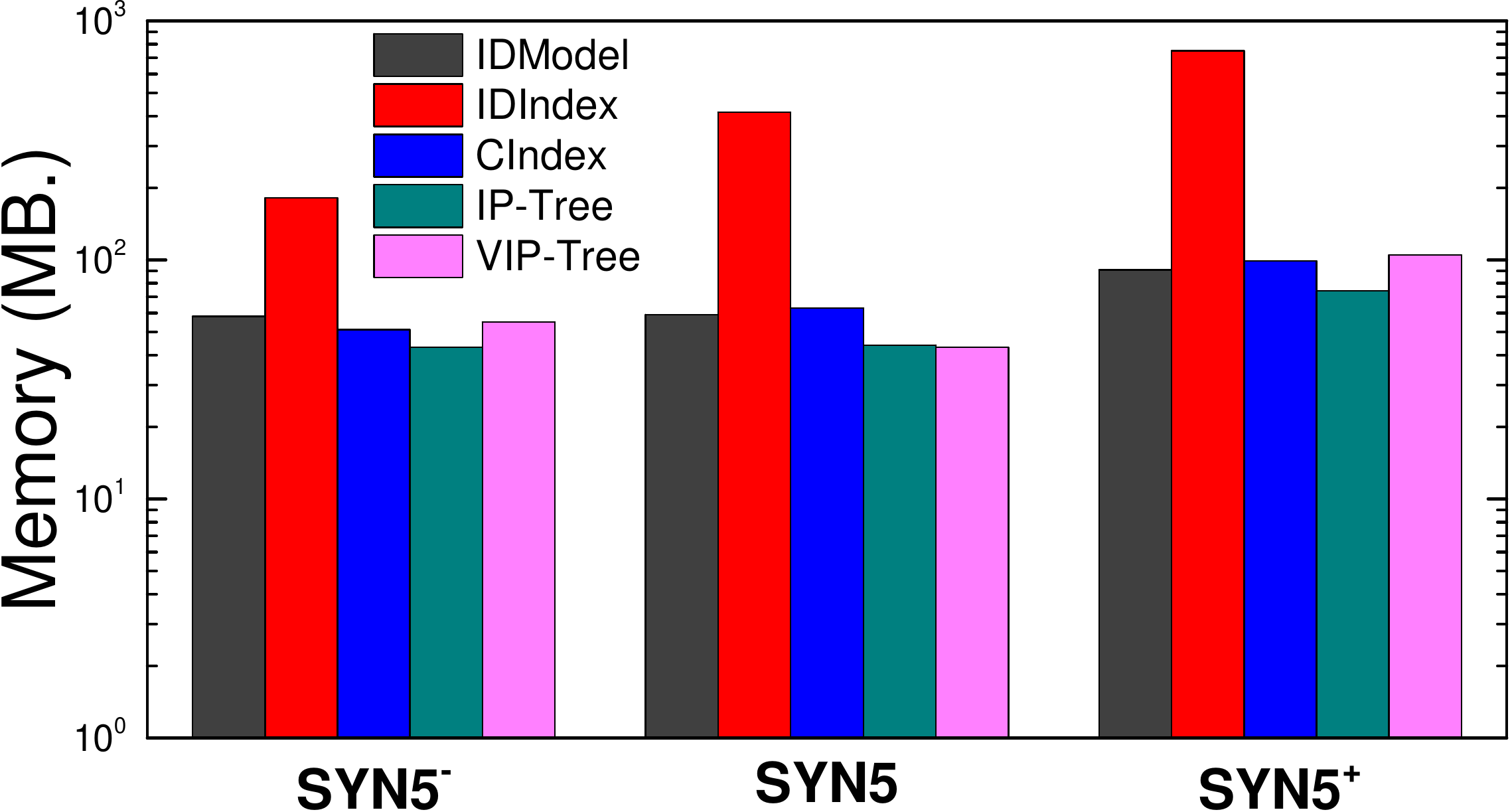}
    \ExpCaption{\textsf{SPDQ} Memory \\vs. Topology}\label{fig:B6_SPDQ_mem}
    \end{minipage}
    \begin{minipage}[t]{0.328\textwidth}
    \centering
    \includegraphics[width=\textwidth, height = 3cm]{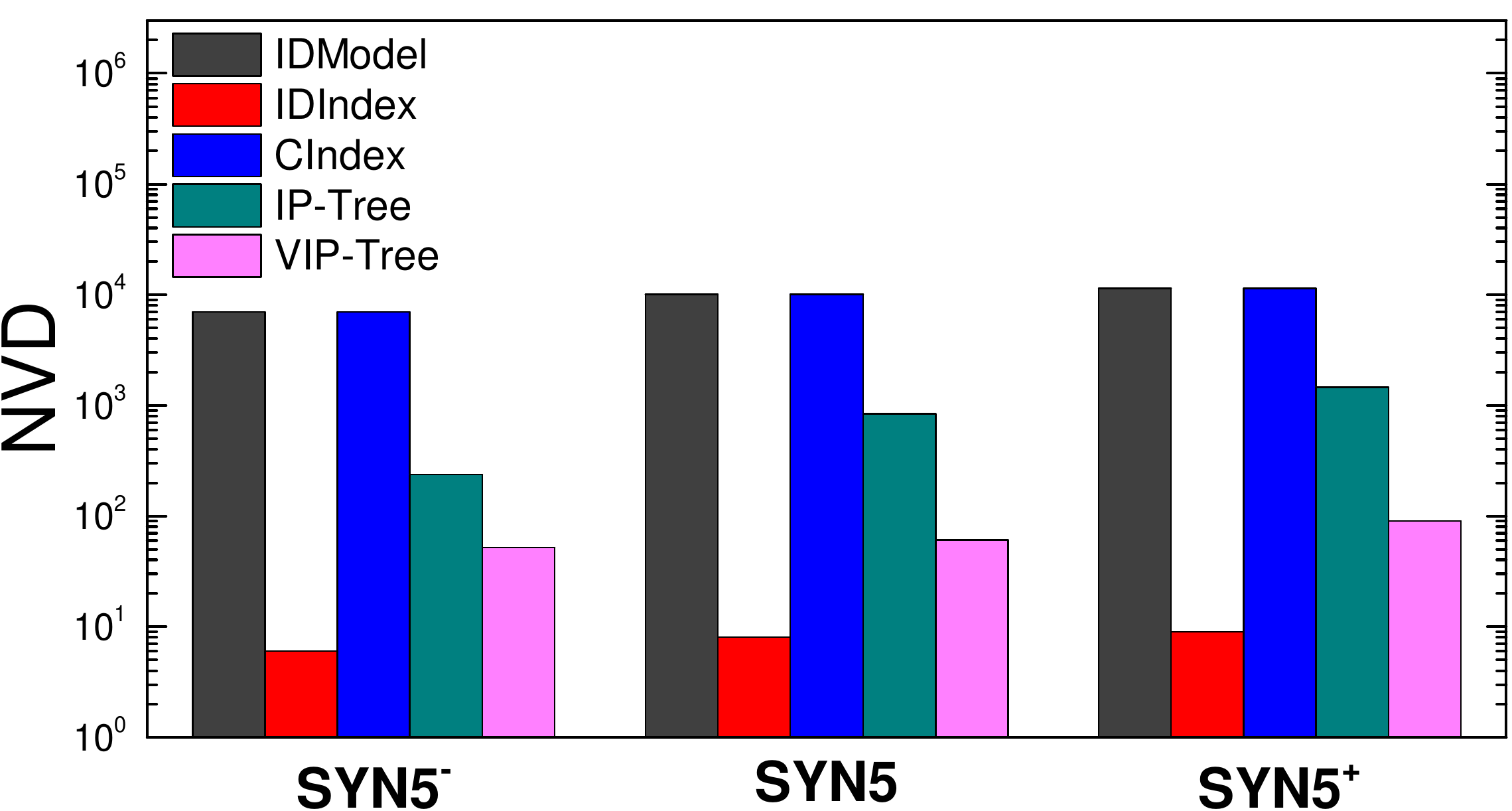}
    \ExpCaption{NVD in \textsf{SPDQ} \\vs. Topology}\label{fig:B6_SPDQ_nvd}
    \end{minipage}
\end{figure*}

\begin{figure*}[!htbp]
    \begin{minipage}[t]{0.24\textwidth}
    \centering
    \includegraphics[width=\textwidth, height = 3cm]{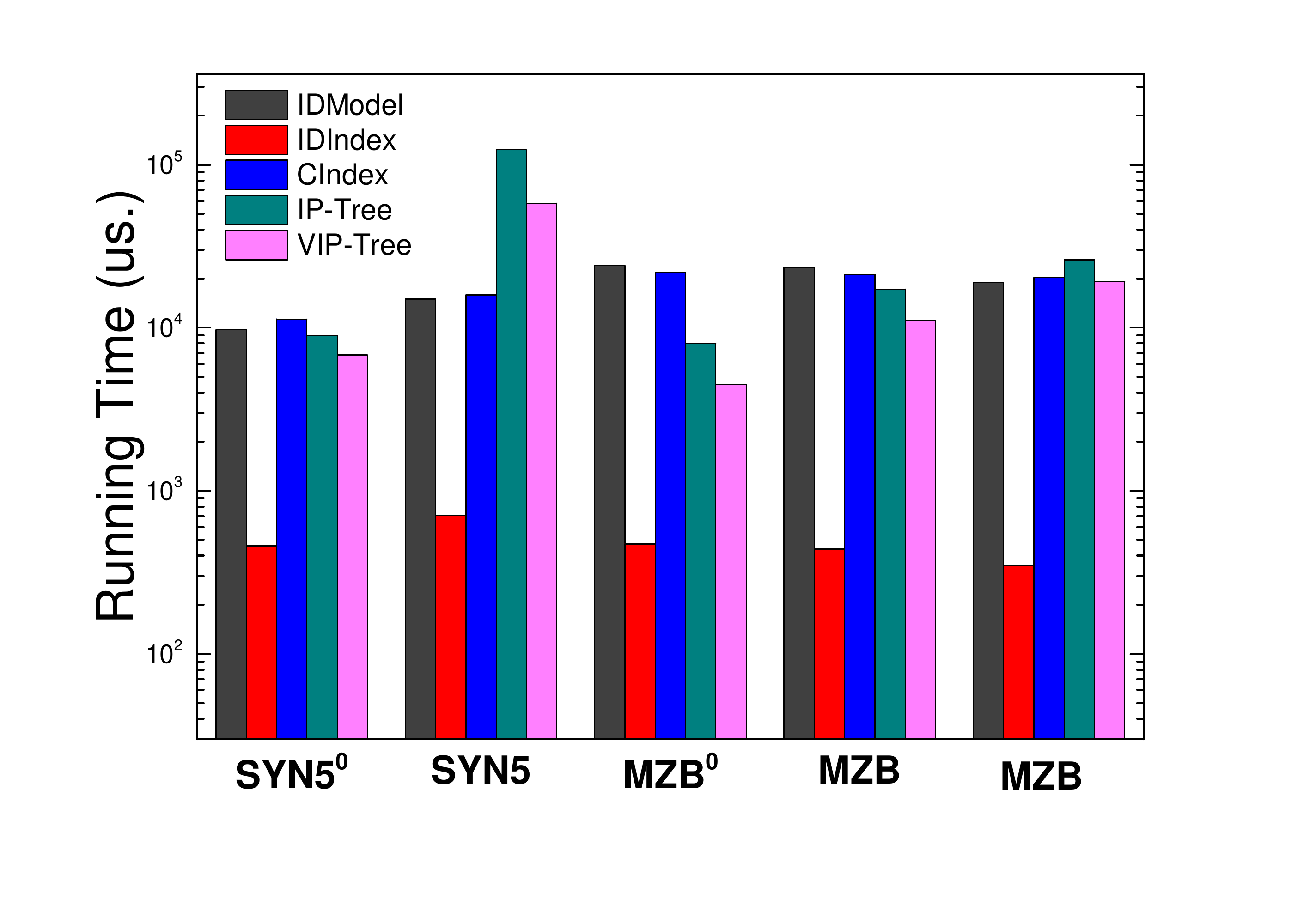}
    \ExpCaption{\textsf{RQ} Time \\vs. Decomposition}\label{fig:B7_RQ_time}
    \end{minipage}
    \begin{minipage}[t]{0.24\textwidth}
    \centering
    \includegraphics[width=\textwidth, height = 3cm]{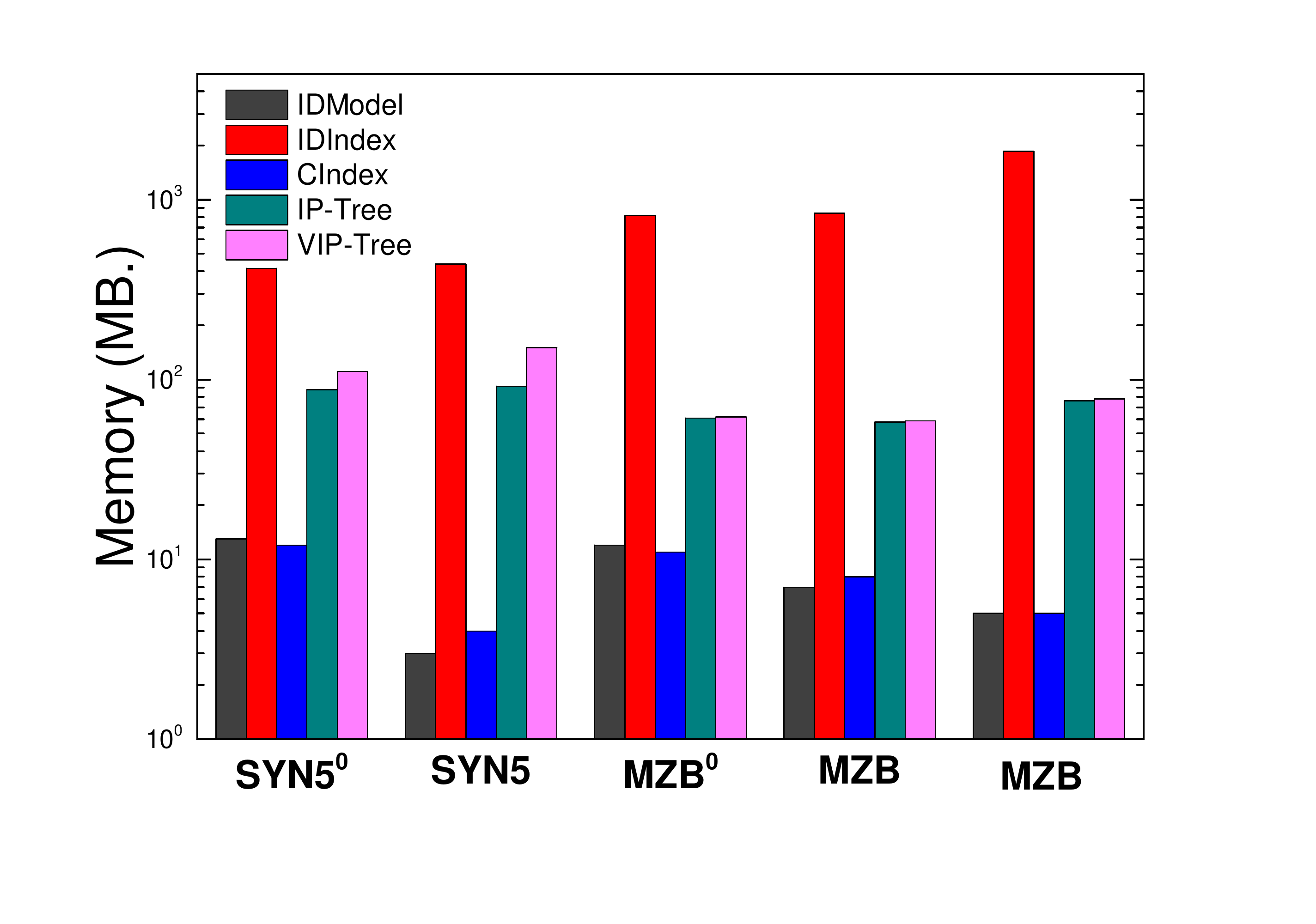}
    \ExpCaption{\textsf{RQ} Memory \\vs. Decomposition}\label{fig:B7_RQ_mem}
    \end{minipage}
    \begin{minipage}[t]{0.24\textwidth}
    \centering
    \includegraphics[width=\textwidth, height = 3cm]{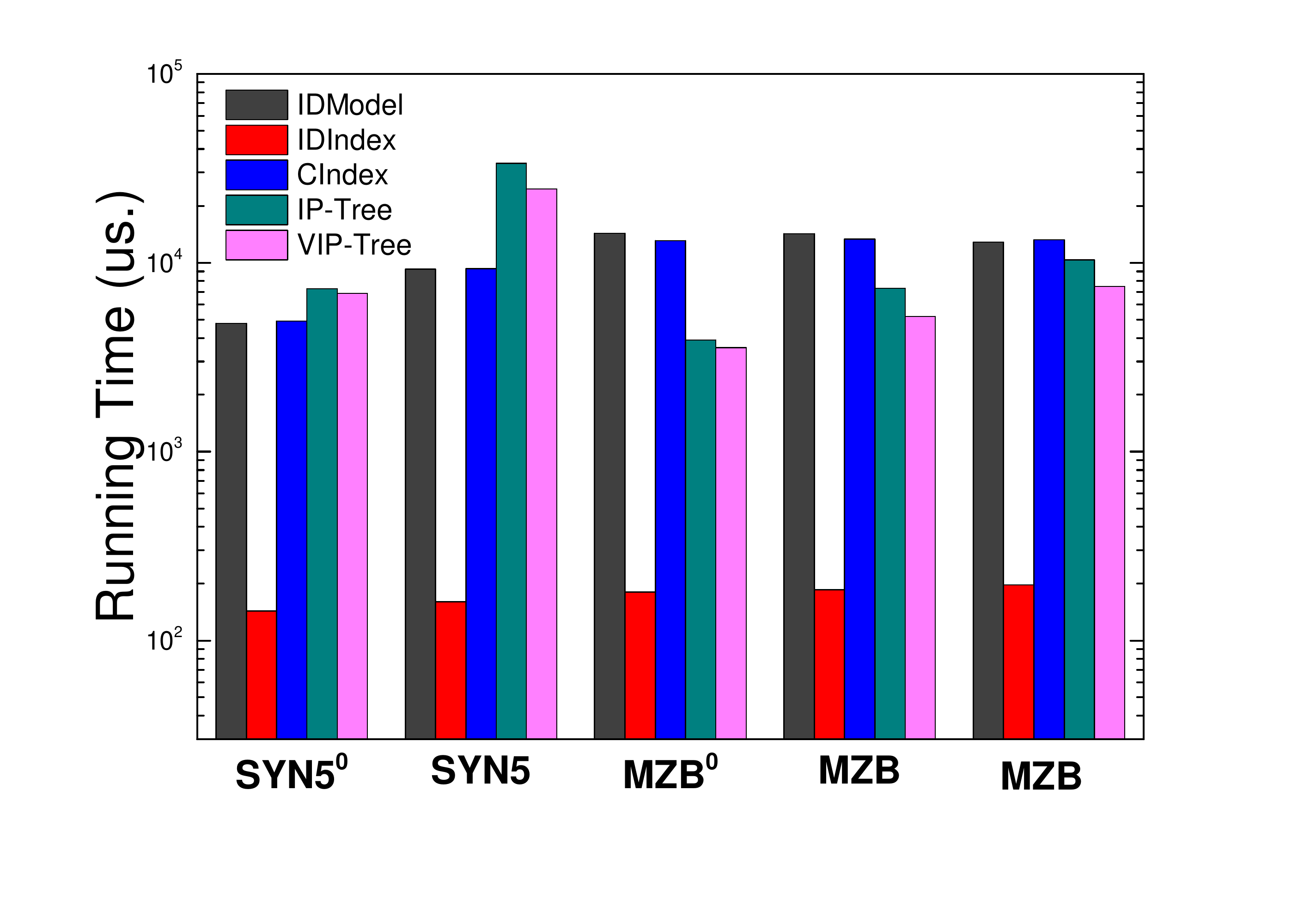}
    \ExpCaption{$k$\textsf{NNQ} Time \\vs. Decomposition}\label{fig:B7_KNNQ_time}
    \end{minipage}
    \begin{minipage}[t]{0.24\textwidth}
    \centering
    \includegraphics[width=\textwidth, height = 3cm]{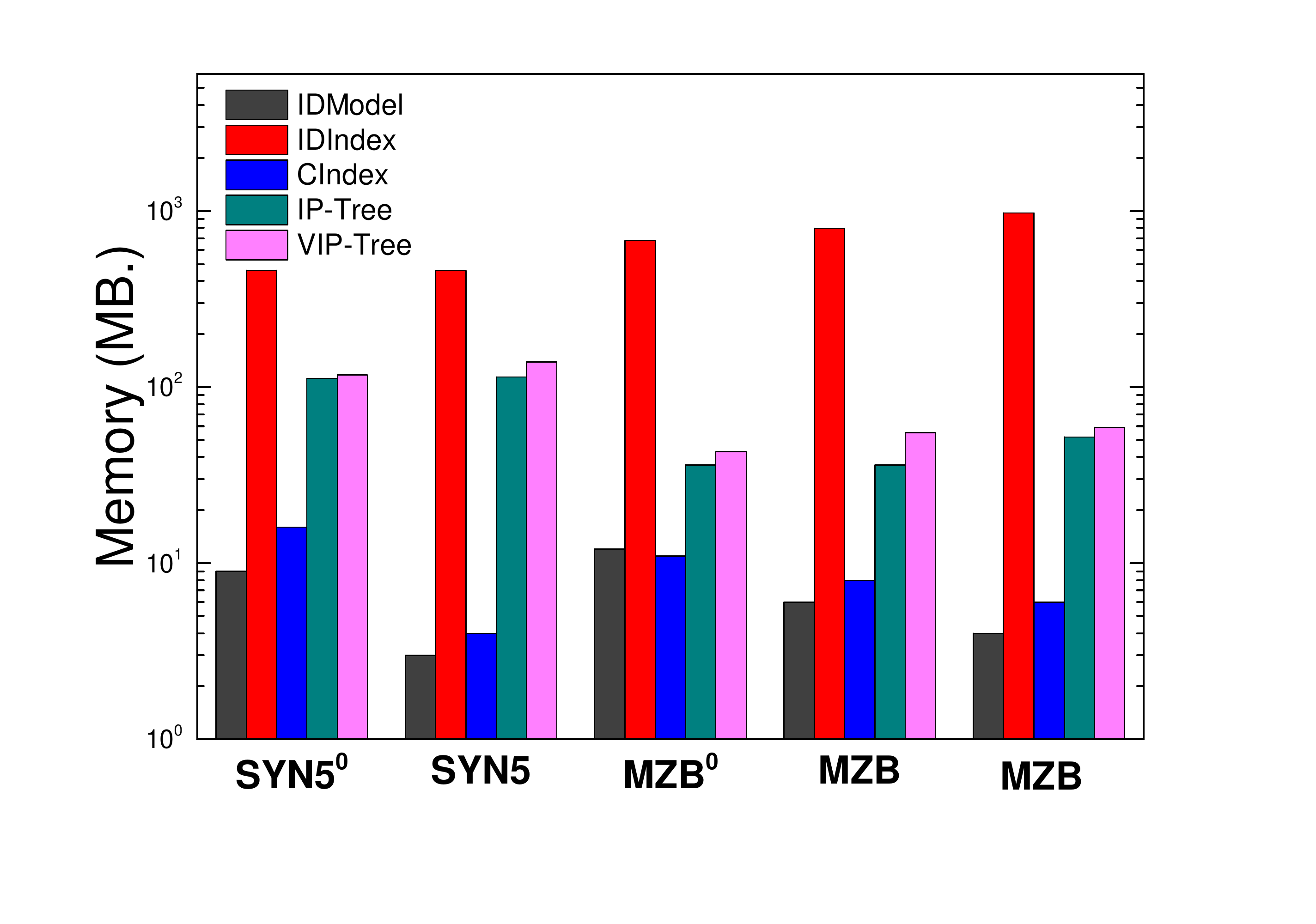}
    \ExpCaption{$k$\textsf{NNQ} Memory \\vs. Decomposition}\label{fig:B7_KNNQ_mem}
    \end{minipage}
\end{figure*}

\begin{figure*}[!htbp]
    \begin{minipage}[t]{0.328\textwidth}
    \centering
    \includegraphics[width=\textwidth, height = 3cm]{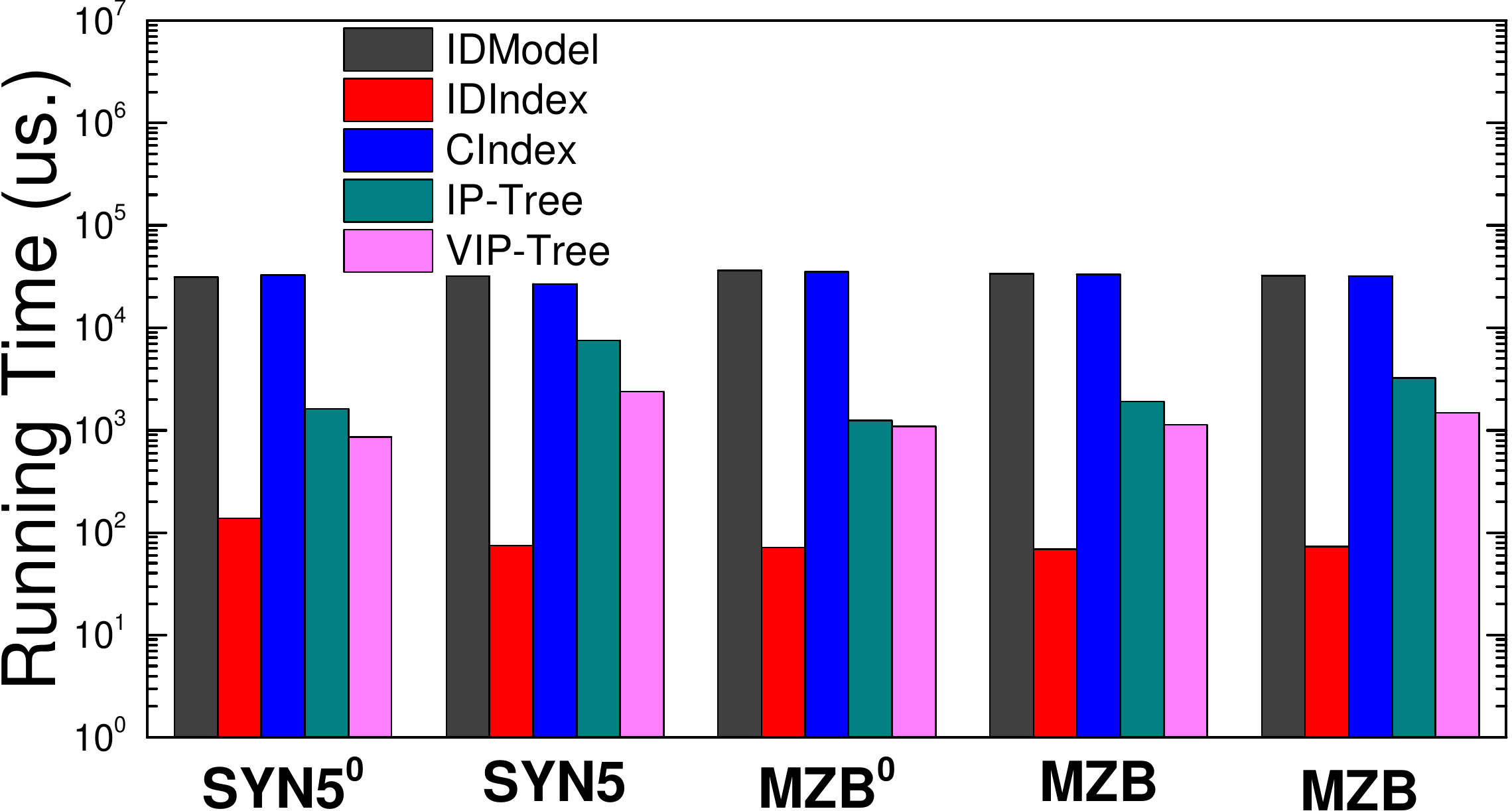}
    \ExpCaption{\textsf{SPDQ} Time \\vs. Decomposition}\label{fig:B7_SPDQ_time}
    \end{minipage}
    \begin{minipage}[t]{0.328\textwidth}
    \centering
    \includegraphics[width=\textwidth, height = 3cm]{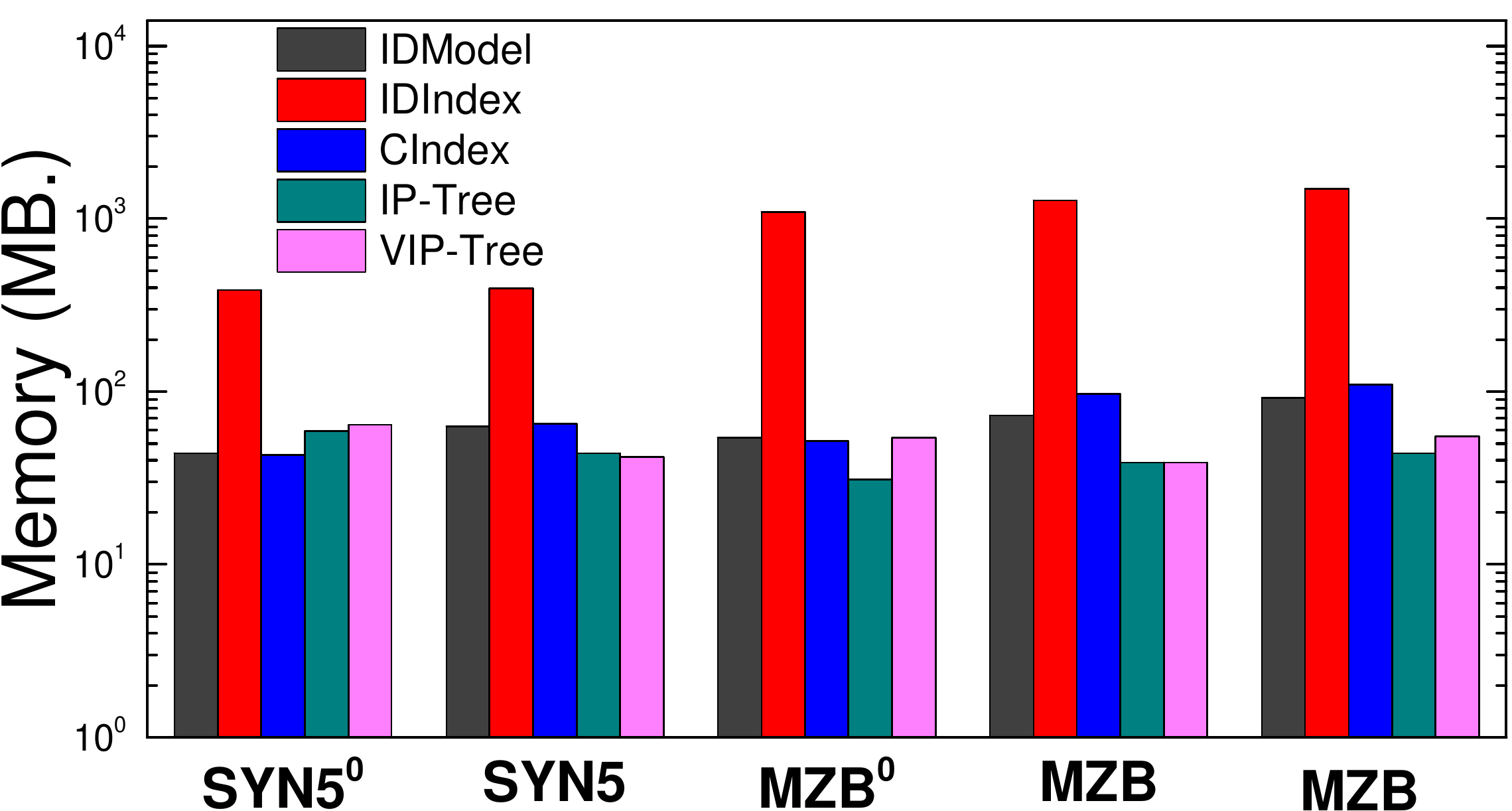}
    \ExpCaption{\textsf{SPDQ} Memory vs. Decomposition}\label{fig:B7_SPDQ_mem}
    \end{minipage}
    \begin{minipage}[t]{0.328\textwidth}
    \centering
    \includegraphics[width=\textwidth, height = 3cm]{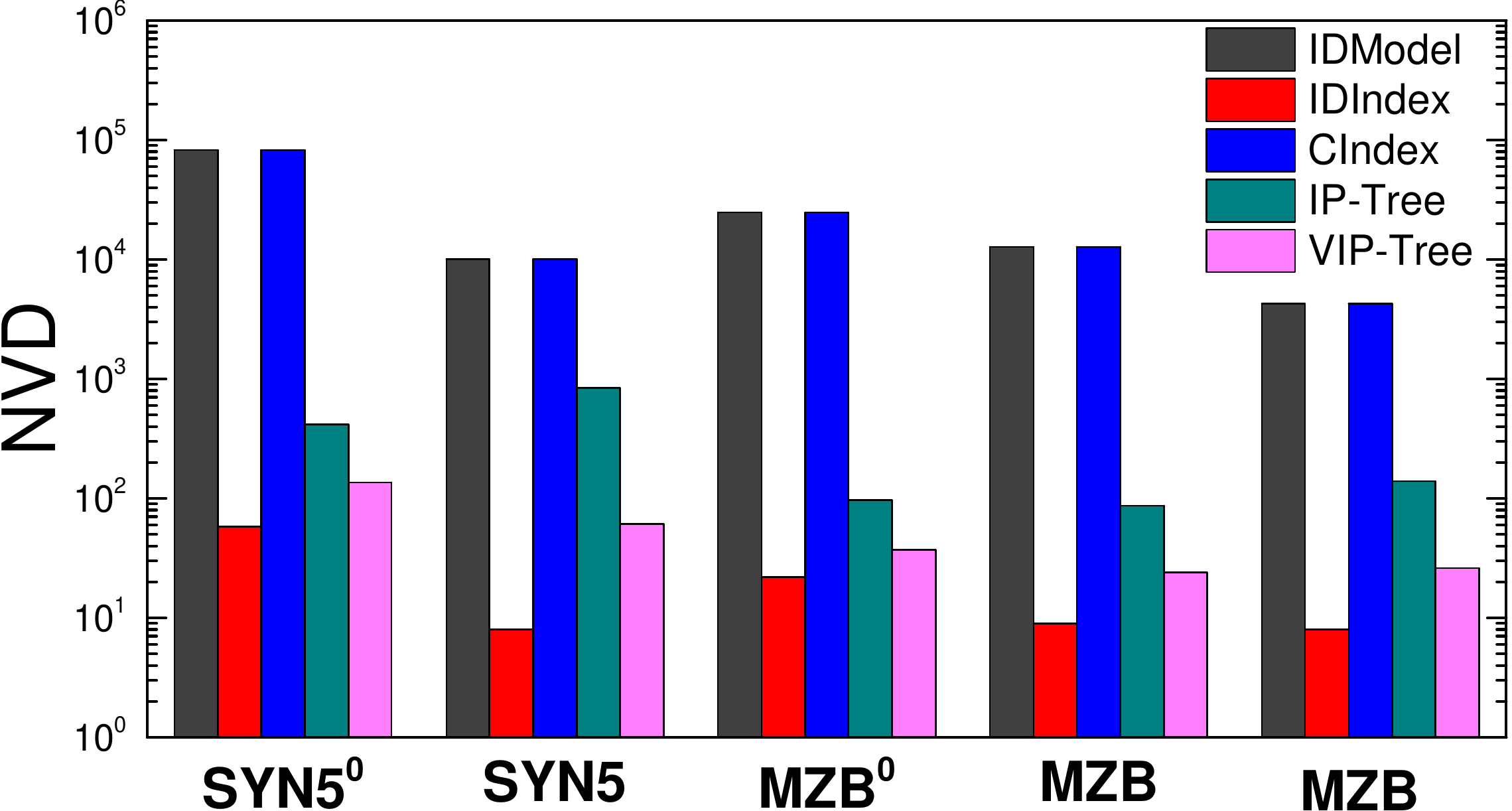}
    \ExpCaption{NVD in \textsf{SPDQ} \\vs. Decomposition}\label{fig:B7_SPDQ_nvd}
    \end{minipage}
\end{figure*}

\ExpHead{B6 Effect of Topological Change}

\noindent\textsf{RQ} and $k$\textsf{NNQ}: The time cost and memory use with respect to different topology characteristics are reported in Figures~\ref{fig:B6_RQ_time}, \ref{fig:B6_RQ_mem}, \ref{fig:B6_KNNQ_time}, and \ref{fig:B6_KNNQ_mem}, respectively.
\begin{itemize}[leftmargin=*]
\item Regarding the time cost, \textsc{IDIndex} runs fastest, but it needs large memory to store the door-to-door distance matrix. With increasing number of doors, its time cost and memory use increase steadily.
\item \textsc{IDModel} and \textsc{CIndex} use the smallest memory when processing \textsf{RQ} and $k$\textsf{NNQ}. Regarding the time cost, they perform medium. When the topology becomes more complex, the memory use keeps stable and the time cost increases slightly.
\item \textsc{IP-Tree} and \textsc{VIP-Tree} cost more time to process \textsf{RQ} and $k$\textsf{NNQ}. Moreover, when the topology  becomes more complex, the time cost rises rapidly. E.g., \textsf{RQ}'s time cost using \textsc{IP-Tree} grows nearly 20 times from SYN5$^-$ to SYN5$^+$ (see Figure~\ref{fig:B6_RQ_time}).
\end{itemize}

\noindent\textsf{SPDQ}: The time cost, memory use and NVD with respect to different topology characteristics are reported in Figures~\ref{fig:B6_SPDQ_time}, \ref{fig:B6_SPDQ_mem}, and \ref{fig:B6_SPDQ_nvd}, respectively.
\begin{itemize}[leftmargin=*]
\item Like in the other cases, \textsc{IDIndex} performs best in terms of the time cost but costs most memory compared with others. When the topology becomes complex, \textsc{IDIndex}'s time cost increases relatively slightly, while the memory use grows fast.
\item Considering time cost and memory use, \textsc{IP-Tree} and \textsc{VIP-Tree} perform best with relatively less time cost and smaller memory use. Regarding the time cost, \textsc{VIP-Tree} is always better than \textsc{IP-Tree} because of the extra precomputation, but it needs more memory. With the doors increasing, the time cost and memory use rise slightly.
\item \textsc{IDModel} and \textsc{CIndex} performs worst in both time and memory costs because they have to visit many doors in search.
\end{itemize}

\ExpHead{B7 Effect of Decomposition Methods for Hallways}

\noindent\textsf{RQ}, $k$\textsf{NNQ} and \textsf{SPDQ}: Regarding \textsf{RQ} and $k$\textsf{NNQ}, the time cost and memory use of different decomposition methods are reported in Figures~\ref{fig:B7_RQ_time}, \ref{fig:B7_RQ_mem}, \ref{fig:B7_KNNQ_time}, and \ref{fig:B7_KNNQ_mem}. For \textsf{SPDQ}, the time cost, memory use and NVD with respect to different decomposition methods are reported in Figures~\ref{fig:B7_SPDQ_time}, \ref{fig:B7_SPDQ_mem}, and \ref{fig:B7_SPDQ_nvd}.
\begin{itemize}[leftmargin=*]
\item \textsc{IDIndex} runs fastest when processing \textsf{RQ} and $k$\textsf{NNQ} but uses most memory. When hallways are decomposed into more partitions, \textsc{IDIndex}'s time cost keeps nearly stable but its memory cost increases. This is because there are more doors connecting increased numbers of partitions, which leads to more door-to-door pairs stored in the distance matrix.
\item \textsc{IDModel} and \textsc{CIndex} use the least memory but runs slowest. With more partitions, both time cost and memory use decrease because hallways are decomposed into more partitions each having less doors to process.
\item \textsc{IP-Tree} and \textsc{VIP-Tree} perform best considering both time cost and memory use. However, when hallways are decomposed into more partitions, the two methods need more time and memory to process \textsf{RQ} and $k$\textsf{NNQ}. Regarding the performance in \textsf{RQ}, \textsc{IP-Tree} and \textsc{VIP-Tree} cost more time than \textsc{IDModel}. There are more nodes in \textsc{IP-Tree} and \textsc{VIP-Tree} when hallways are decomposed into more partitions, which entails more on-the-fly computations to prune tree nodes when processing \textsf{RQ} and $k$\textsf{NNQ}. Moreover, the time cost of \textsc{IP-Tree} and \textsc{VIP-Tree} rises faster when processing \textsf{RQ} and $k$\textsf{NNQ} than processing \textsf{SPDQ}. That is because there is some extra cost to prune nodes when processing \textsf{RQ} and $k$\textsf{NNQ}. As the nodes increase, this extra cost increases fast.
\end{itemize}

\ExpHead{Summary of Findings}
%
\begin{itemize}[leftmargin=*]
\setlength\itemsep{0em}
\item \textsc{IDModel} requires minimum time and space costs in construction. Moreover, it works well for \textsf{RQ} and $k$\textsf{NNQ} due to its good balance in execution time and memory use. However, it is inferior to \textsc{IP-Tree} and \textsc{VIP-Tree} for \textsf{SPQ} and \textsf{SDQ}.
The more partitions the hallways are decomposed into, the better \textsc{IDModel} performs, while \textsc{IP-Tree} and \textsc{VIP-Tree} the worse.
\item \textsc{IDIndex} runs fastest for all types of indoor spatial queries. However, it requires significantly large time to construct offline and high memory consumptions during query processing. In general, it is suitable for indoor spaces with relatively small numbers of doors.
\item \textsc{CIndex} is intended for handling indoor moving objects and therefore it performs only comparably to \textsc{IDModel} when processing the four queries on static space and objects.
\item \textsc{IP-Tree} and \textsc{VIP-Tree} are optimized for \textsf{SPQ}/\textsf{SDQ} tasks.
In particular, their advantages stand out when the indoor space accommodates a certain number of crucial partitions that are connected by so-called access doors.
\textsc{VIP-Tree} gains better search efficiency than \textsc{IP-Tree} at the cost of extra storage overhead for precomputed local distances.
Their construction costs are in the middle among all five proposals under investigation.
\end{itemize}


In summary, \textsc{IDIndex} is preferred for small-scale spaces. \textsc{VIP-Tree} is recommended if routing is the task or the space accommodates crucial partitions.
For other scenarios, \textsc{IDModel} is recommended due to its low construction cost and good balance in terms of storage and query time costs.

\section{Extensibility Analysis}
\label{sec:extensibility}

Table~\ref{tab:extensibility} summarizes the extensibility of all model/indexes.


\begin{table}[ht]
\centering
\caption{Extensibility Analysis}\label{tab:extensibility}
\scriptsize
\begin{tabular}{|c|cccc|} 
\hline
                 & \textsc{IDModel} & \textsc{IDIndex} & \textsc{CIndex} & \textsc{IP/VIP-Tree} \\
\hline
\tabincell{c}{Temporal Variation} & \checkmark & \text{\sffamily X} & \checkmark & \text{\sffamily X} \\
\hline
\tabincell{c}{Moving Objects} & \checkmark & \checkmark & \checkmark & \checkmark \\
\hline
\tabincell{c}{Uncertain Locations}  & \text{\sffamily X} & \text{\sffamily X} & \checkmark & \text{\sffamily X} \\
\hline
Keywords & \checkmark  & \checkmark & \checkmark & \checkmark \\
\hline
\end{tabular}
\end{table}

\noindent\textbf{Temporal Variation.} Indoor topology may feature temporal variations, e.g., doors may have open and close hours. To support indoor spatial queries in such cases, temporal variation information like open and close time for doors can be maintained as a table attached to the accessibility base graph of \textsc{IDModel} or the topological layer of \textsc{CIndex}~\cite{DBLP:conf/icde/2020}.
However, frequent temporal variations are very hard to handle for \textsc{IDIndex} and \textsc{IP-Tree}/\textsc{VIP-Tree} because they need to precompute door-to-door distances globally or locally.


\noindent\textbf{Moving Objects.}
All model/indexes can index moving objects by maintaining dynamic object buckets attached to indoor partitions in the way similar to how we handle the static objects. Nevertheless, the buckets need to be updated appropriately for indoor moving objects.

\noindent\textbf{Uncetain Locations.}
In some settings, indoor points or objects are represented as uncertain regions.
To process indoor spatial queries over uncertain locations, a model/index should support geometric operations on partitions.
As a result, only \textsc{CIndex} with partition R*-tree excels at handling uncertain locations~\cite{Xie2013, Xie2015}.

\noindent\textbf{Keywords.} A spatial keyword query~\cite{Chen2013} returns objects or paths that are spatially and textually relevant to user-specified location(s) and keyword(s).
Such queries can be supported if we extend the model/indexes by additionally maintaining mappings between partitions/objects and keywords.
%
Specially, top-$k$ keyword-aware shortest path queries have been supported based on \textsc{IDModel}~\cite{Feng2019}, and boolean $k$NN spatial keyword queries have been supported based on \textsc{VIP-Tree}~\cite{Shao2020}.

\vspace*{-5pt}
\section{Conclusion and Future Work} 
\label{sec:summary}

This work reports on an extensive experimental evaluation of five indoor space model/indexes that support four typical indoor spatial queries, namely range query (\textsf{RQ}), $k$ nearest neighbor query ($k$\textsf{NNQ}), shortest path query (\textsf{SPQ}), and shortest distance query (\textsf{SDQ}). Our evaluation concerns the costs in model/index construction and query processing using a model/index. By analyzing the results, we summarize the pros and cons of all techniques and suggest the best choice for typical scenarios.

For future work, an interesting direction is to consider more semantic information in the model/indexes. It is also interesting to adapt the existing model/indexes to handle indoor moving objects and support more query types on moving objects.

\begin{spacing}{0.8}
\bibliographystyle{plain}
\bibliography{main}
\end{spacing}

\begin{appendix}
\section{APPENDIX}


\textsf{RQ} based on \textsc{IDModel} is formalized in Algorithm~\ref{alg:RQIDModel}, generally based on Dijkstra's algorithm.
The function $\textit{rangeSearch}$ $(B, p, r)$ searches an object bucket $B$ for those objects each having its distance from $p$ not larger than $r$~\cite{Lu2012}.
Instead of keeping the last-hop door, we maintain last-hop partition for each door in an array $\mathit{prev}[]$, as we assume that a door always connects two partitions.\\

\begin{algorithm}[!htbp]
\caption{{RQ\_IDModel}(Point $p$, distance $r$)} \label{alg:RQIDModel}
\begin{algorithmic}[1]
\State $v \gets \textit{getHostPartition}(p)$; $B_v \gets \textit{getObjectBucket}(v)$
\State $O^{*} \gets \textit{rangeSearch}(B_v, p, r)$
\State initialize distance array $\mathit{dist}[]$ for all doors
\State initialize last-hop partition array $\mathit{prev}[]$ for all doors
\For {each door $d \in \mathit{P2D}_\sqsubset(v)$}
	\State $\mathit{dist}[d] \gets ||p, d||_{v}$
\EndFor
\State initialize a min-heap $H$
\For {each door $d_i \in D$}
	\If {$d_i \notin \mathit{P2D}_\sqsubset(v)$}
		$\mathit{dist}[d_i] \gets \infty$; $\mathit{prev}[d_i] \gets v$
	\EndIf
	\State \text{enheap}($H$, $\left \langle d_i, \mathit{dist}[d_i] \right \rangle$); $\mathit{prev}[d_i] \gets$ \textit{null}
\EndFor
\While {$H$ is not empty}
	\State $\left \langle d_i, \mathit{dist}[d_i] \right \rangle \gets$ \text{deheap}($H$)
	\If {$\mathit{dist}[d_i] > r$}
		\textbf{return} $O^{*}$
	\EndIf
	\For {each partition $v_i \in \mathit{D2P}_\sqsupset(d_j)$ $\wedge$ $v_i \neq \mathit{prev}[d_i]$}
		\State $r \gets r - \mathit{dist}[d_i]$; $B_{i} \gets \textit{getObjectBucket}(v_i)$
		\State $O_i \gets \textit{rangeSearch}(B_{i}, d_i, r)$; add $O_i$ to $O^{*}$
	\EndFor
	\State mark $d_i$ and $v_i$ as visited
	\For {each unvisited door $d_j \in \mathit{P2D}_\sqsubset(v_i)$}
		\State $v'_i \gets \mathit{D2P}_\sqsupset(d_j) \backslash v_i$; $\mathit{dist}_j \gets \mathit{dist}[d_i] + f_\text{d2d}(v_i,d_i,d_j)$
		\If {$\mathit{dist}_j < \mathit{dist}[d_j]$}
			\State $\mathit{dist}[d_j] \gets \mathit{dist}_j$; \text{enheap}($H$, $\left \langle d_j, \mathit{dist}[d_j] \right \rangle$); $\mathit{prev}[d_j] \gets v_i$
		\EndIf
	\EndFor
\EndWhile
\State \textbf{return} $O^{*}$
\end{algorithmic}
\end{algorithm}


Algorithm~\ref{alg:kNNQIDModel} processes $k$\textsf{NNQ} based on \textsc{IDModel}.
A function $\textit{knnSearch}(O, B, p, k, \textit{kBound})$ is designed to update the top-$k$ result set $O^{*}$ with objects from a bucket $B$.
$\textit{kBound}$ is used to prune unpromising objects and is updated after a calling of $\textit{knnSearch}$.\\

\begin{algorithm}[ht]
\caption{{$k$NNQ\_IDModel}(Point $p$, integer $k$)} \label{alg:kNNQIDModel}
\begin{algorithmic}[1]
\State $v \gets \textit{getHostPartition}(p)$; $B_v \gets \textit{getObjectBucket}(v)$; $\textit{kBound} \gets \infty$
\State $O^{*} \gets \varnothing$; $\textit{knnSearch}(O^{*}, B_v, p, k, \textit{kBound})$
\State initialize distance array $\mathit{dist}[]$ for all doors
\State initialize last-hop partition array $\mathit{prev}[]$ for all doors
\For {each door $d \in D$}
	\State $\mathit{dist}[d] \gets ||p, d||_{v}$
\EndFor
\State initialize a min-heap $H$
\For {each door $d_i \in D$}
	\If {$d_i \notin \mathit{P2D}_\sqsubset(v)$}
		$\mathit{dist}[d_i] \gets \infty$; $\mathit{prev}[d_i] \gets v$
	\EndIf
	\State \text{enheap}($H$, $\left \langle d_i, \mathit{dist}[d_i] \right \rangle$); $\mathit{prev}[d_i] \gets$ \textit{null}
\EndFor
\While {$H$ is not empty}
	\State $\left \langle d_i, \mathit{dist}[d_i] \right \rangle \gets$ \text{deheap}($H$)
	\If {$\mathit{dist}[d_i] > \textit{kBound}$}
		\textbf{return} $O^{*}$
	\EndIf
	\For {each partition $v_i \in \mathit{D2P}_\sqsupset(d_j)$ $\wedge$ $v_i \neq \mathit{prev}[d_i]$}
    \State $B_{i} \gets \textit{getObjectBucket}(v_i)$; $\textit{knnSearch}(O^{*}, B_{i}, d_i, k, \textit{kBound})$
	\EndFor
	\State mark $d_i$ and $v_i$ as visited
	\For {each unvisited door $d_j \in \mathit{P2D}_\sqsubset(v_i)$}
		\State $v'_i \gets \mathit{D2P}_\sqsupset(d_j) \backslash v_i$; $\mathit{dist}_j \gets \mathit{dist}[d_i] + f_\text{d2d}(v_i,d_i,d_j)$
		\If {$\mathit{dist}_j < \mathit{dist}[d_j]$}
			\State $\mathit{dist}[d_j] \gets \mathit{dist}_j$; \text{enheap}($H$, $\left \langle d_j, \mathit{dist}[d_j] \right \rangle$); $\mathit{prev}[d_j] \gets v_i$
		\EndIf
	\EndFor
\EndWhile
\State \textbf{return} $O^{*}$
\end{algorithmic}
\end{algorithm}

\vfill

\end{appendix}

\balance

\end{document}